\documentclass[12pt]{article}
\usepackage{booktabs}
\usepackage[margin=2cm]{geometry}
\usepackage{amsmath,array}
\usepackage{arydshln}
\usepackage[T1]{fontenc}
\usepackage{lmodern,microtype}
\usepackage{amsmath,amssymb,amsthm,bm,mathtools}
\usepackage{thm-restate,booktabs,tabularx,enumitem}
\usepackage[hidelinks]{hyperref}
\usepackage{thm-restate}
\usepackage{xcolor}
\usepackage{etoolbox}
\usepackage{tikz}
\usepackage{graphicx}
\usetikzlibrary{arrows.meta}
\usetikzlibrary{positioning,fit,decorations.pathreplacing}

\newcommand{\Q}{\mathbb Q}

\newcommand{\N}{\mathbb N}
\newcommand{\Z}{\mathbb Z}

\newcommand{\gl}{\mathrm{GL}}
\DeclareMathOperator{\rk}{rank}
\DeclareMathOperator{\im}{im}

\DeclareMathOperator{\End}{End}

\DeclareMathOperator{\mat}{Mat}

\newcommand{\cl}[1]{\overline{#1}}

\newtheorem{lemma}{Lemma}

\newtheorem{claim}{Claim}[section]
\theoremstyle{definition}

\declaretheorem[
  name=Example,
  qed=\ensuremath{\clubsuit}
]{example}

\title{Burnside Problems for Matrix Semigroup Quotients\\  with Applications to Border Complexity}

\author{
\begin{tabular}{cc}
  \begin{tabular}{c}
    Roland Guttenberg\\
   University of Warsaw, Poland  \\
  \end{tabular}
  &
  \begin{tabular}{c}
    Nathan Lhote\\
    Aix-Marseille Universit\'e, LIS, France  \\
  \end{tabular}
  \\[1.5em]
  \begin{tabular}{c}
    Mahsa Shirmohammadi\\
    CNRS, IRIF, France
  \end{tabular}
  &
  \begin{tabular}{c}
    James Worrell\\
    University of Oxford, UK
  \end{tabular}
\end{tabular}
}

\date{}

\begin{document}

\maketitle

\begin{abstract}
The  Burnside theorem for matrix semigroups states
that a finitely generated matrix semigroup is finite whenever each element has only finitely many distinct powers. 
We study analogous finiteness questions for quotients of matrix semigroups by congruences, which need not themselves admit faithful finite-dimensional matrix representations.
We show three  finiteness criteria based on uniform bounds on the quotient images of (rank-)stable subsemigroups and groups in the Zariski closure.
Our proofs use exterior algebra and the Ramsey theorem to 
develop variations of Simon's factorisation forest to matrix semigroups modulo congruences.
We also present applications of our results to border complexity.
One such involves separating homogeneous algebraic branching programs of fixed width and degree from its Zariski closure.

 \vspace{3em}
\textbf{AI Use: }  Discussions with GPT-6 Astra suggested the explicit polynomial used to separate homogeneous algebraic branching programs (ABPs) of fixed width and degree from the Zariski closure of this class.
This polynomial was found after the 
matrix encoding of polynomials through translations of universal ABPs and our previously established variants  of Burnside Theorems for semigroup congruences  were given to the model.  The proofs of Theorem~\ref{theorem2} and Corollary~\ref{cor:closed-torsion-quotient} were developed through an interaction with GPT-6.
We also used GPT-6 to help develop several examples.

\vspace{3em}
\textbf{Funding: } 
 Shirmohammadi is supported by the ERC Synergy grant Project 101224640 (VePaSS). Worrell is supported by EPSRC Fellowship EP/X033813/1. Guttenberg is supported by the ERC Consolidator grant Polaris 101231447.
\end{abstract}

\newpage
\section{Introduction}

    

 In 1902, Burnside asked whether a finitely generated group is necessarily finite if every element has finite order, that is, if some positive power of each element equals the identity. 
 Novikov and Adian showed in~1968 that the answer is
negative even when the orders of all elements are
bounded by a common constant~\cite{NovikovAdian68}. 
 By contrast, Schur proved that the answer is positive for groups of invertible matrices over the complex numbers~\cite{SchurPeriodic11}.
 Kaplansky generalised Schur's result from the complex numbers to arbitrary fields (see~\cite{kaplansky1972fields}).
 McNaughton and Zalcstein further extended this conclusion to matrix semigroups over arbitrary fields~\cite{McNaughtonZalcstein75}. 
 They proved that a finitely generated matrix semigroup is finite whenever each element is torsion, meaning that its powers take only finitely many distinct values. 
Among other things, this shows that infiniteness of finitely generated matrix semigroups is semi-decidable (a witness of infiniteness is a matrix of infinite order) and hence
finiteness of finitely generated matrix semigroups is decidable.  

The positive answer to the Burnside problem for matrix groups generalises to quotients of matrix groups by Zariski closed normal subgroups, since 
such quotients can be embedded in matrix groups in higher ambient dimension. 
This finiteness result is a key ingredient in the first proposed
algorithm for computing the Zariski closure of a finitely
generated matrix group~\cite{DerksenJeandelKoiran05},
resolving an open problem in quantum computation.
In subsequent joint work, again relying on such a Burnside finiteness theorem, we obtained an elementary algorithm for computing this closure~\cite{NosanEtAl22}.  

In this paper, we study Burnside-type questions for quotients of matrix semigroups.
Our approach is based on congruences, that is, equivalence relations of matrix semigroups that are compatible with matrix multiplication.
Given a finite set $X$ of rational matrices and a congruence~$\sim$ on the semigroup $\langle X\rangle$ they generate, we ask 
for sufficient conditions that 
\begin{itemize}
    \item  the quotient $\langle X\rangle/{\sim}$ be finite, and
    \item there be a uniform bound on the length of the shortest product of generators from $X$ representing each equivalence class.
\end{itemize} 
The classical results on the Burnside problem for matrix semigroups 
do not apply here, since the quotient need not admit a faithful finite-dimensional 
representation in a monoid of matrices with entries in a field.\footnote{The matrices $A=\begin{pmatrix} 1 & 1 \\0 &1 \end{pmatrix}$ and
$B=\begin{pmatrix}
    1&0\\1&1 
\end{pmatrix}$ generate a free monoid.  For the congruence $\sim$ generated by the relation $AB\sim I$ we have $BA\not\sim I$.  Then, in the quotient $\langle A,B\rangle /{\sim}$, the image of $B$ is a right inverse of the image of $A$, but not a left inverse.  Hence
the quotient is not embeddable as a submonoid of a monoid of square matrices with entries in a field.}
We establish three quantitative finiteness criteria for such quotients and present two applications to border complexity.

Border complexity, also known as approximate complexity, studies  polynomials and tensors obtained as  limits  of
parametrised families of  bounded complexity.
The distinction between exact  and approximate computability is central to geometric complexity theory, where questions about computational power are expressed through polynomial and tensor orbit closures~\cite{MulmuleySohoni01,BlaserIkenmeyer25,dutta2026border}.
 This distinction is already visible  in small-width  algebraic branching programs, as approximation can  increase the expressive power in these models~\cite{BringmannIkenmeyerZuiddam18}.
Our two applications study these distinctions through
semigroup congruences:
\begin{enumerate}
    \item First, for homogeneous algebraic branching programs, we separate the class of polynomials computable by algebraic branching programs of fixed width and degree from its Zariski closure.
    \item Second, we construct a finitely generated matrix monoid under whose action the orbit closure of the permanent is not contained in that of the determinant. Curiously, under the congruence of  mutual reducibility,
    the orbit closure of the determinant has a finite quotient whereas that of the permanent is not finite. 
\end{enumerate}
 Both of these applications use the fact that a finitely generated matrix semigroup can have a finite quotient under a congruence, while its Zariski closure  has infinitely many classes under the same congruence. 
These applications connect semigroup theory with border complexity and provide a new toolkit for studying a broader range of algebraic computational models.

The central tool in our proof is the notion of factorisation forest for matrix semigroups, which we developed in recent works~\cite{OurVASS26,Finiteness26}.  
 The original notion of factorization forest for semigroups
 was introduced by Imre Simon~\cite{Simon90} and was later used in~\cite{Simon1994} to 
 give an algorithm for deciding finiteness of matrix semigroups with entries in the tropical semiring.
The notion has become central in the algebraic theory of formal languages (see, for example,~\cite{bojanczyk9F,Colcombet21}) and has
played a  significant role in the resolution of a number of open problems (e.g., the proof of Courcelle's conjecture that
for a class of graphs of bounded treewidth a property is MSO-definable if and only if it is recognisable by a tree automaton~\cite{BojanczykP16}).
In particular, factorisation forests provide a means to translate results on groups to the setting of semigroups.
Our variants generalise Simon's factorisation from finite semigroups to (infinite) matrix semigroups~\cite{OurVASS26} and matrix semigroup quotients.
In our factorisation trees, the role of idempotent elements in the classical
version is replaced by (rank-)stable matrices and generalisations. 
We recently used one of these variants in the context of finite matrix semigroups, to give a polynomial-space algorithm for the finiteness problem for finitely generated matrix semigroups, improving a previous  NExp-upper bound. This was a corollary of our
 polynomial bit-size bound on matrix entries in finite monoids~\cite{Finiteness26}.

\subsection*{Our Contribution}
In this section we state our 
main technical contributions, which concern congruences on finitely generated semigroups $\langle X \rangle$.
Each result assumes a uniform bound $N$ on the number of equivalence classes of a certain family of \emph{group-like} sub-semigroups, and infers
a bound on the number of classes of the ambient semigroup. The bound  depends on $N$ and the ambient dimension $d$, with the first result additionally depending on the cardinality of the set $X$ of generators.  
There are three variants, according to the kind of sub-semigroups on which the uniform bound is applied.  Theorem~\ref{TheoremMainQuotient} considers \emph{subgroups} of $\overline{\langle X\rangle}$, 
Theorem~\ref{theorem-1} considers \emph{stable sub-semigroups} of $\langle X\rangle$ (that is, sub-semigroups in which all matrices share the same kernel and the same image that moreover have intersection zero),
finally Theorem~\ref{theorem2} considers \emph{rank-stable sub-semigroups} of $\langle X\rangle$, that is, sub-semigroups in which all matrices have the same rank.
Every subgroup of a matrix semigroup is a stable sub-semigroup and
a stable semigroup of rank $r$ is isomorphic to a sub-semigroup of the general linear group in dimension $r$.  
We state a Burnside theorem for quotients of matrix semigroups by closed congruences in Corollary~\ref{cor:closed-torsion-quotient}.

For positive integers $N,d,m$, define
\begin{equation*} F_d(N,m):=\bigl[m(N+1)^2\bigr]^{(2^{d+3}+1)^{d-1}}\,
\end{equation*}
This yields the asymptotic bound  $(m(N+1))^{2^{O(d^2)}}$, which is
polynomial in $m$ and $N$ for  fixed~$d$.
\begin{restatable}{theorem}{StableTheoremQuo}\label{theorem-1}
Let $X\subseteq\mat_d(\Q)$ be finite, and  $\sim$
be a congruence on $\langle X\rangle$.
Then $\langle X\rangle/{\sim}$ is finite if and only if there is
an $N\ge1$ such that $|S/{\sim}|\le N$ for every stable
subsemigroup $S\le\langle X\rangle$.
The cardinality and the diameter of
the quotient are at most $F_d(N,|X|)$.
\end{restatable}

We say that a congruence~$\sim$ defined on $\langle X\rangle$ is 
relatively Zariski closed in $\langle X\rangle\times\langle X\rangle$ when
 there exists a Zariski-closed set
$R\subseteq\mat_d(\overline{\Q})\times\mat_d(\overline{\Q})$
such that, for all $a,b\in\langle X\rangle$,
$a\sim b$ if and only if $(a,b)\in R$.
The following finiteness result for torsion quotients follows from Theorem~\ref{theorem-1}. 
This gives a Burnside theorem for quotients of matrix semigroups by closed congruences.

\begin{restatable}{corollary}{corclosedtorsionquotient}\label{cor:closed-torsion-quotient}
Let $X\subseteq\mat_d(\Q)$ be finite, and  $\sim$
be a congruence on $\langle X\rangle$.
Suppose that the relation~$\sim$ is
relatively Zariski closed in $\langle X\rangle\times\langle X\rangle$.
If $\langle X\rangle/{\sim}$ is torsion, then it is finite.
\end{restatable}

For positive integers $N,d$, define
\begin{equation*}
 G_d(N):=(N+1)(H+1)^d-1
 \ \le\ \min\bigl\{2^{O(d^2N)},\ (N+1)^{2^{O(d)}}\bigr\}.
\end{equation*}
where
\[
H:=\binom{N+D^2+1}{D^2}
 \qquad \text{and}\qquad 
 D:=\binom d{\lfloor d/2\rfloor}.
\]
\begin{restatable}{theorem}{RankCriterion}\label{theorem2}
Let $X\subseteq\mat_d(\Q)$ be finite  and  $\sim$
be a congruence on $\langle X\rangle$.
Then $\langle X\rangle/{\sim}$ is finite if and only if there is
an $N\ge1$ such that $|S/{\sim}|\le N$ for every rank-stable
subsemigroup $S\le\langle X\rangle$.
The diameter of
the quotient is at most $G_d(N)$.
\end{restatable}

The uniform boundedness assumption in Theorem~\ref{theorem2} is stronger than in Theorem~\ref{theorem-1}, since it applies to all rank-stable semigroups, not just stable semigroups.
Correspondingly, the theorem yields a better bound on the diameter of $\langle X \rangle / {\sim}$: while the bound in Theorem~\ref{theorem-1} is doubly exponential in $d^2$ 
and depends on the number $|X|$ of generators, the bound in Theorem~\ref{theorem2} is doubly exponential only in $d$ and is independent of the number of generators.

The next result,
Theorem~\ref{TheoremMainQuotient}, obtains a diameter bound 
on $\langle X \rangle / {\sim}$ that is only singly exponential in $d$,
but assumes that 
 the congruence~$\sim$
is defined on the  \emph{Zariski closure} of~$\langle X \rangle$, with
 a uniform bound on the number of classes of this closure met by each subgroup.

\begin{restatable}{theorem}{TheoremNameMainQuotient}
\label{TheoremMainQuotient}
Let $X\subseteq\mat_d(\Q)$ be finite and   $\sim$
be a congruence on $\cl{\langle X\rangle}$.
Assume that there exists~$N\ge1$ such that $|G/{\sim}|\le N$
for every group $G\le\cl{\langle X\rangle}$.
Then $\langle X\rangle/{\sim}$ is finite and has diameter at most
$2^{2d(d+3)}N^d$.
\end{restatable}

We now apply the viewpoint of matrix semigroup congruences and quotients
to questions in border complexity, focusing on
 two lines of applications.
In each application, we construct a finitely generated matrix semigroup and a congruence on its
Zariski closure. We show that 
any group in the closure lies in a single congruence
class, so Theorem \ref{TheoremMainQuotient} applies  and gives a
finite quotient of the generated semigroup.
We then show that the closure has infinitely many
classes, and relate this distinction or the congruence classes to the polynomial
families  (or their orbits) under study.

\paragraph{Separation of homogeneous (commutative) ABPs and their closures.} 
Let $w\ge2$ and~$d\ge3$.
Recall that a homogeneous algebraic branching program (ABP) of width
at most $w$ and degree $d$ computes a polynomial through matrix multiplication, that is,
\[
 f(\bm{x})=L_1(\bm{x})\cdots L_d(\bm{x}),
\]
where $L_1$ is a $1\times w$ matrix, $L_d$ is a
$w\times1$ matrix, and the remaining matrices are
$w\times w$.
The entries of the matrices~$L_i$ are homogeneous linear forms in the
input variables $\bm{x}=(x_1,\ldots,x_m)$, with coefficients in~$\overline{\Q}$.
ABPs  and their border complexity have
been studied in recent years~\cite{BringmannIkenmeyerZuiddam18,AllenderWang16,
BlaserEtAl20,DuttaEtAlSTACS24}.
Homogeneous models and their connections with algebraic
geometry have also received recent
attention~\cite{GesmundoEtAl22,
DuttaGesmundoIkenmeyerJindalLysikov24}.
In our constructions, we keep both the width $w$ and the degree $d$
fixed and use homogeneous linear forms as used in~\cite{DuttaGesmundoIkenmeyerJindalLysikov24}.

Write $\overline{\Q}[\bm{x}]_d$ for the vector space
of homogeneous degree-$d$ polynomials in the variables
$\bm{x}$, with coefficients in~$\overline{\Q}$.
Denote by $\mathcal A_{w,d}$ the set of polynomials $\overline{\Q}[\bm{x}]_{d}$
computed by the ABPs defined above. 
As usual, viewing each polynomial as its vector
of coefficients, we 
take the Zariski closure of the set~$\mathcal A_{w,d}$ of polynomials in the vector space of
homogeneous degree-$d$ polynomials over $\overline{\Q}$; we
denote this closure by~$\cl{\mathcal A_{w,d}}$.

Below, we prove the separation of $\mathcal A_{w,d}$ and its closure through a semigroup congruence. Roughly speaking, we use translation of the degree-$d$ and width-$w$ universal ABP  to encode polynomials (in $\mathcal A_{w,d}$) 
in a finitely generated matrix semigroup.
We construct a congruence on its  closure whose
zero class consists exactly of the encodings of
polynomials in $\mathcal A_{w,d}$.
We  give an explicit 
polynomial~$F_{w,d}$  with integer coefficients whose encoding is not congruent with zero, implying that~$F_{w,d} \in \overline{\mathcal A_{w,d}}\setminus\mathcal A_{w,d}$. 


\begin{restatable}{theorem}{HomogeneousABPSeparation}
\label{theo:homogeneous-abp-separation}
Let $w\ge2$ and $d\ge3$ and  $m\geq 2w+7$. The class  $\mathcal A_{w,d}$ of width-$w$ degree-$d$ homogenous ABPs is separated from
~$\cl{\mathcal A_{w,d}}$.
\end{restatable}

\paragraph{Separation of $\mathrm{Det}_n$  from $\mathrm{Perm}_n$ under  a subclass of reductions}

Let $n\geq 3$. Consider the matrix
$[x_{i,j}]_{1\leq i,j\leq n}$
whose entries are the formal variables $x_{i,j}$; and 
$\bm{X}:=(x_{1,1},\ldots,x_{n,n})$
be the vector of these $n^2$ variables.
We denote by $\mathrm{Det}_n$ and $\mathrm{Perm}_n$ the determinant and permanent polynomials, respectively, defined by
\[
\mathrm{Det}_n(\bm{X})
:=
\sum_{\sigma\in\mathfrak S_n}
\operatorname{sgn}(\sigma)
\prod_{i=1}^n x_{i,\sigma(i)}
\qquad 
\text{and}
\qquad
\mathrm{Perm}_n(\bm{X})
:=
\sum_{\sigma\in\mathfrak S_n}
\prod_{i=1}^n x_{i,\sigma(i)},
\]
where $\mathfrak S_n$ is the set of permutations of
$\{1,\ldots,n\}$ and $\operatorname{sgn}(\sigma)$ denotes the sign of $\sigma$.

Below, we use the standard action of
$\mat_{n^2}(\overline{\Q})$ on the polynomial ring
$\overline{\Q}[\bm{X}]$,
defined by 
\[
 (h\triangleleft L)(\bm{x})
 :=
 h(\bm{x}L)\,,
\]
where $L\in \mat_{n^2}(\overline{\Q})$ and $h\in \overline{\Q}[\bm{X}]$. 
This is the action of linear substitutions in the
$n^2$ variables; as defined in ~\cite[Section~3.2]{BlaserIkenmeyer25}.
 
We  construct a finitely generated monoid
$\langle Y_n\rangle\leq\mat_{n^2}(\Q)$ such that
\[
 \overline{\mathrm{Perm}_n  \triangleleft \langle Y_n\rangle
}
 \not\subseteq
 \mathrm{Det}_n \triangleleft 
   \overline{\langle Y_n\rangle.}
\]
Specifically, these orbit closures behave differently
under the mutual reducibility congruence~$\approx$. 
We write $f\approx g$ if there exist
$L,M\in\cl{\langle Y_n\rangle}$ such that
$f=  g \triangleleft L$ and $g=f \triangleleft M$.
Using Theorem~\ref{TheoremMainQuotient}, we show that
both generated orbits have finite quotients under~$\approx$. However,  the orbit closure of the determinant has a
finite quotient, whereas that of the permanent has
an infinite quotient.

\bigskip

All proofs and details omitted from the main text are provided in the appendix.

\section{Preliminaries}
Write $\N:=\{0,1,2,\ldots\}$, $\Z$, and $\Q$ for the sets of
natural numbers, integers, and rational numbers, respectively.
We denote by $\overline{\Q}$ the field of algebraic numbers, that is,
the roots of nonzero polynomials with rational coefficients.

Let $S$ be a semigroup. Given $s\in S$, the subsemigroup generated
by $s$ is $\langle s\rangle:=\{s,s^2,s^3,\ldots\}$.
We call $s$ \emph{torsion} if $\langle s\rangle$ is finite.
In this case, the \emph{idempotent order} (or simply order) of $s$
is the least positive integer $n$ such that $s^n=s^{2n}$, or
equivalently, such that $s^n$ is idempotent.
The semigroup $S$ is called \emph{torsion} if all its elements
are torsion, and \emph{uniformly torsion of order $n$} if
$s^n=s^{2n}$ holds for all $s\in S$.

\paragraph{Matrix semigroups and their congruences.}
Given a positive integer $d$ and a field $F$, write $\mat_d(F)$
for the set of $d\times d$ matrices with entries in $F$, and
$\gl_d(F)$ for the group of invertible matrices.
 Given $a\in\mat_d(F)$, write
$\ker(a)$ and
 $\im(a)$ for the kernel and image of the linear transformation~$\bm{x}\mapsto \bm{x}a$ on $F^d$, respectively.
We call $a$ \emph{rank-stable} if~$\rk(a)=\rk(a^2)$.
A set $X\subseteq\mat_d(F)$ is called \emph{rank-stable} if
$\rk(a)=\rk(b)=\rk(ab)$ for all~$a,b\in X$.
 A rank-stable set is called \emph{stable} if all its
elements have the same kernel and the same image.

\begin{example}\label{example1prelim}
Let
\[
 A=\begin{pmatrix}0&1\\0&0\end{pmatrix},\qquad
 A_x=\begin{pmatrix}1&x\\0&0\end{pmatrix}.
\]
For $x\ne y$, the set $\{A_x,A_y\}$ is rank-stable but not stable, as
the two matrices have the same kernel but different images.
The matrix $A$ is nilpotent, since $A^2=0$.
Although singleton sets trivially satisfy the common kernel and
image condition, $\{A\}$ is not stable because $A$ is not rank-stable.
\end{example}

Given a finite set $X\subseteq\mat_d(F)$,  write
$\langle X\rangle$ for the multiplicative semigroup of all nonempty
products of matrices from $X$.
The \emph{length of an element} $a\in\langle X\rangle$, with respect
to $X$, is the least positive integer $k$ such that
$a=a_1\cdots a_k$ with $a_i\in X$; 
when the generating set is clear, we simply refer to this quantity as the length of $a$.
The \emph{diameter} of $\langle X\rangle$ with respect to $X$ is
the least $n\in\N$ such that every element has length at most $n$,
and is infinite if no such $n$ exists.
By the extension of the Burnside--Schur theorem to semigroups
\cite{McNaughtonZalcstein75}, every finitely generated uniformly torsion
subsemigroup of $\mat_d(F)$ is finite. In particular, its diameter
with respect to any finite generating set is finite.

A \emph{congruence} on a semigroup is an equivalence relation
compatible with multiplication, that is,  $a_1 \sim b_1$ and $a_2 \sim b_2$ implies $a_1 a_2 \sim b_1 b_2$. If $\sim$ is a congruence on
$\langle X\rangle$,  its quotient is a semigroup  generated by $X/{\sim}$.
The length of a congruence class is the smallest length of an
element in that class: the class of~$a$ has length equal to the
least positive integer~$k$ such that $a\sim a_1\cdots a_k$ with  $a_i\in X$.
The diameter of $\langle X\rangle/{\sim}$ with respect to
$X/{\sim}$ is the least $n\in\N$ such that every class has length
at most $n$, and is infinite if no such $n$ exists.
For a subsemigroup $S\le\langle X\rangle$, we write $S/{\sim}$ for the quotient
by the restriction of $\sim$ to $S$; its size counts the classes
of $\langle X\rangle/{\sim}$ that meet $S$.

\paragraph{Zariski Topology.}
We recall basic notions from algebraic geometry, and refer the reader to~\cite{CoxLittleOShea15,humphreys1975linear} for an extended
background.

An affine algebraic variety (or simply a variety) in
$\overline{\Q}^{\,m}$ is the common zero set of
a collection of polynomials in
$\overline{\Q}[x_1,\ldots,x_m]$.
The \emph{Zariski topology} on $\overline{\Q}^{\,m}$
is the topology whose closed sets are 
these varieties.
Given a subset~$S\subseteq\overline{\Q}^{\,m}$, its
Zariski closure, denoted by $\cl{S}$,
is the smallest Zariski-closed set containing~$S$.

We recall several facts following standard
properties~\cite{CoxLittleOShea15}.
Finite sets and vector spaces are Zariski closed,
and the  set $\Z^m$ is Zariski dense in
$\overline{\Q}^{\,m}$, meaning that
$\cl{\Z^m}=\overline{\Q}^{\,m}$.
Moreover, closure commutes with finite unions.
Polynomial maps are continuous in this  topology, meaning that, 
\[\varphi(\cl{S}) \subseteq \cl{\varphi(S)}\]
for every polynomial map $\varphi:\overline{\Q}^{\,m} \to \overline{\Q}^{\,n}$ and 
$S\subseteq\overline{\Q}^{\,m}$ (the equality holds when  $\varphi$ is an injective linear map). 
By  continuity of matrix multiplication, the Zariski closure of a matrix semigroup is again
a semigroup.
Moreover, the closure of a nonempty subsemigroup
(inside a linear algebraic group) is a
group~\cite[Fact~9]{OurVASS26}.

\section{Burnside Problems for Matrix Semigroup Quotients}
The classical Burnside problem asks whether finitely generated torsion
groups must be finite. The answer is negative for arbitrary groups,
even with bounded exponent~\cite{NovikovAdian68}, but positive for
linear groups in characteristic zero by Schur's theorem~\cite{Schur11}.
McNaughton and Zalcstein extended the latter conclusion to matrix
semigroups~\cite{McNaughtonZalcstein75}. We are interested in such
Burnside questions concerning quotients of matrix semigroups by
congruences, which in general may not themselves be linear.

Throughout this section, fix a finite nonempty set
$X\subseteq\mat_d(\Q)$ with size $|X|$.
We measure lengths of matrices with respect to $X$, and lengths
of congruence classes with respect to $X/{\sim}$.
Our first two theorems concern congruences on $\langle X\rangle$, whereas
for the last theorem, the congruence is defined on its Zariski
closure in $\mat_d(\overline{\Q})$.

\paragraph{Factorisation of matrix products.}
For a  matrix product $a_1\cdots a_k$, a \emph{factor}
is a consecutive subproduct $a_i\cdots a_j$ with
$1\le i\le j\le k$; it is proper if it is not the full product. 
Given such a matrix product~$a_1\cdots a_k$, 
a \emph{factorisation tree} groups its factors recursively. The leaves of the tree are labelled with the matrices $a_1$ to $a_k$ in the exact order  from left to right, whereas each
internal node of the tree is labelled by the product of its children's labels. The  height of the tree is the length of the longest path from the root to a leaf.

The original notion
of factorization tree was introduced by Imre Simon~\cite{Simon90} in his resolution of the star-height problem
for regular languages (see also~\cite{bojanczyk9F}).
Simon's original factorisation tree   establishes  a bound on the height of factorisation trees over finite monoids, which depends on the size of the monoid.
In a recent work~\cite{OurVASS26}, we adapt this  
notion  from finite monoids to the infinite monoid~$\mat_d(\Q)$.  In our variant, the role of idempotent elements in the classical
version is replaced by rank-stable matrices\footnote{In~\cite{OurVASS26}, the rank-stable matrices are simply called stable.}. We have proved in particular that every matrix product in~$\mat_d(\Q)$, no matter its length,  
has a factorization tree of height
bounded only by $d(d+3)$.
We then refine this notion slightly, in~\cite{Finiteness26}, for finite
matrix monoids in $\mat_d(\Q)$, 
replacing rank-stability with stability of the children (see more details below).
We combined the height bound on such factorisation trees,
together with Artin–Wedderburn decomposition, 
to obtain the first singly exponential bound on the cardinality
of finite rational matrix monoids, in terms of the total encoding
size of their generators.

\subsection{Uniform bound on stable subsemigroups}
For positive integers $N,d,m$, define
\begin{equation}\label{eq-def-Fd}
 F_d(N,m):=\bigl[m(N+1)^2\bigr]^{(2^{d+3}+1)^{d-1}}\,
\end{equation}
This yields the asymptotic bound  $(m(N+1))^{2^{O(d^2)}}$, which is
polynomial in $m$ and $N$ for  fixed~$d$.

\StableTheoremQuo*

Before explaining the proof, we give a family in dimension three
showing that the diameter must depend on both~$N$ and $|X|$.

\begin{restatable}{example}{GeneratorLower}
\label{exm:generator-lower}\label{prop:generator-lower}
Let $N,m\geq 2$ with  
 $B:=m^2+1$.
 Let $X_m=\{A_1,\ldots,A_m\} \subseteq \mat_3(\Q)$  where 
\[
 A_i=\begin{pmatrix}B&0&Bi\\0&1&0\\0&i&0\end{pmatrix}\,.
\]
We define the congruence $\sim$ by the following matrix product
\[W=(A_1\cdots A_m)^N
=\underbrace{(A_1\cdots A_m)\cdots(A_1\cdots A_m)}_{N\text{ repetitions}}\,.\]
Let each distinct matrix appearing as a factor of $W$ be in a separate singleton class, and group every other product in $\langle X_m \rangle$ in a single class (we  refer to it as the \emph{large} class).

The crux of the construction is that 
the $A_i$ freely generate the semigroup 
$\langle X_m\rangle$, in other words, 
 every matrix  $a\in \langle X_m\rangle$  has a  unique matrix product $a_1 \cdots a_k$ with the $a_i$'s in $X_m$
(the uniqueness comes from the fact that the product encodes a number in  the  base $B$). The kernel and image of a product comes from $a_1\in X_m$ and $a_k\in X_m$, respectively.
Intuitively speaking, by the above,  every stable subsemigroup has $\ker(A_i)$ and  $im(A_j)$ for some pair $(i,j)$ with $1\leq i,j\leq m$. Notice that
 there are at most $N$ distinct factors of $W$
starting with $A_i$ and ending with $A_j$.
This implies that the stable subsemigroup quotient with $\sim$ meets at most $N$ singleton classes
and possibly the large class, giving the bound $N+1$. Moreover, the stable  subsemigroup
generated by the pair $(1,m)$ will meet this exact bound. The diameter bound $N|X_m|$  follows from the freeness of $\langle X \rangle$ and the length of the 
singleton class
of $W$. The details are provided in Appendix~\ref{app:lower}.
\end{restatable}

\begin{proof}[Proof Sketch of Theorem~\ref{theorem-1}]
We outline the main ideas of the proof here; the full details, including proofs of the claims, are provided in Appendix~\ref{app:theorem-1}.
We focus on the non-trivial direction of the proof, obtaining the stated bound on $|\langle X \rangle /{\sim}|$ from the uniform bound
on $N$ on $|S/{\sim}|$ for stable $S\leq \langle X\rangle$.

The proof of the theorem relies on the notion of forest factorisation from~\cite{OurVASS26}. Since we will also use a second variant used in~\cite{Finiteness26}, we refer to this one as the \emph{rank-stable forest factorisation}.
Given a  matrix product, in its
 rank-stable forest factorisation every node with more than two children is rank-stable. 
It is shown in~\cite[Proposition~3]{OurVASS26} that every matrix product $a_1\cdots a_k$ in $\mat_d(\Q)$ has a  rank-stable factorization tree of height at most  $d(d+3)$.
To obtain this bound, we have shown in~\cite[Proposition~7]{OurVASS26}
that if 
$\rk(a_1)=\cdots=\rk(a_k)=\rk(a_1\cdots a_k)$ holds,  such a tree can be chosen with height at most~$d+2$. 
The following claim follows from a truncation procedure on such trees, see Figure~\ref{fig:fixed-rank-factors}.

\begin{figure}[t]
\centering
\begin{tikzpicture}[
    font=\small,
    branch/.style={
        circle, fill=black, inner sep=1.3pt
    },
    stable/.style={
        draw, rectangle,
        minimum width=9mm,
        minimum height=6mm,
        inner sep=2pt
    }
]

\node[branch] (root) at (0.2,0) {};
\node[stable] (P1) at (-1.6,-0.7) {$P_1$};
\node[branch] (v) at (1.3,-0.7) {};
\node (a4) at (0.4,-1.4) {$a_4$};
\node[stable] (P2) at (2.3,-1.4) {$P_2$};

\draw (root) -- (P1);
\draw (root) -- (v);
\draw (v) -- (a4);
\draw (v) -- (P2);

\foreach \i/\x in {
    1/-2.4, 2/-1.6, 3/-0.8,
    5/1.5, 6/2.3, 7/3.1
} {
    \node (a\i) at (\x,-2.3) {$a_{\i}$};
}

\foreach \i in {1,2,3} {
    \draw[dashed] (P1.south) -- (a\i.north);
}
\foreach \i in {5,6,7} {
    \draw[dashed] (P2.south) -- (a\i.north);
}

\node at (0.35,-3.1) {
    $\underbrace{a_1a_2a_3}_{P_1}
     \;\underbrace{a_4}_{\text{single matrix}}\;
     \underbrace{a_5a_6a_7}_{P_2}$
};

\end{tikzpicture}
\caption{In the factorisation tree of matrix product, along each path from the root to a leaf, we truncate the  tree at the first nodes with more than
two children. Here, the factors $P_1$ and $P_2$ are rank-stable.
}\label{fig:fixed-rank-factors}
\end{figure}
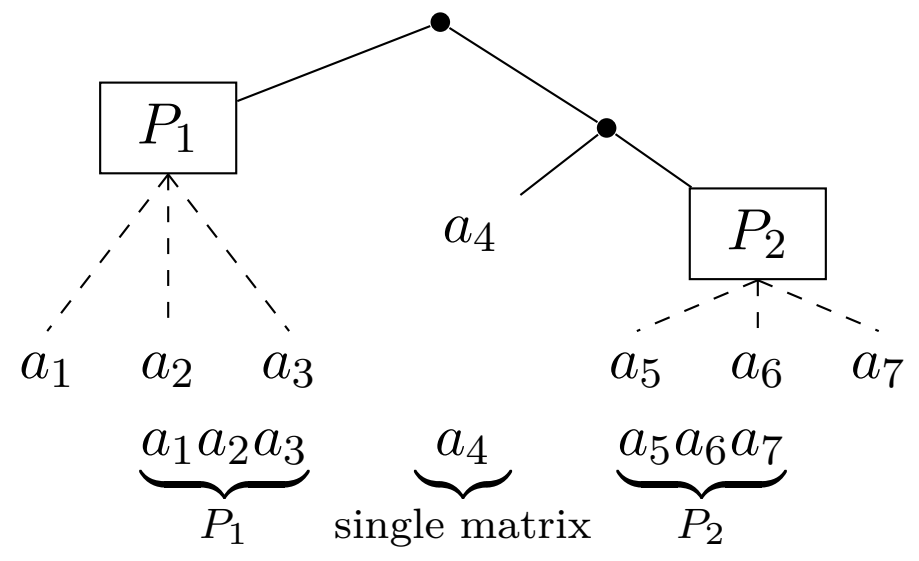

\begin{restatable}{claim}{FixedRankFactors}\label{clm:fixed-rank-factors}
Let $a_1,\ldots,a_k\in\mat_d(\Q)$ be such that 
$\rk(a_1\cdots a_k)=\rk(a_i)$ for all $1\leq i\leq k$. 
Then $a_1\cdots a_k$ can be partitioned into at most $2^{d+2}$
factors, each of which is either a single  matrix among the $a_i$'s
or a rank-stable matrix product.
\end{restatable}

To prove finiteness of the quotient, we bound its number of classes by decreasing
induction on rank.
Let  $T\subseteq\langle X\rangle$ be finite and all its matrices have rank~$r$.
The essential step is to count the number of congruence classes
of  matrix products over~$T$, whose resulting product has rank~$r$.
Such rank-preserving products keep the kernel of the first matrix and the image
of  the last matrix, so  at  most $|T|^2$  pairs $(K,U)$ of kernel and image can occur. Observe that, whenever $K\cap U=\{0\}$, the matrices in
$\langle T\rangle$ with kernel $K$ and image $U$ form a stable
subsemigroup, whose quotient by assumption has at most $N$ classes. 
  Applying Claim~\ref{clm:fixed-rank-factors}
now gives the following bound.

\begin{restatable}{claim}{FixedRankClasses}\label{clm:fixed-rank-classes}
Let $T\subseteq\langle X\rangle$ be finite and all its matrices
have rank $r$.  The rank-$r$ matrices in $\langle T\rangle$ meet
at most $(1+|T|+N|T|^2)^{2^{d+2}}$ congruence classes.
\end{restatable}

We note in passing that congruent matrices can have different ranks. For each class $c$ in the congruence, we define
$\rho(c):=\min\{\rk(a):a\in c\}$.
We count classes with $\rho(c)\ge r$ by decreasing induction on $r$.
For the base case, observe that the invertible matrices in $\langle X\rangle$ form a stable subsemigroup whose quotient has at most
$N$ classes. Hence there are at most $N$ classes with $\rho(c)=d$.

Suppose there are at most $n$ classes $c$ with $\rho(c)>r$.
Let $a_1\cdots a_k$ with $a_i\in X$ be a matrix product, whose
class $[a_1\cdots a_k]_{\sim}$ has minimum rank $r$. 
 Recall that matrix 
multiplication cannot increase  rank, so
$\rho(bc)\le\min\{\rho(b),\rho(c)\}$ for all congruence classes~$b,c$. This implies that every factor of $a_1\cdots a_k$
belongs to a class with minimum rank at least $r$.
Write $a_1\cdots a_k$ as $y_1c_1\cdots y_\ell c_\ell y_{\ell+1}$
with each 
$c_i\in X$ 
and $\rho([y_ic_i]_{\sim})=r$, and the 
  $y_i$ are either an empty product  or 
$\rho([y_i])>r$. See Figure~\ref{fig:rank-representatives} for an illustration.

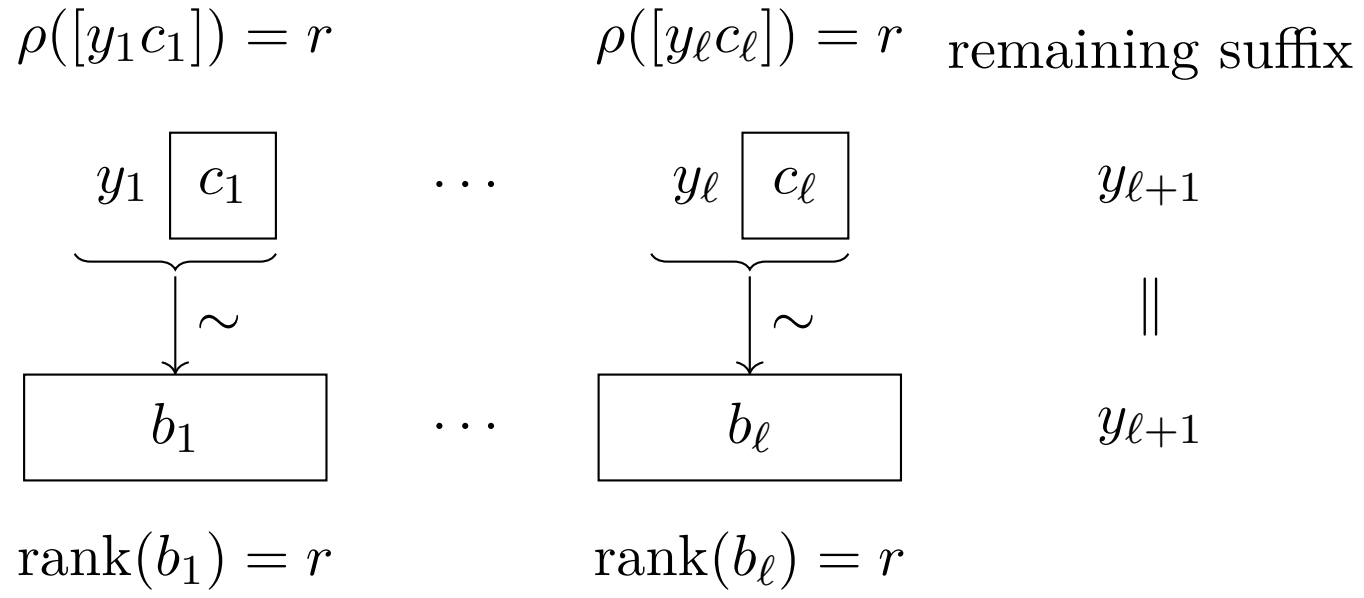
\begin{figure}[t]
\centering
\begin{tikzpicture}[
    font=\small,
    every node/.style={outer sep=0pt},
    factor/.style={
        draw,
        minimum height=7mm,
        inner sep=2pt
    },
    prefix/.style={factor, minimum width=13mm},
    generator/.style={factor, minimum width=7mm},
    representative/.style={factor, minimum width=20mm}
]

\node (y1) at (0,0) {$y_1$};
\node[generator, right=0pt of y1] (c1) {$c_1$};
\node[fit=(y1)(c1), inner sep=0pt] (f1) {};
\node[above=3mm of f1] {$\rho([y_1c_1])=r$};

\node (yl) at (3.8,0) {$y_\ell$};
\node[generator, right=0pt of yl] (cl) {$c_\ell$};
\node[fit=(yl)(cl), inner sep=0pt] (fl) {};
\node[above=3mm of fl] {$\rho([y_\ell c_\ell])=r$};

\node (dots) at (2.3,0) {$\cdots$};

\node[ minimum width=17mm] (suffix)
    at (6.8,0) {$y_{\ell+1}$};
\node[above=3mm of suffix] {remaining suffix};

\node[representative, below=9mm of f1] (b1) {$b_1$};
\node[below=2mm of b1] {$\rk(b_1)=r$};

\node[representative, below=9mm of fl] (bl) {$b_\ell$};
\node[below=2mm of bl] {$\rk(b_\ell)=r$};

\node at (dots |- b1) {$\cdots$};

\node[ minimum width=17mm] (suffixbottom)
    at (suffix |- b1) {$y_{\ell+1}$};

\foreach \factorname/\repname in {f1/b1,fl/bl} {
    \draw[
        decorate,
        decoration={brace,mirror,amplitude=3pt}
    ]
        ([yshift=-1mm]\factorname.south west) --
        ([yshift=-1mm]\factorname.south east);

    \draw[->]
        ([yshift=-2.5mm]\factorname.south) --
        node[right] {$\sim$}
        (\repname.north);
}

\path (suffix.south) --
    node {$\Vert$} (suffixbottom.north);

\end{tikzpicture}
\caption{The factor $y_ic_i$ can be replaced by a
rank-$r$ congruent matrix $b_i$.}
\label{fig:rank-representatives}
\end{figure}

Since factors $y_ic_i$  belong to a  class
of the form  $[c_i]_{\sim}$ or $[y_i]_{\sim}[c_i]_{\sim}$, where $c_i\in X$ and $\rho([y_i]_{\sim})>r$,   there are at most
$t:=|X|(n+1)$ such classes.
We choose one matrix product $b_i$ of 
 rank $r$ in each of these classes $[y_ic_i]_{\sim}$ to  
 form a  finite set
$T\subseteq\langle X\rangle$ where $|T|\le t$.
 Then
\[
 a_1\cdots a_k\sim b_1\cdots b_\ell \, y_{\ell+1}.
\]
Since each $b_i$ has rank $r$ and  $\rho(b_1\cdots b_\ell \, y_{\ell+1})=r$ we get  
$r\le \rk(b_1\cdots b_\ell \, y_{\ell+1})
   \le \rk(b_1\cdots b_\ell)
   \le r\,$,
which means that~$\rk(b_1\cdots b_\ell)=r$.
Claim~\ref{clm:fixed-rank-classes} now bounds the number
of possible classes  $[b_1\cdots b_\ell]_{\sim}$ by
$(1+t+Nt^2)^{2^{d+2}}$.
There are at most $n+1$ choices for the suffix class,
including the empty suffix. Hence the number of
classes with minimum rank exactly $r$ is at most
\[
 (n+1)(1+t+Nt^2)^{2^{d+2}}\,,
\]
where $t=|X|(n+1)$. Adding the at most $n$ classes with larger minimum rank
completes the induction step.
By a summation over ranks, we arrive to the bound 
given in~\eqref{eq-def-Fd} for the cardinality of the quotient (and thus for its diameter as well).
\end{proof}

\subsection{Uniform bound on rank-stable subsemigroups.}

The next theorem differs from Theorem~\ref{theorem-1} by requiring a uniform bound on rank-stable
subsemigroups rather than stable subsemigroups. The next example shows that these uniform bounds can be different.

\begin{restatable}{example}{ShiftSeparation}\label{exm:shift-separation}
Let $m,d\ge2$. 
Let $X=\{A_1,\ldots,A_m\}\subseteq\mat_d(\Q)$ where
 \[
A_i=
\begin{pmatrix}
1 & \mathbf{0} & i\\
0 & \mathbf{0} & 0\\
\mathbf{0} & I_{d-2} & \mathbf{0}
\end{pmatrix}\,,
\]
where $I_{d-2}$ is the identity matrix in dimension~$d-2$. Let $\sim$ be the 
 equality  congruence.
Since  $(x_1,\ldots,x_d)A_i=(x_1,x_3,\ldots,x_d,ix_1)$,  
every product of $k$ generators has rank $\max\{1,d-k\}$.
Hence, the 
rank-stable subsemigroups contain only
rank one matrices. 
The  rank-one matrices in $\langle X \rangle $, say obtained by products $a_1\cdots a_k$ with  $a_i\in X$, are equal to their suffix factor  of length $d-1$, as shown below; given $k\geq d-1$ we have
\[
 a_1\cdots a_k=a_{k-d+2} \cdots a_k
   =A_{i_1}\cdots A_{i_{d-1}}=
 \begin{pmatrix}
  1&i_1&\cdots&i_{d-1}\\
  0&0&\cdots&0\\
  \vdots&\vdots&&\vdots\\
  0&0&\cdots&0
 \end{pmatrix}
\]
Observe that  $S=\langle a_1\cdots a_{d-1} \,:\, a_i \in X\rangle$
is a rank-stable subsemigroup of $\langle X\rangle$; moreover, 
 every rank-stable (and stable)
subsemigroup is contained in~$S$. Hence the  bound on the cardinalities of quotients of rank-stable subsemigroups is~$|X|^{d-1}$.
Notice that the  stable 
  subsemigroups are all singletons, as each matrix in~$S$ is idempotent and has a distinct image:
\[
 \im\bigl(A_{i_1}\cdots A_{i_{d-1}}\bigr)
 =\operatorname{Span}_{\Q}\{(1,i_1,\ldots,i_{d-1})\}\,\,
\]
This means that the smallest  bound on the quotient images
of stable subsemigroups in this example is~$1$. We provide more details 
 in Appendix~\ref{app:lower}.
\end{restatable}

For positive integers $N,d$, define
\begin{equation}\label{eq:rank-bound}
 G_d(N):=(N+1)(H+1)^d-1
 \ \le\ \min\bigl\{2^{O(d^2N)},\ (N+1)^{2^{O(d)}}\bigr\}.
\end{equation}
where
\[
H:=\binom{N+D^2+1}{D^2}
 \qquad \text{and}\qquad 
 D:=\binom d{\lfloor d/2\rfloor}.
\]

\RankCriterion*

\begin{proof}[Proof Sketch of Theorem~\ref{theorem2}]
The proof of this theorem relies on exterior algebra. We provide a gentle introduction to the necessary background, together with the detailed proof, in Appendix~\ref{app:rank}.

We bound class lengths  using the following claim.

\begin{restatable}{claim}{RankClique}
\label{clm:rank-clique}
Let $a_1,\ldots,a_k\in\langle X\rangle$ be such that $k>H$
and $\rk(a_1\cdots a_k)=\rk(a_i)$ for all $1\leq i\leq k$. 
There exist \(i,j\) with $1 \leq i+1 <j \leq k$ such that
$a_1\cdots a_i\,a_j\cdots a_k
 \sim a_1\cdots a_k$.
\end{restatable}

To prove Claim~\ref{clm:rank-clique}, let $r$ be the common
rank of the matrices $a_1,\ldots,a_k$ and their product.
We call the position between $a_i$ and $a_{i+1}$ cut~$i$,
and form a graph whose vertices are the $k-1$ cuts.
For two distinct cuts $i,j$, consider the products
$a_i a_{j+1}$ and $a_j a_{i+1}$.
We connect $i$ and $j$ by an edge  when both products
are rank preserving; see Figure~\ref{fig:cut-graph} for an illustration.
An edge between cuts $i<j$ means that the middle factor
$b=a_{i+1}\cdots a_j$ can be deleted or repeated  while preserving
the rank of the  matrix product.

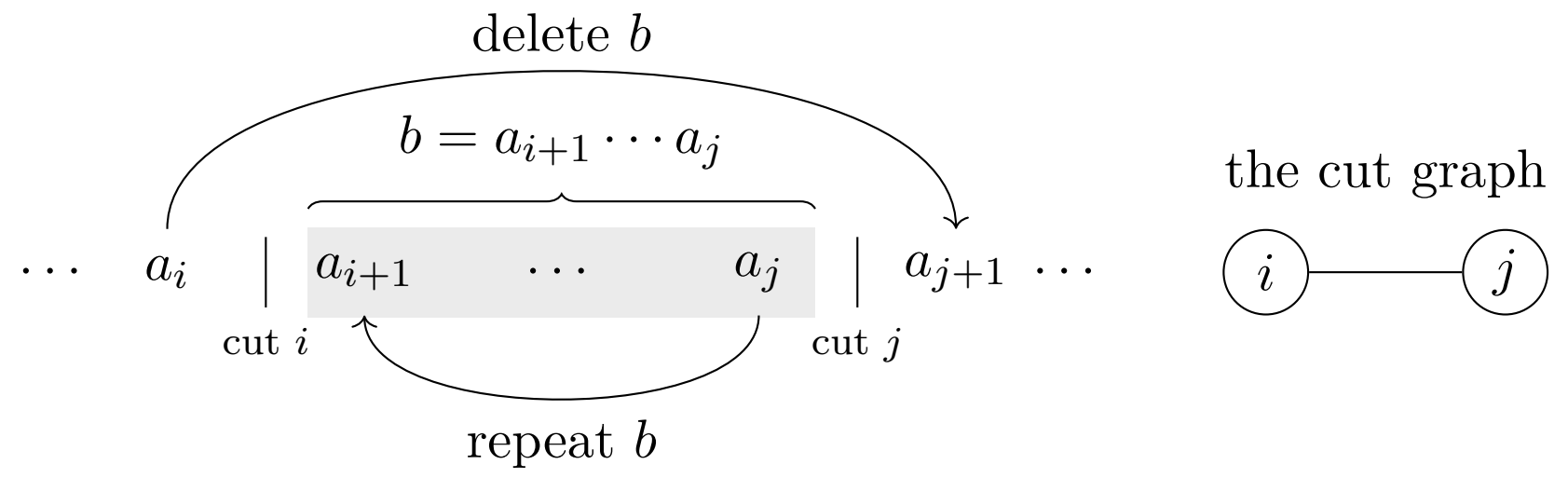
\begin{figure}[t]
\centering
\begin{tikzpicture}[
    font=\small,
    entry/.style={
        minimum width=8mm,
        minimum height=6mm,
        inner sep=1pt
    },
    vertex/.style={
        circle, draw,
        minimum size=6mm,
        inner sep=1pt
    }
]

\fill[black!8]
    (1.0,-0.32) rectangle (4.6,0.32);
    
\node at (-0.8,0) {$\cdots$};
\node[entry] (ai)  at (0,0)   {$a_i$};
\node[entry] (ain) at (1.4,0) {$a_{i+1}$};
\node at (2.8,0) {$\cdots$};
\node[entry] (aj)  at (4.2,0) {$a_j$};
\node[entry] (ajn) at (5.6,0) {$a_{j+1}$};
\node at (6.4,0) {$\cdots$};

\draw (0.7,-0.25) -- (0.7,0.25);
\node[below] at (0.7,-0.25) {\scriptsize cut $i$};

\draw (4.9,-0.25) -- (4.9,0.25);
\node[below] at (4.9,-0.25) {\scriptsize cut $j$};

\draw[
    decorate,
    decoration={brace,amplitude=3pt}
]
    (1.0,0.45) -- (4.6,0.45)
    node[midway,above=4pt] {$b=a_{i+1}\cdots a_j$};

\draw[->]
    (ai.north)
    .. controls (0,1.8) and (5.6,1.8) ..
    (ajn.north)
    node[midway,above] {delete $b$};

\draw[->]
    (aj.south)
    .. controls (4.2,-1.1) and (1.4,-1.1) ..
    (ain.south)
    node[midway,below] {repeat $b$};

\node[vertex] (vi) at (7.8,0) {$i$};
\node[vertex] (vj) at (9.5,0) {$j$};
\draw (vi) -- (vj);

\node at (8.65,0.7) {the cut graph};

\end{tikzpicture}
\caption{
There is an edge between cuts $i$ and $j$ when 
$\rk(a_ia_{j+1})=\rk(a_ja_{i+1})=r$.
For cuts~$i<j$, such an edges means the middle
factor $b$ can be deleted or repeated while preserving rank~$r$.
The deletion uses
$\im(a_i)\cap\ker(a_{j+1})=\{0\}$,
and repetition uses
$\im(a_j)\cap\ker(a_{i+1})=\{0\}$.}
\label{fig:cut-graph}
\end{figure}

We first show that in the resulting graph a set of cuts with no edges between them, also known as an independent set,   
cannot be too large.

\begin{restatable}{claim}{IndependentCuts}
\label{clm:independent-cuts}
Each independent set of the cut graph has size at most $\binom dr^2\le D^2$.
\end{restatable}

The proof of this claim is the heart of the argument, and this is where we use exterior algebra.
The idea is to encode the image and kernel at each cut by a rank-one linear map on $W:=\bigwedge^r\Q^d$.
For each cut $i$, denote this map by $T_i:W\to W$.
Write $\End(W)$ for the vector space of all linear maps
from $W$ to itself, so $T_i\in \End(W)$.
Our construction also uses a linear functional
$f_i:\End(W)\to\Q$, which assigns a rational number to
each such map.
The  functionals $f_i$ are constructed such that~$f_i(T_j)\ne0$ if and only if
$\rk(a_i a_{j+1})=\rk(a_j a_{i+1})=r$.

Now let $I$ be an independent set.
For each $i\in I$, the functional $f_i$ therefore vanishes
on every map~$T_j$ with $j\in I\setminus\{i\}$ 
but not on~$T_i$. To obtain the claimed bound on $|I|$, it suffices to show
that the maps $T_i$ with $i\in I$, are linearly independent in $\End(W)$.
As usual, this means that a relation
$\sum_{i\in I}\alpha_iT_i=0$
can hold only when each $\alpha_i$ is zero.
The bound will then follow from the fact that
$\End(W)$ has dimension $\binom dr^2$.

Consider such a relation $\sum_{i\in I}\alpha_iT_i=0$ and fix $j\in I$.
Evaluating $f_j$ at both sides and using its linearity gives
\[
 0
 =f_j\left(\sum_{i\in I}\alpha_iT_i\right)
 =\sum_{i\in I}\alpha_i f_j(T_i)
 =\alpha_j f_j(T_j).
\]
The last equality holds because $f_j(T_i)=0$ for every
$i\in I\setminus\{j\}$.
Since $f_j(T_j)\ne0$, we obtain $\alpha_j=0$.
This holds for every $j\in I$, proving linear independence and hence the claimed bound in Claim~\ref{clm:independent-cuts}.

To bound the diameter of the quotient, let the matrix product
$a_1\cdots a_k$ be such that 
$a_i\in X$ and~$k$ is 
equal to the length of the class
$[a_1\cdots a_k]_{\sim}$. 
Define $L_d:=N$ and
$L_r:=(H+1)L_{r+1}+H$ for $0\le r<d$.
We prove by decreasing induction on $r$ that every matrix product $a_1\cdots a_k$,
chosen as above, with $\rk(a_1\cdots a_k)\ge r$ satisfies
$k\le L_r$. (Here the induction is on the rank of the chosen matrix product,
compared to Theorem~\ref{theorem-1} where  the minimum rank of its congruence class was used.)

For the base case with rank $d$, all prefixes lie in the rank-stable subsemigroup
of invertible matrices in $\langle X\rangle$.
Similar to Claim~\ref{clm:rank-clique}, we have  $k\leq N=L_d$, as otherwise two prefixes would be congruent, contradicting the minimality of~$k$. 

For the induction step, suppose $\rk(a_1\cdots a_k)=r<d$.
Decompose the product as $y_0(c_1y_1)\cdots(c_\ell y_\ell)$,
where $c_i\in X$, each nonempty $y_i$ has rank greater than $r$,
and each $c_i y_i$ has rank $r$.
Observe that each nonempty factor $y_i$ is itself shortest in its class
$[y_i]_{\sim}$, as otherwise, replacing it by a shorter congruent
product would contradict the minimality of~$k$.
The induction hypothesis therefore bounds each $y_i$ by
$L_{r+1}$ matrices.
Since the factors $c_i y_i$ and their product all have rank $r$, so $\ell\le H$ must hold by 
Claim~\ref{clm:rank-clique}. Again as otherwise,
it would contradict the minimality of~$k$.
This implies 
\[
 k\le L_{r+1}+\ell(L_{r+1}+1)
   \le L_{r+1}+H(L_{r+1}+1)=L_r\,,
\]
where solving this recurrence gives
$L_0=(N+1)(H+1)^d-1$, proving the diameter bound.
\end{proof}

\begin{restatable}{example}{CounterLower}\label{prop:counter}
Let $k\ge1$.  Let $X_k=\{A_1,\ldots,A_k\}\subseteq\mat_{2k}(\Q)$, where
\[
 A_i=\operatorname{diag}\!\left(
 \underbrace{E_{10},\ldots,E_{10}}_{i-1\text{ blocks}},
 \underbrace{E_{01}}_{\text{block }i},
 \underbrace{I_2,\ldots,I_2}_{k-i\text{ blocks}}
 \right) \qquad \text{with }   E_{10}:=\begin{pmatrix}0&0\\1&0\end{pmatrix}
 \text{ and }
 E_{01}:=\begin{pmatrix}0&1\\0&0\end{pmatrix}.
\]
Each $A_i$ consists of $k$ diagonal blocks of size
$2\times2$, and all its entries belong to $\{0,1\}$. Let the congruence be 
 equality.
We show that $\langle X_k\rangle$ is finite and every
rank-stable subsemigroup is a singleton, while the matrix
\[
 T_k:=\operatorname{diag}\!\left(
 \underbrace{E_{01},\ldots,E_{01}}_{k\text{ blocks}}
 \right)
\]
has length $2^k-1$.
Hence, in dimension $d=2k$, the quotient has diameter at least
$2^{d/2}-1$, even though the uniform bound on rank-stable
subsemigroups is $N=1$.

To understand the construction, represent the two bit values
by the row vectors $e_0=(1,0)$ and $e_1=(0,1)$.
Each diagonal block records one bit, with the first block
representing the least significant bit.
The two matrices act as follows:
\[
 e_1E_{10}=e_0,\qquad e_0E_{01}=e_1,
 \qquad e_0E_{10}=e_1E_{01}=\mathbf0.
\]
So multiplication by $A_i$ changes the bit pattern
\[
 \bigl(\underbrace{1,\ldots,1}_{i-1},\,0,\,
       b_{i+1},\ldots,b_k\bigr)
 \quad\text{to}\quad
 \bigl(\underbrace{0,\ldots,0}_{i-1},\,1,\,
       b_{i+1},\ldots,b_k\bigr).
\]

Now consider a product representing $P_k$, starting with
$e_0$ in every block. Every tracked row must end at $e_1$,
and none can become zero along the way, since subsequent
multiplication cannot restore a zero row.
The next generator must therefore correspond to the least
significant bit that is still zero.
Its action is exactly a binary increment, as it resets the
preceding $1$-bits to $0$, changes this $0$-bit to $1$,
and leaves the remaining bits unchanged.
Reaching the all-one pattern from the all-zero pattern thus
requires $2^k-1$ increments.
Following these increments gives a product equal to $P_k$,
so its length is exactly $2^k-1$.

The rank-stable matrices in $\langle X_k\rangle$ are  diagonal, with $0$ and $1$ along the diagonal. For two such matrices $a,b$, the condition
$\rk(a)=\rk(b)=\rk(ab)$ forces $a=b$.
Thus every rank-stable subsemigroup is a singleton, giving
the uniform bound $N=1$.
The details are given in
Appendix~\ref{app:lower}.
\end{restatable}

\subsection{Uniform bound on groups in the Zariski closure}
The next theorem assumes a uniform bound on the quotient images
of groups in the Zariski closure of $\langle X\rangle$, where
 the congruence is also defined on the  closure.
Write $\cl{\langle X\rangle}$ for this closure in~$\mat_d(\overline{\Q})$.

\TheoremNameMainQuotient*

To obtain the diameter bound, we return to the factorisation
trees used in Theorem~\ref{theorem-1}.
There, we retained a bounded number of factors and counted their
possible classes. Here, we use the full tree to bound the length
of each class.
Our starting point is the refinement of rank-stable
factorisation trees 
in~\cite[Theorem~5]{Finiteness26}.
There, we strengthened the rank-stability condition to
stability for finite matrix monoids.
In this finite setting, the children at every node with
at least three children generate a group.
Below, we adapt this construction by using the closure
and allowing matrix products to change within their
congruence classes (more details below).

Before we proceed, we note that Example~\ref{prop:counter} provides a lower bound for this theorem as well. 

\begin{example}
\label{exam:counter-cont}
Consider the matrices $X_k\subseteq\mat_{2k}(\Q)$ from
Example~\ref{prop:counter}, with equality as the congruence.
Since $\langle X_k\rangle$ is finite, it is Zariski closed.
Recall that every matrix group is a rank-stable subsemigroup.
As all rank-stable subsemigroups in this example are
singletons (see Example~\ref{prop:counter}), every group in the closure is trivial.
Thus Theorem~\ref{TheoremMainQuotient} applies with $N=1$. However, as 
shown in Example~\ref{prop:counter},  the matrix $T_k$ has length $2^k-1$.
Writing $d=2k$, the quotient therefore has diameter at least
$2^{d/2}-1$.
Hence an exponential dependence on $d$ is necessary,
even when all groups in the closure are trivial.

We can also obtain an infinite Zariski closure for the semigroup closure by adding one
coordinate to each generator, while preserving the quotient
and the bound $N=1$. Let $Y_k=\{B_1,\ldots,B_k\}
 \subseteq\mat_{2k+1}(\Q)$
 be such that 
$B_i=diag(2,A_i)$.
We have that 
$B_{i_1}\cdots\widehat B_{i_\ell}
 =diag\bigl(2^\ell,A_{i_1}\cdots A_{i_\ell}\bigr)$.
Observe that both $\langle  Y_k\rangle$ and its Zariski closure
$C_k:=\cl{\langle Y_k\rangle}$ are infinite.

We define a congruence $\sim$ by forgetting the added coordinate.
Since $\langle X_k\rangle$ is finite, the set
\[
 \{diag(t,B):
       t\in\overline{\Q},\ B\in\langle X_k\rangle\}
\]
is Zariski closed and contains $C_k$.
Hence projection onto the lower-right block defines a
semigroup homomorphism
$\pi:C_k\to\langle X_k\rangle$.
Define $a\sim b$ if $\pi(a)=\pi(b)$.
This gives a congruence on the  closure.
For every group $G\le C_k$, its image $\pi(G)$ is a group
in $\langle X_k\rangle$.
By Example~\ref{prop:counter}, every such group is trivial.
Consequently, $|G/{\sim}|=1$, and
Theorem~\ref{TheoremMainQuotient} applies with $N=1$.

Moreover, projection induces an isomorphism
$\langle Y_k\rangle/{\sim}\cong\langle X_k\rangle$,
sending the class of $B_i$ to $A_i$.
By the above, the class mapping to $T_k$ has length
$2^k-1$. Again,
writing $d=2k+1$, the quotient thus has diameter at least $2^{(d-1)/2}-1$.
\end{example}

\begin{proof}[Proof Sketch of Theorem~\ref{TheoremMainQuotient}]
We outline the main ideas here; the full details, including
proofs of the claims, are provided in Appendix~\ref{app:closure}.

To prove finiteness of the quotient and bound its diameter,
it suffices to bound the length of every congruence class.
For each $a\in\langle X\rangle$, we will construct a tree whose
leaves belong to $X$ and whose leaf product is congruent to $a$.
We then shorten this tree and bound its number of leaves.
We will need the following claim.

\begin{restatable}{claim}{ClosureAbsorption}
\label{clm:closure-absorption}
Given $a,b\in\langle X\rangle$,  the condition
$\im(a)=\im(ab)$ implies $
a\sim ab^n$  for some $n\ge1$.
\end{restatable}

We use the fact that the Zariski closure of a subsemigroup
inside an algebraic group is itself a group.
Let $a\in\langle X\rangle$ be rank-stable.
The matrices with kernel $\ker(a)$ and image $\im(a)$
form an algebraic group, whose identity~$I_{\im(a)}$  is the
projection onto $\im(a)$ along $\ker(a)$.
Therefore, the Zariski closure of $\langle a\rangle$
inside this group is a group contained in
$\cl{\langle X\rangle}$.
By assumption, its quotient has at most $N$ elements, so~$a^n\sim I_{\im(a)}$ for some $1\le n\le N$.

Now suppose $\im(a)=\im(ab)$. This means that $b$ maps
$\im(a)$ onto itself, so intuitively speaking
 $b$ restricts to an invertible map on $\im(a)$. This gives $\im(a) =\im (a)b^d \subseteq \im(b^d)$.
But $b^d$ is rank-stable, and its identity $I_{\im(b^d)}$
fixes every vector in $\im(a)$.
Consequently,
$ab^{dn}=a(b^d)^n\sim a I_{\im(b^d)}=a$,
which proves our Claim~\ref{clm:closure-absorption}.

We next use Claim~\ref{clm:closure-absorption} to modify
a rank-stable factorisation tree.
We follow the strategy of the proof
of~\cite[Theorem~5]{Finiteness26}, where rank-stable
factorisation trees for finite matrix monoids are modified
so that the children at every node with at least three
children generate a group.
Here, we use Claim~\ref{clm:closure-absorption} to modify
products within their congruence classes, so that these
children generate a stable subsemigroup.

Following the strategy, the starting tree for this procedure is 
defined in~\cite[Theorem~7]{Finiteness26}. Such a tree 
is a rank-stable factorisation tree, where  every node with at least three children,
besides the rank-stability of the parent,  must have 
 children with the same rank as the parent.
Thus, if the  children are $a_1,\ldots,a_s$ and the
parent is $a=a_1\cdots a_s$, then
$\rk(a_1)=\cdots=\rk(a_s)=\rk(a)=\rk(a^2)$.
Such a tree has height at most $d(d+3)$ and \emph{reduced height}
 at most $d$, where the reduced height counts the
 maximum number of nodes with at least three children  on a  path from the root to a leaf~\cite[Theorem 7]{Finiteness26}.

We now explain how to modify a node $a$
with at least three children; see Figure~\ref{fig:closure-tree-modification} for an illustration.
Recall that the children all have the same rank as $a$, however, they
may have different kernels and images.
Our aim is to replace their product by a congruent
product whose factors all have the kernel and image
of $a$.
Since $a$ is rank-stable, these factors will generate
a stable subsemigroup.

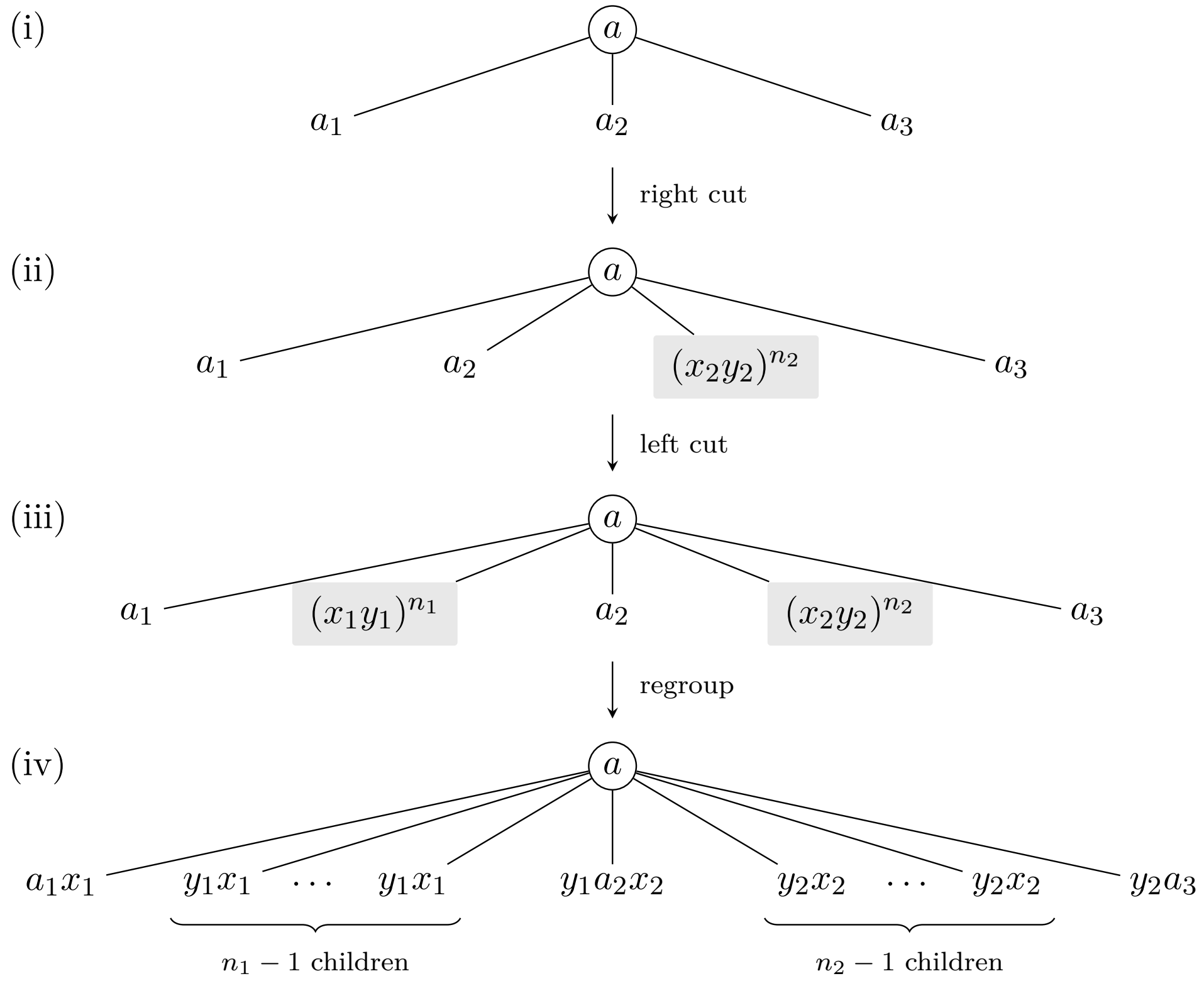
\begin{figure}[t]
\centering
\begin{tikzpicture}[
  font=\small,
  factor/.style={inner sep=2.5pt},
  parent/.style={
    circle,draw,inner sep=2pt,minimum size=5mm
  },
  inserted/.style={
    factor,fill=black!9,rounded corners=1pt,
    inner xsep=5pt,inner ysep=4pt
  },
  repeat/.style={
    decorate,
    decoration={brace,mirror,amplitude=4pt}
  },
  line width=0.45pt,
  >=stealth
]

\node[anchor=west] at (-6.5,0) {(i)};
\node[parent] (root1) at (0,0) {$a$};

\foreach \j/\x in {1/-3,2/0,3/3} {
  \node[factor] (one\j) at (\x,-1) {$a_\j$};
  \draw (root1)--(one\j);
}

\draw[->] (0,-1.45)--(0,-2.05)
  node[midway,right=4pt,font=\scriptsize] {right cut};

\node[anchor=west] at (-6.5,-2.55) {(ii)};
\node[parent] (root2) at (0,-2.55) {$a$};

\node[factor] (two1) at (-4.2,-3.55) {$a_1$};
\node[factor] (two2) at (-1.6,-3.55) {$a_2$};
\node[inserted] (twoP) at (1.3,-3.55)
  {$(x_2y_2)^{n_2}$};
\node[factor] (two3) at (4.2,-3.55) {$a_3$};

\foreach \v in {two1,two2,twoP,two3}
  \draw (root2)--(\v);

\draw[->] (0,-4.05)--(0,-4.65)
  node[midway,right=4pt,font=\scriptsize] {left cut};

\node[anchor=west] at (-6.5,-5.15) {(iii)};
\node[parent] (root3) at (0,-5.15) {$a$};

\node[factor] (three1) at (-5,-6.15) {$a_1$};
\node[inserted] (threeP) at (-2.5,-6.15)
  {$(x_1y_1)^{n_1}$};
\node[factor] (three2) at (0,-6.15) {$a_2$};
\node[inserted] (threeQ) at (2.5,-6.15)
  {$(x_2y_2)^{n_2}$};
\node[factor] (three3) at (5,-6.15) {$a_3$};

\foreach \v in {three1,threeP,three2,threeQ,three3}
  \draw (root3)--(\v);

\draw[->] (0,-6.65)--(0,-7.25)
  node[midway,right=4pt,font=\scriptsize] {regroup};

\node[anchor=west] at (-6.5,-7.75) {(iv)};
\node[parent] (root4) at (0,-7.75) {$a$};

\node[factor] (four1) at (-5.8,-9) {$a_1x_1$};
\node[factor] (four2) at (-4.15,-9) {$y_1x_1$};
\node[factor] (four3) at (-2.1,-9) {$y_1x_1$};
\node[factor] (four4) at (0,-9) {$y_1a_2x_2$};
\node[factor] (four5) at (2.1,-9) {$y_2x_2$};
\node[factor] (four6) at (4.15,-9) {$y_2x_2$};
\node[factor] (four7) at (5.8,-9) {$y_2a_3$};

\node at (-3.125,-9) {$\cdots$};
\node at (3.125,-9) {$\cdots$};

\foreach \v in {four1,four2,four3,four4,four5,four6,four7}
  \draw (root4)--(\v);

\draw[repeat] (-4.65,-9.35)--(-1.6,-9.35)
  node[midway,below=6pt,font=\scriptsize]
  {$n_1-1$ children};

\draw[repeat] (1.6,-9.35)--(4.65,-9.35)
  node[midway,below=6pt,font=\scriptsize]
  {$n_2-1$ children};

\end{tikzpicture}

\caption{We illustrate the insertions and regrouping procedure at a node
$a=a_1a_2a_3$ with three children, where
$\rk(a_i)=\rk(a)
=\rk(a^2)$.
At cut $i$ with $i\in\{1,2\}$, the  products $y_i$ and $x_i$
have the kernel and image of the prefix
$a_1\cdots a_i$ and suffix $a_{i+1}\cdots a_3$,
respectively.
The shaded powers are inserted first at the right cut,
then at the left.
Every child in (iv) has kernel $\ker(a)$ and
image $\im(a)$.
Throughout, the child product is congruent to the
unchanged root label $a$.}
\label{fig:closure-tree-modification}
\end{figure}

To obtain these factors, we insert suitable products
between the children and then regroup. Roughly speaking, 
consider a cut that divides $a_1\cdots a_s$ into
a prefix product $c_1$ and a suffix product $c_2$,
so that~$a=c_1c_2$.
To construct a suitable insertion, we focus on  the kernels
and images of  $c_1$ and $c_2$.
The key observation is that these subspaces can 
be obtained from a bounded number of their children.
In particular, by~\cite[Lemma~8]{Finiteness26} 
there are products $y$ and $x$,
each using at most $2^d+1$ children, such that
\[
 \ker(y)=\ker(c_1) \quad \im(y)=\im(c_1),
 \qquad \text{and} \qquad 
 \ker(x)=\ker(c_2)\quad \im(x)=\im(c_2).
\]
The rank conditions on the node imply that $b:=xy$
satisfies $\im(c_1b)=\im(c_1)$.
Claim~\ref{clm:closure-absorption} therefore gives
$c_1\sim c_1b^n$ for some $n\ge1$, and hence
$a=c_1c_2\sim c_1(xy)^nc_2$.
We insert such powers $(xy)^n$ starting at the rightmost cut and moving left, as shown in Figure~\ref{fig:closure-tree-modification}.
At each step, the powers already inserted lie to
the right of the current cut, so the prefix is still
 $c_1$ and we can apply the relation $c_1\sim c_1b^n$.
 We then regroup the resulting product into factors
whose kernel and image are those of $a$.
As explained, the construction is such that each new factor consists of at most two auxiliary
products and some (original) child $a_i$, and therefore
uses at most $2(2^d+1)+1\le 2^{d+2}$
matrices among the children $a_i$'s.
Since $a$ is rank-stable, these factors generate
a stable subsemigroup.
The details of this regrouping are given
in Appendix~\ref{app:closure}.

Applying this modification throughout the tree results a tree notion as follows.
Let $\sim$ be a congruence on $\langle X\rangle$.
A \emph{stable $\sim$-factorisation tree} of
$a\in\langle X\rangle$ is a finite ordered tree
with nodes  in~$\langle X\rangle$.
The root of the tree is 
$a$ and leaves lie  in $X$.
The nodes $c$ of the tree, except leaves, have children~$c_1,\ldots,c_s$ with \(s \geq 2\) such that 
(1) 
 $c\sim c_1\cdots c_s$ and
(2) the matrices $c_i$
generate a stable subsemigroup when $s\ge3$.
A \emph{stable $\sim$-factorisation forest} is a choice
of tree for every $a\in\langle X\rangle$.

The construction sketched above gives the following bounds.

\begin{restatable}{lemma}{StableForest}
\label{lem:forest}\label{thm:factorization}
Suppose $\sim$ is a congruence on $\langle X\rangle$
such that, for all $a,b\in\langle X\rangle$,
the equality $\im(a)=\im(ab)$ implies
$a\sim ab^n$ for some $n\ge1$.
Then $\langle X\rangle$ admits a stable
$\sim$-factorisation forest whose trees have height
at most $2d(d+3)$ and reduced height at most $d$.
\end{restatable}


To obtain these bounds, observe that each new factor
obtained after regrouping  is built from at most
$2^{d+2}$ matrices among the children $a_i$ of $a$,
using a binary tree of height at most $d+2$.
Here we were modifying an arbitrary node $a$ of the
tree, so its children $a_i$ may themselves be products
of matrices in $X$.
We process the  tree from the leaves upwards, so that for each  $a_i$ (re)appearing while processing $a$, we attach a copy of the already modified subtree rooted at $a_i$.
All in all, each such local modification adds at most $d+3$ levels without increasing the reduced height.
Since a path in the starting tree contains at most $d$
nodes with at least three children, the resulting height
is at most $d(d+3)+d(d+3)=2d(d+3)$,
and the reduced height remains at most $d$.

As the last step, we shorten these trees using the assumed group bound.
Consider a node whose children generate a stable
subsemigroup $S$. Taking the Zariski closure of $S$ inside the group of
matrices with its common kernel and image gives a group $G\le\cl{\langle X\rangle}$ containing $S$.
By assumption, $|G/{\sim}|\le N$, so $S$ meets at most
$N$ congruence classes.
Clearly, if this node had more than $N$ children, two nonempty prefix products of its children are congruent.
We can therefore delete the children between these
prefixes while preserving the congruence class, and repeat until the number of children is at most~$N$.

The remaining binary nodes contribute at most a factor
of two per level to the number of leaves.
The shortened nodes contribute at most a factor of $N$,
and at most $d$ such nodes occur on any  path  from the root to a leaf.
These two facts combined together yield the bound
$2^{2d(d+3)}N^d$ for the number of leaves in the final tree.
These leaves all lie in  $X$, and notably their product is
congruent to the root of the tree.
This establishes that every congruence class has length at most
$2^{2d(d+3)}N^d$.
This proves the diameter bound and, since $X$ is finite,
the finiteness of the quotient.
\end{proof}

\section{Application to Border Complexity}

\subsection{Separation of homogeneous ABPs and their closures}

Let $w\ge2$ and $d\ge3$.
A homogeneous algebraic branching program (ABP) of width
at most $w$ and degree $d$ computes a polynomial
\[
 f(\bm{x})=L_1(\bm{x})\cdots L_d(\bm{x}),
\]
where $L_1$ is a $1\times w$ matrix, $L_d$ is a
$w\times1$ matrix, and the remaining matrices are
$w\times w$.
The entries of the $L_i$ are homogeneous linear forms in the
input variables $\bm{x}=(x_1,\ldots,x_m)$, with coefficients in~$\overline{\Q}$.
We note in passing that, in this definition, we have exactly $d$ matrix factors.
ABPs  and their border complexity have
been studied in recent years~\cite{BringmannIkenmeyerZuiddam18,AllenderWang16,
BlaserEtAl20,DuttaEtAlSTACS24}.
Homogeneous models and their connections with algebraic
geometry have also received recent
attention~\cite{GesmundoEtAl22,
DuttaGesmundoIkenmeyerJindalLysikov24}.
In our constructions, we keep both the width $w$ and the degree $d$
fixed and use homogeneous linear forms as used in~\cite{DuttaGesmundoIkenmeyerJindalLysikov24}.

Write $\overline{\Q}[\bm{x}]_d$ for the vector space
of homogeneous degree-$d$ polynomials in the variables
$\bm{x}$, with coefficients in~$\overline{\Q}$.
Denote by $\mathcal A_{w,d}$ the set of polynomials $\overline{\Q}[\bm{x}]_{d}$
computed by the ABPs defined above. 
As usual, viewing each polynomial as its vector
of coefficients, we 
take the Zariski closure of the set~$\mathcal A_{w,d}$ of polynomials in the vector space of
homogeneous degree-$d$ polynomials over $\overline{\Q}$; we
denote this closure by~$\cl{\mathcal A_{w,d}}$.

Below, we prove the following separation and describe
it through a semigroup congruence.

\HomogeneousABPSeparation*

We will give an explicit 
polynomial $F_{w,d}$  with integer coefficients such that that 
$F_{w,d} \in \overline{\mathcal A_{w,d}}\setminus\mathcal A_{w,d}$.

\begin{proof}[Proof Sketch of
Theorem~\ref{theo:homogeneous-abp-separation}]
We first construct a matrix semigroup and a congruence
whose zero class consists solely of an encoding
of polynomials in $\mathcal A_{w,d}$.
We then give a polynomial in
$\overline{\mathcal A_{w,d}}\setminus\mathcal A_{w,d}$
and show that its encoding belongs to a different class.
The details including the proofs of the claims are provided in
Appendices~\ref{app:abp-separation}.

To construct the semigroup,  we use a notion of universal ABPs. 
For this aim, we replace each coefficient
of each linear form in  entries of the ABP by an independent
parameter~$\theta_i$; 
let the vector~$\bm{\theta}=(\theta_1,\ldots,\theta_\nu)$
denote the introduced formal parameters 
in this way, where $\nu=m\bigl(2w+(d-2)w^2\bigr)$.
Denote the polynomial computed by this universal
ABP by~$U(\bm{\theta},\bm{x})$, which is linear in each $\theta_i$.  Choosing values for the parameters produces all ABPs
in~$\mathcal A_{w,d}$:

\[
 \mathcal A_{w,d}
 =
 \{U(\bm{t},\bm{x}):
       \bm{t}\in\overline{\Q}^{\,\nu}\}.
\]

Below, we use translations to describe the choices of parameters~$\bm \theta$.
For $\bm{t}\in\overline{\Q}^{\,\nu}$, the translation
$\bm{\theta}\mapsto\bm{\theta}+\bm{t}$ replaces each
$\theta_i$ by $\theta_i+t_i$, leaving the input variables $\bm{x}$
unchanged.
Starting from the zero assignment, this gives the
assignment $\bm{\theta}=\bm{t}$.
Our aim is to represent these translations by matrices
and then recover the polynomial computed by the
corresponding ABP.

Recall that each parameter $\theta_i$ belongs to one layer of the universal
ABP, and each produced monomial will have total degree $d$ in $\bm \theta$. The combination of translations might however produce monomials with smaller total degree, for example, 
$(\theta_i+t_i)(\theta_j+t_j)
 =
 \theta_i\theta_j+t_j\theta_i+t_i\theta_j+t_it_j$.
We therefore collect all monomials obtained by choosing
at most one parameter from each layer into a row vector
$v(\bm{\theta})$.
The number of such monomials is  
$L_{\theta}=(1+mw)^2(1+mw^2)^{d-2}$;
we then fix an order on them,  placing the constant monomial $1$ at  first.
This allows us define
$\rho_{\bm{t}}\in\mat_{L_{\theta}}(\overline{\Q})$, for~$\bm{t}\in\overline{\Q}^{\,\nu}$, 
by
$v(\bm{\theta})\rho_{\bm{t}}
 =v(\bm{\theta}+\bm{t})$.
Composing translations gives
\[
 \rho_{\bm{0}}=I_{L_{\theta}},
 \qquad
 \rho_{\bm{s}}\rho_{\bm{t}}
 =\rho_{\bm{s}+\bm{t}},
 \qquad
 \rho_{\bm{t}}^{-1}=\rho_{-\bm{t}}.
\]

Recall that the universal
polynomial  is homogenous of degree $d$ in variables $\bm x$. There are $L_{x}:=\binom{m+d-1}{d}$ many such homogenous monomials, we again fix an order on them.  
Write $\operatorname{coeff}$ for the coefficient map in this order:
\[
 \operatorname{coeff}:
 \overline{\Q}[\bm{\theta}][\bm{x}]_d
 \longrightarrow
 \overline{\Q}[\bm{\theta}]^{1\times L_x}\,
\]
which extracts  coefficients with respect to
$\bm{x}$ in the fixed order of monomials in~$\bm{x}$.
By the above, since the universal polynomial lie in
$\Q[\bm{\theta}][\bm{x}]_d$, and since each of its coefficients is a $\Q$-linear combination of coordinates in
$v(\bm{\theta})$, there is a fixed matrix
$C_U\in\Q^{L_{\theta}\times L_x}$ such that
$\operatorname{coeff}
 \bigl(U(\bm{\theta},\bm{x})\bigr)
 =v(\bm{\theta})C_U$.
Translating the parameters gives
$\operatorname{coeff}
 \bigl(U(\bm{\theta}+\bm{t},\bm{x})\bigr)
 =
 v(\bm{\theta}+\bm{t})C_U
 =
 v(\bm{\theta})\rho_{\bm{t}}C_U$. Moreover, evaluating at $\bm{\theta}=\bm{0}$  yields
\begin{equation}\label{eq:abp-coefficient-evaluation}
 v(\bm{0})\rho_{\bm{t}}C_U
 =
 \operatorname{coeff}\bigl(U(\bm{t},\bm{x})\bigr)\,,
\end{equation} 
where  $v(\bm{0})=(1,\bm{0})$ is the value of
the monomial row at the zero parameter assignment. Intuitively speaking, multiplication of $v(\bm{0})$ by $\rho_{\bm{t}}$ changes it to
$v(\bm{t})$, and then multiplication by $C_U$  gives
the coefficient vector of the resulting ABP; see Figure~\ref{fig:abp-parameter-example} for an example. 

\begin{figure}[t]
\centering
\scalebox{.8}{
\begin{tikzpicture}[
  font=\large,
  line width=0.65pt,
  >={Stealth[length=5pt,width=4.5pt]}
]
  \foreach \i/\xx in {0/-0.3,1/3.65,2/7.6,3/11.5}{
    \node[
      circle,draw,fill=white,
      inner sep=0pt,minimum size=5pt
    ] (q\i) at (\xx,3.55) {};
  }
  \draw[->] (q0) -- node[above=6pt] {$\theta_1x$} (q1);
  \draw[->] (q1) -- node[above=6pt] {$\theta_2x+y$} (q2);
  \draw[->] (q2) -- node[above=6pt] {$x$} (q3);

  \node at (-0.3,2.35) {$v(\bm{0})$};
  \node at (5.0,2.35) {$v(\bm{t})$};
  \node at (11.5,2.35)
    {$\operatorname{coeff}(U(\bm{t},x,y))$};

  \node[inner sep=4pt] (initial) at (-0.3,1.70)
    {$(1,0,0,0)$};
  \node[inner sep=4pt] (translated) at (5.0,1.70)
    {$(1,2,3,6)$};
  \node[inner sep=4pt] (output) at (11.5,1.70)
    {$(6,2,0,0)$};

  \draw[->] (initial.east)
    -- node[above=5pt] {$\times\rho_{\bm{t}}$}
    (translated.west);
  \draw[->] (translated.east)
    -- node[above=5pt] {$\times C_U$}
    (output.west);

  \node at (1.7,-0.22) {
    $\rho_{\bm{t}}=
    \begin{pmatrix}
      1&2&3&6\\
      0&1&0&3\\
      0&0&1&2\\
      0&0&0&1
    \end{pmatrix}$
  };
  \node at (8.5,-0.22) {
    $C_U=
    \begin{pmatrix}
      0&0&0&0\\
      0&1&0&0\\
      0&0&0&0\\
      1&0&0&0
    \end{pmatrix}$
  };
\end{tikzpicture}%
}
\caption{A small instance of the universal ABP with width~$1$ and degree~$3$, with input variables $x,y$. For simplicity of the example, we fix all parameters to 0 or 1, except $\theta_1,\theta_2$.
This three-layer ABP computes
$U(\theta_1,\theta_2,x,y)=(\theta_1x)(\theta_2x+y)x$.
We use
$v(\bm{\theta})=(1,\theta_1,\theta_2,\theta_1\theta_2)$
and order the input monomials as $(x^3,x^2y,xy^2,y^3)$.
At $\bm{t}=(2,3)$, multiplication by $\rho_{\bm{t}}$
gives $v(\bm{t})=(1,2,3,6)$, whose coordinates are the values of
$1,\theta_1,\theta_2,\theta_1\theta_2$ at $\bm\theta=\bm t$.
The multiplication by $C_U$
extracts the coefficients of $6x^3+2x^2y$ as shwon above.}
\label{fig:abp-parameter-example}
\end{figure}

By Equation~\eqref{eq:abp-coefficient-evaluation} in the above, we can 
expresses the evaluation of the universal ABP at  a parameter
as a matrix product; 
below, we  place the defined matrices  into
 one common dimension, so that
we can work within a matrix semigroup.
To this aim, we consider 4 blocks of size $1$, $L_\theta$, $L_\theta$, $L_x+L_\theta$, and write 
$L_{\mathrm{tot}}:=1+3L_\theta+L_x$ for the total length of these blocks. We define a linear map 
$\iota:\overline{\Q}[\bm{x}]_d
 \to \mat_{L_{\mathrm{tot}}}(\overline{\Q})$ as follows, for all polynomials
$f\in\overline{\Q}[\bm{x}]_d$, 
\begingroup
\setlength{\arraycolsep}{2pt}
\renewcommand{\arraystretch}{1.7}
\[
\iota(f):=
\left(
\begin{array}{c:c:c:c}
 \bm{0}_{1\times1}
 &
 \bm{0}_{1\times L_\theta}
 &
 \bm{0}_{1\times L_\theta}
 &
 \bigl(\operatorname{coeff}(f),
       \bm{0}_{1\times L_\theta}\bigr)
 \\ \hdashline
 \bm{0}_{L_\theta\times1}
 &
 \bm{0}_{L_\theta\times L_\theta}
 &
 \bm{0}_{L_\theta\times L_\theta}
 &
 \bm{0}_{L_\theta\times(L_x+L_\theta)}
 \\ \hdashline
 \bm{0}_{L_\theta\times1}
 &
 \bm{0}_{L_\theta\times L_\theta}
 &
 \bm{0}_{L_\theta\times L_\theta}
 &
 \bm{0}_{L_\theta\times(L_x+L_\theta)}
 \\ \hdashline
 \bm{0}_{(L_x+L_\theta)\times1}
 &
 \bm{0}_{(L_x+L_\theta)\times L_\theta}
 &
 \bm{0}_{(L_x+L_\theta)\times L_\theta}
 &
 \bm{0}_{(L_x+L_\theta)\times(L_x+L_\theta)}
\end{array}
\right)\,,
\]
\endgroup
where as defined above $\operatorname{coeff}(f)$ is a row vector with dimension
$L_x$.
Observe that the map $\iota$ is injective, since the coefficient
vector determines the polynomial. 

We need to encode $v(\bm{0})\rho_{\bm{t}}C_U
 =
 \operatorname{coeff}\bigl(U(\bm{t},\bm{x})\bigr)$ given in
Equation~\eqref{eq:abp-coefficient-evaluation} in this higher dimensional space, where we  represent these  operations by square
matrices of dimension $L_{\mathrm{tot}}$, with
the following correspondence:
\[
 \begin{array}{ccc}
 v(\bm{0}) & \rho_{\bm{t}} & C_U\\
 \downarrow & \downarrow & \downarrow\\
 V_0 & \Gamma_{\bm{t}} & C
 \end{array}
\]
where $V_0$ contains the initial row $v(\bm{0})$,
$\Gamma_{\bm{t}}$ contains the translation matrix,
and $C$ contains the coefficient matrix $C_U$. Using the same four block sizes as for $\iota(f)$, we
define
$\Gamma_{\bm{t}}
 :=
 \operatorname{diag}
 \bigl(1,\rho_{\bm{t}},\rho_{\bm{t}},
             I_{L_x+L_\theta}\bigr)$,
and
\begingroup
\setlength{\arraycolsep}{3pt}
\renewcommand{\arraystretch}{1.4}
\[
V_0:=
\left(
\begin{array}{c:c:c:c}
 \bm{0}&v(\bm{0})&\bm{0}&\bm{0}
 \\ \hdashline
 \bm{0}&\bm{0}&\bm{0}&\bm{0}
 \\ \hdashline
 \bm{0}&\bm{0}&\bm{0}&\bm{0}
 \\ \hdashline
 \bm{0}&\bm{0}&\bm{0}&\bm{0}
\end{array}
\right)
\qquad \text{and} 
\qquad 
C:=
\left(
\begin{array}{c:c:c:c}
 \bm{0}&\bm{0}&\bm{0}&\bm{0}
 \\ \hdashline
 \bm{0}&\bm{0}&\bm{0}&(C_U,\bm{0})
 \\ \hdashline
 \bm{0}&\bm{0}&\bm{0}&(\bm{0},I_{L_\theta})
 \\ \hdashline
 \bm{0}&\bm{0}&\bm{0}&\bm{0}
\end{array}
\right).
\]
\endgroup
It is straightforward to verify that multiplying $V_0$ by $\Gamma_{\bm{t}}$ replaces
its nonzero row block by
$v(\bm{0})\rho_{\bm{t}}=v(\bm{t})$. Then
multiplication by $C$  places
$\bigl(v(\bm{t})C_U,\bm{0}\bigr)
 =
 \bigl(\operatorname{coeff}
       (U(\bm{t},\bm{x})),\bm{0}\bigr)$
in the top-right block.
All other blocks are zero.
Consequently,
\begin{equation}\label{eq:universal-abp-evaluation}
 V_0\Gamma_{\bm{t}}C
 =
 \iota\bigl(U(\bm{t},\bm{x})\bigr).
\end{equation}
In particular, since $U(\bm{0},\bm{x})=0$,
we have $V_0C=\bm{0}$; see Figure~\ref{fig:abp-matrix-example} for an illustration.

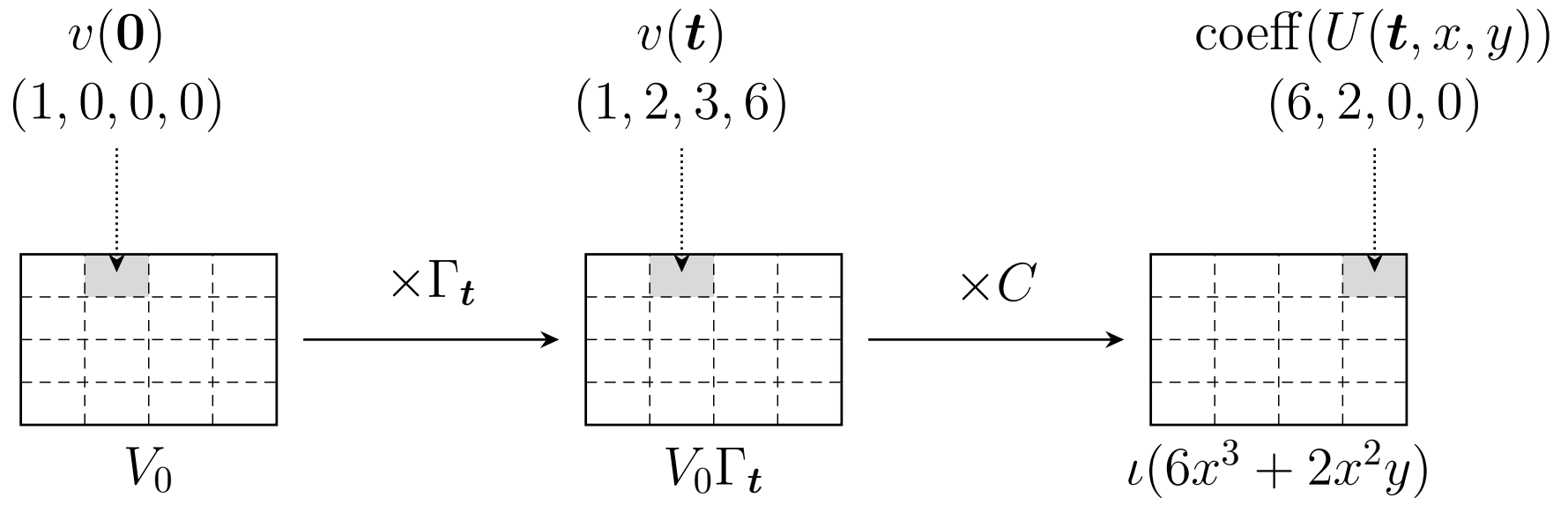
\begin{figure}[t]
\centering
\scalebox{.8}{
\begin{tikzpicture}[
  font=\large,
  line width=0.65pt,
  >={Stealth[length=5pt,width=4.5pt]}
]
  \node at (-0.3,2.85) {$v(\bm{0})$};
  \node at (5.0,2.85) {$v(\bm{t})$};
  \node at (11.5,2.85)
    {$\operatorname{coeff}(U(\bm{t},x,y))$};

  \node[inner sep=4pt] (initial) at (-0.3,2.20)
    {$(1,0,0,0)$};
  \node[inner sep=4pt] (translated) at (5.0,2.20)
    {$(1,2,3,6)$};
  \node[inner sep=4pt] (output) at (11.5,2.20)
    {$(6,2,0,0)$};

  \foreach \ox/\active in {0/1,5.3/1,10.6/3}{
    \fill[black!15]
      (\ox-1.2+0.6*\active,0.4)
      rectangle
      (\ox-0.6+0.6*\active,0.8);

    \draw
      (\ox-1.2,-0.8)
      rectangle
      (\ox+1.2,0.8);

    \foreach \j in {1,2,3}{
      \draw[densely dashed,line width=0.35pt]
        (\ox-1.2+0.6*\j,-0.8)
        --
        (\ox-1.2+0.6*\j,0.8);

      \draw[densely dashed,line width=0.35pt]
        (\ox-1.2,-0.8+0.4*\j)
        --
        (\ox+1.2,-0.8+0.4*\j);
    }
  }

  \draw[->,densely dotted]
    (initial.south) -- (-0.3,0.63);
  \draw[->,densely dotted]
    (translated.south) -- (5.0,0.63);
  \draw[->,densely dotted]
    (output.south) -- (11.5,0.63);

  \draw[->] (1.45,0)
    -- node[above=5pt] {$\times\Gamma_{\bm{t}}$}
    (3.85,0);
  \draw[->] (6.75,0)
    -- node[above=5pt] {$\times C$}
    (9.15,0);

  \node at (0,-1.22) {$V_0$};
  \node at (5.3,-1.22) {$V_0\Gamma_{\bm{t}}$};
  \node at (10.6,-1.22) {$\iota(6x^3+2x^2y)$};
\end{tikzpicture}%
}
\caption{The  ABP from
Figure~\ref{fig:abp-parameter-example}
with $\bm{t}=(2,3)$ is encoded by square
matrices.
Here $L_\theta=L_x=4$, so the four block sizes
are $1,4,4,8$ and the matrices have dimension~$17$.
The shaded blocks contain the rows displayed above,
with four trailing zeros in the final block;
all unshaded blocks are zero.
Thus $V_0\Gamma_{\bm{t}}C=\iota(6x^3+2x^2y)$.}
\label{fig:abp-matrix-example}
\end{figure}

We now use these matrices to construct a finitely
generated semigroup.
The translation matrices form a group, since
\[
 \Gamma_{\bm{0}}=I_{L_{\mathrm{tot}}},
 \qquad
 \Gamma_{\bm{s}}\Gamma_{\bm{t}}
 =\Gamma_{\bm{s}+\bm{t}},
 \qquad
 \Gamma_{\bm{t}}^{-1}=\Gamma_{-\bm{t}}.
\]
To obtain finitely many generators, we take the
translations by the standard basis vectors and
their negatives, together with $V_0$ and $C$.
Thus, writing $\bm{e}_1,\ldots,\bm{e}_\nu$ for
the standard basis vectors, define
\[
 X:=
 \{\Gamma_{\bm{e}_i},\Gamma_{-\bm{e}_i}:
       1\le i\le\nu\}\cup\{V_0,C\}.
\]
All these matrices belong to
$\mat_{L_{\mathrm{tot}}}(\Q)$.
Indeed, $C_U$ is rational, and the translation
matrices are rational at rational parameter values. Moreover, $\langle \Gamma_{\bm{e}_i},\Gamma_{-\bm{e}_i}:
       1\le i\le\nu \rangle$ give
$\Gamma_{\bm{n}}$ for every $\bm{n}\in\Z^\nu$.
The non-zero blocks  of $V_0$ and $C$ restrict wrong encodings, as 
\[
 \Gamma_{\bm{t}}V_0=V_0 \qquad \text{and}
 \qquad
 C\Gamma_{\bm{t}}=C \qquad \text{and}
 \qquad
 V_0^2=C^2=CV_0=\bm{0}.
\]
Together with
Equation~\eqref{eq:universal-abp-evaluation},
these identities show that every element of
$\langle X\rangle$ has one of the forms
\[
 \Gamma_{\bm{n}}\qquad \text{and} \qquad
 V_0\Gamma_{\bm{n}} \qquad \text{and}\qquad
 \Gamma_{\bm{n}}C \qquad \text{and}\qquad
 \iota\bigl(U(\bm{n},\bm{x})\bigr)\,,
\]
with $\bm{n}\in\Z^\nu$. 
Conversely, every matrix in this list belongs
to $\langle X\rangle$.
We now take the Zariski closure of $\langle X\rangle$
in $\mat_{L_{\mathrm{tot}}}(\overline{\Q})$.
The map
$\bm{t}\mapsto\iota(U(\bm{t},\bm{x}))$
is polynomial, and $\Z^\nu$ is Zariski dense in~$\overline{\Q}^{\,\nu}$.
Consequently,
\[
 \cl{
 \{\iota(U(\bm{n},\bm{x})):\bm{n}\in\Z^\nu\}}
 =
 \cl{\iota(\mathcal A_{w,d})}
 =
 \iota\bigl(\overline{\mathcal A_{w,d}}\bigr),
\]
where the last equality uses that $\iota$ is a
linear embedding. The proof of the following claim is in Appendix~\ref{app:abp-separation}.
\begin{restatable}{claim}{ABPClosureDescription}
\label{clm:abp-closure-description}
$\cl{\langle X\rangle}
 =
 \{\Gamma_{\bm{t}},V_0\Gamma_{\bm{t}},\Gamma_{\bm{t}}C:
       \bm{t}\in\overline{\Q}^{\,\nu}\}
 \;\dot\cup\;
 \iota\bigl(\overline{\mathcal A_{w,d}}\bigr)$.
\end{restatable}
Informally speaking, each of these 
families $\{\Gamma_{\bm{t}},V_0\Gamma_{\bm{t}},\Gamma_{\bm{t}}C:
       \bm{t}\in\overline{\Q}^{\,\nu}\}$ is already Zariski closed. For example, in the case of 
 $\Gamma_{\bm{t}}$ and $V_0\Gamma_{\bm{t}}$,
the parameter values can be recovered from
matrix entries, and all remaining entries
are polynomial functions of them.

We are now in a position to  define the congruence~$\sim$.
Our aim is to identify translations with the
identity, so that the parameter assignment
disappears when passing to the quotient.
Let $\sim$ be the least  semigroup congruence
on~$\cl{\langle X\rangle}$, with respect to inclusion, satisfying
$\Gamma_{\bm{t}}\sim I_{L_{\mathrm{tot}}}$ 
for every $\bm{t}\in\overline{\Q}^{\,\nu}$.
The 
compatibility of $\sim$ with multiplication  gives
\[
 \iota\bigl(U(\bm{t},\bm{x})\bigr)
 =
 V_0\Gamma_{\bm{t}}C
 \sim V_0C=\bm{0}\,
\]
implying that  the encoding of every  computable
polynomial with ABPs belongs to the zero class.

\begin{restatable}{claim}{UniversalABPCongruence}
\label{clm:universal-abp-congruence}
The  class of zero matrix consists  of the encoding of polynomials computable by ABPs, that is, 
\[
 [\bm{0}]_\sim=\iota(\mathcal A_{w,d}).
\]
\end{restatable}

We next count the classes that meet $\langle X\rangle$.

\begin{restatable}{claim}{UniversalABPQuotient}
\label{clm:universal-abp-quotient}
The quotient $\langle X\rangle/{\sim}$ has exactly
four classes, represented by
$I_{L_{\mathrm{tot}}}$, $V_0$, $C$, and $\bm{0}$.
\end{restatable}

This finiteness here  follows from
Theorem~\ref{TheoremMainQuotient}.
Indeed,  every group in
$\cl{\langle X\rangle}$ is either contained in the
translation group or consists only of $\bm{0}$.
By the definition of $\sim$, each such group has
a singleton quotient, so the theorem applies with~$N=1$.
Our particular construction is also such that  the finiteness and the count on the classes also directly follows from the definition and the  multiplication rules.

We now turn to the separation of $\mathcal A_{w,d}$ and $\overline{\mathcal A_{w,d}}$.
Claim~\ref{clm:universal-abp-congruence} says that a computable polynomial with ABPs belongs to $[\bm0]_\sim$; below, 
we  construct a polynomial in
$\overline{\mathcal A_{w,d}}$ whose encoding does not
belong to this class.
(We show an example in Figure~\ref{fig:abp-closure-congruence-example} showing the underlying phenomenon, following Figures~\ref{fig:abp-parameter-example} and \ref{fig:abp-matrix-example}.)

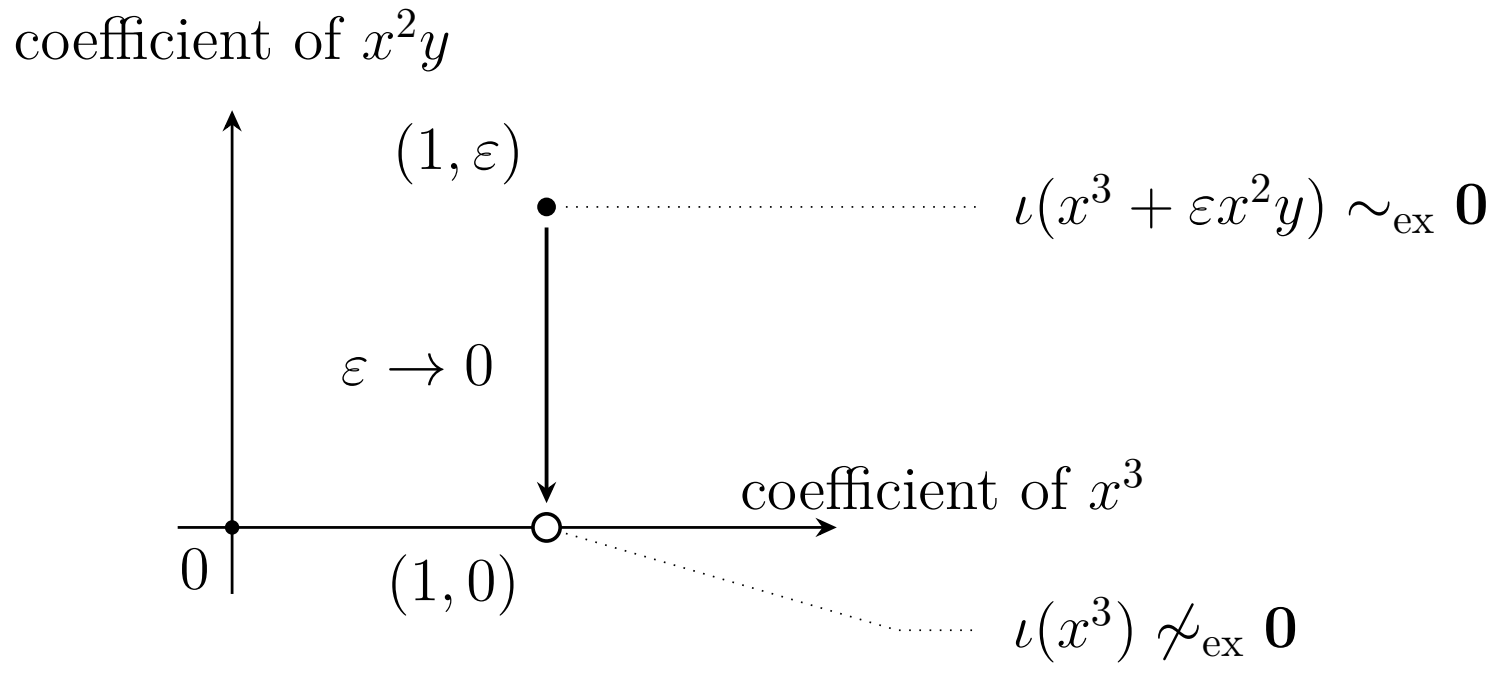
\begin{figure}[t]
\centering
\scalebox{.8}{
\begin{tikzpicture}[
  font=\large,
  line width=0.65pt,
  >={Stealth[length=5pt,width=4.5pt]}
]

\draw[->] (-0.45,0) -- (5.0,0);
\draw[->] (0,-0.55) -- (0,3.45);

\node[above=5pt] at (0,3.45)
  {coefficient of $x^2y$};
\node[above right,inner sep=4pt] at (4.05,0)
  {coefficient of $x^3$};

\fill (0,0) circle[radius=1.7pt];
\node[below left,inner sep=5pt] at (0,0) {$0$};

\coordinate (before) at (2.6,2.65);
\coordinate (after) at (2.6,0);

\draw[->,line width=0.9pt]
  (2.6,2.48) --
  node[left=8pt] {$\varepsilon\to0$}
  (2.6,0.20);

\fill (before) circle[radius=2.2pt];
\node[above left,inner sep=5pt] at (before)
  {$(1,\varepsilon)$};

\draw[fill=white,line width=0.85pt]
  (after) circle[radius=3.2pt];
\node[below left,inner sep=6pt] at (after)
  {$(1,0)$};

\draw[dotted,line width=0.4pt]
  (2.76,2.65) -- (6.15,2.65);
\node[anchor=west,inner sep=4pt] at (6.3,2.65)
  {$\iota(x^3+\varepsilon x^2y)
    \sim_{\mathrm{ex}}\bm{0}$};

\draw[dotted,line width=0.4pt]
  (2.76,-0.05) -- (5.50,-0.85) -- (6.15,-0.85);
\node[anchor=west,inner sep=4pt] at (6.3,-0.85)
  {$\iota(x^3)\not\sim_{\mathrm{ex}}\bm{0}$};

\end{tikzpicture}%
}
\caption{%
Closure and congruence for
$U(\theta_1,\theta_2,x,y)=\theta_1\theta_2x^3+\theta_1x^2y$ given in Figures~\ref{fig:abp-parameter-example} and \ref{fig:abp-matrix-example}.
For $\varepsilon\ne0$, the parameter assignment 
$(\theta_1,\theta_2)=(\varepsilon,\varepsilon^{-1})$
gives $x^3+\varepsilon x^2y$.
In the diagram, we see that the   pair
$(1,\varepsilon)$ approaching $(1,0)$.
The  point $(1,\varepsilon)$ represents an output
of the family, while the  point $(1,0)$ is present only in its closure.
Recall that we construct the  congruence $\sim_{\mathrm{ex}}$ to be the least congruence on
the closure of the matrix semigroup constructed from
this family satisfying
$\Gamma_{\bm{t}}\sim_{\mathrm{ex}}I_{17}$
for every $\bm{t}\in\overline{\mathbb Q}^{\,2}$.
The  zero class consists exactly of the encoding of polynomials computable by this two-parameter family of ABPs. Indeed,
$\iota(x^3+\varepsilon x^2y)
 =V_0\Gamma_{(\varepsilon,\varepsilon^{-1})}C
 \sim_{\mathrm{ex}}V_0C=\bm{0}$.
At $\varepsilon=0$, this encoded output specializes
to $\iota(x^3)$, which belongs to the matrix closure
but lies outside the zero class (producing $x^3$ would require both $\theta_1=0$
and $\theta_1\theta_2=1$).
This is due to the fact  that the zero class need not be Zariski closed.
}
\label{fig:abp-closure-congruence-example}
\end{figure}

For simplicity of construction,  we rename  the input
variables by 
$x,y,a,b,u,v,z,p,q,r,s$ and 
 $c_i,z_i$ with $1\leq i\leq w-2$ (for $w=2$, there is  no $c_i$ and $z_i$).
 We first define the polynomials for width two as follows, for all $k\ge0$, 
\[
\begin{aligned}
 F_{2,2k+3}
 &:=
 a\xi_k^2+b\eta_k^2
 +z\Delta^k(ux+vy),\\
 F_{2,2k+4}
 &:=
 a\xi_{k+1}\xi_k+b\eta_{k+1}\eta_k
 +\Delta^{k+1}(ux+vy),
 \qquad k\ge0\,
\end{aligned}
\]
where $\Delta:=ps-qr$ and 
\[
 \binom{\xi_k}{\eta_k}:=G^k\binom{x}{y}
 \qquad 
 \text{with}
 \qquad 
 G:=\begin{pmatrix}
 p&q\\
 r&s
 \end{pmatrix}
 .
\]
For general width $w$, define
\begin{equation}\label{eq:abp-witness}
 F_{w,d}:=
 F_{2,d}+\sum_{i=1}^{w-2}c_i z_i^{d-1}.
\end{equation}
Observe that polynomials~$F_{w,d}$ are homogeneous of degree $d$ and
have integer coefficients.
For example,
\[
 F_{w,3}
 =
 ax^2+by^2+z(ux+vy)
 +\sum_{i=1}^{w-2}c_i z_i^2.
\]

We first prove that these polynomials belong to the closure~$\overline{\mathcal A_{w,d}}$.

\begin{restatable}{claim}{ABPApproximation}
\label{clm:abp-approximation}
$F_{w,d}\in\overline{\mathcal A_{w,d}}$.
\end{restatable}

The main idea in proving Claim~\ref{clm:abp-approximation} is to give an ABP depending on a nonzero parameter $\varepsilon$, which specializes
to~$F_{w,d}$ when~$\varepsilon=0$. Below, we  consider width $2$. For the odd degree $d=2k+3$, consider the matrix product
\[
 (x+\varepsilon v,\ y-\varepsilon u)
 (G^{\mathsf T})^k
 \begin{pmatrix}
 a&z/\varepsilon\\
 -z/\varepsilon&b
 \end{pmatrix}
 G^k
 \binom{x}{y}\,
\]
and for even degree $d=2k+4$, use
\[
 (x+\varepsilon v,\ y-\varepsilon u)
 (G^{\mathsf T})^{k+1}
 \begin{pmatrix}
 a+r/\varepsilon&s/\varepsilon\\
 -p/\varepsilon&b-q/\varepsilon
 \end{pmatrix}
 G^k
 \binom{x}{y}.
\]
Observe that these two given  products  define  width-$2$
ABPs with degree $d$  whenever $\varepsilon\ne0$;  morover,  
the result of the products  have the form
$F_{2,d}+\varepsilon H_{2,d}$, where $H_{2,d}$ is
homogeneous of degree $d$ and independent of $\varepsilon$.
For example, at degree $3$ we have
\[
 (x+\varepsilon v,\ y-\varepsilon u)
 \begin{pmatrix}
 a&z/\varepsilon\\
 -z/\varepsilon&b
 \end{pmatrix}
 \binom{x}{y}
 =
 F_{2,3}+\varepsilon(avx-buy).
\]
Consequently, $F_{2,d}$ belongs to
$\overline{\mathcal A_{2,d}}$, proving
Claim~\ref{clm:abp-approximation} for $w=2$; see 
Appendix~\ref{app:abp-separation} for details and discussion for other widths.

We have shown that $F_{w,d}\in \overline{\mathcal A_{w,d}}$.
It remains to prove that $F_{w,d}\not \in \mathcal A_{w,d}$, for which 
 we will use 
Claim~\ref{clm:universal-abp-congruence} about the zero congruent class.

\begin{restatable}{claim}{ABPWitnessClass}
\label{clm:abp-witness-class}
$\iota(F_{w,d})\not\sim\bm0$.
\end{restatable}

Suppose, towards a contradiction, that
$\iota(F_{w,d})\sim\bm0$.
By Claim~\ref{clm:universal-abp-congruence},
$F_{w,d}$ is computed by an ABP of width at most $w$
and degree $d$.
We examine the first and last layers of this ABP, where 
each consists of at most $w$ homogeneous linear forms in $\bm{x}$. The polynomial
 $F_{w,d}$ vanishes whenever all such forms
in either layer vanish.
The main algebraic step is to show that 
each of these forms is  a linear combination of variables
$x,y$ and $c_i, z_i$ with $1\leq w-2$.
This follows from a property of the linear subspaces
on which $F_{w,d}$ vanishes, proved in
Appendix~\ref{app:abp-separation}.
Since every terms produced in  the ABP contains a form from
the first layer and a form from the last layer, 
 every monomial  must contain
at least two occurrences of the   variables $x,y$ and $c_i, z_i$.
However, the terms involving $ux+vy$ in $F_{w,d}$
contain exactly one of these variables.
They are nonzero and cannot be cancelled by the
other summands,  a contradiction.
\color{black}

We conclude that $\iota(F_{w,d})\not\sim\bm0$,
proving Claim~\ref{clm:abp-witness-class}.
By Claim~\ref{clm:universal-abp-congruence}, this implies
$F_{w,d}\notin\mathcal A_{w,d}$.
Together with Claim~\ref{clm:abp-approximation}, we obtain
\[
 F_{w,d}\in
 \overline{\mathcal A_{w,d}}\setminus\mathcal A_{w,d},
\]
which completes the proof of Theorem~\ref{theo:homogeneous-abp-separation}.
\end{proof}

\subsection{Separation of $\mathrm{Det}_n$  from $\mathrm{Perm}_n$ under  a subclass of reductions}

Let $n\geq 3$. Consider the matrix
\[
[x_{i,j}]_{1\leq i,j\leq n}
=
\begin{pmatrix}
x_{1,1} & \cdots & x_{1,n}\\
\vdots  & \ddots & \vdots\\
x_{n,1} & \cdots & x_{n,n}
\end{pmatrix},
\]
whose entries are the formal variables $x_{i,j}$.
We write $\bm{x}_i:=(x_{i,1},\ldots,x_{i,n})$ for the
$i$-th row of this matrix, and let
$\bm{X}:=(\bm{x}_1,\ldots,\bm{x}_n)$
be the vector of its $n^2$ variables.
We denote by $\mathrm{Det}_n$ and $\mathrm{Perm}_n$ the determinant and permanent polynomials, respectively, defined by
\[
\mathrm{Det}_n(\bm{X})
:=
\sum_{\sigma\in\mathfrak S_n}
\operatorname{sgn}(\sigma)
\prod_{i=1}^n x_{i,\sigma(i)}
\qquad 
\text{and}
\qquad
\mathrm{Perm}_n(\bm{X})
:=
\sum_{\sigma\in\mathfrak S_n}
\prod_{i=1}^n x_{i,\sigma(i)},
\]
where $\mathfrak S_n$ is the set of permutations of
$\{1,\ldots,n\}$ and $\operatorname{sgn}(\sigma)$ denotes the sign of $\sigma$.

Below, we use the standard action of
$\mat_{n^2}(\overline{\Q})$ on the polynomial ring
$\overline{\Q}[\bm{X}]$,
defined by
\[
 (h \triangleleft L)(\bm{x})
 :=
 h(\bm{x}L)\,,
\]
where $L\in \mat_{n^2}(\overline{\Q})$ and $h\in \overline{\Q}[\bm{X}]$. 
This is the action of linear substitutions in the
$n^2$ variables; as defined in ~\cite[Section~3.2]{BlaserIkenmeyer25}.
It satisfies
\[
 h \triangleleft(L_1L_2)
 =
(h \triangleleft L_1) \triangleleft L_2 \qquad \text{and}
 \qquad
 h\triangleleft I_{n^2}=h\,,
\]
where $I_{n^2}$ is the identity matrix of dimension~$n^2$.
Curiously, under the congruence of  mutual reducibility,
    the determinant orbits closure has a finite quotient whereas that of permanent is not finite. 

We  construct a finitely generated monoid
$\langle Y_n\rangle\leq\mat_{n^2}(\Q)$ such that
\[
 \overline{\mathrm{Perm}_n \triangleleft \langle Y_n\rangle
 }
 \not\subseteq
 \overline{\mathrm{Det}_n \triangleleft\langle Y_n\rangle}.
\]
Notably, under the congruence of  mutual reducibility,
    the determinant orbits closure has a finite quotient whereas that of permanent is not finite.
For simplicity, to construct the finite set
$Y_n\subseteq\mat_{n^2}(\Q)$, we first define  
 $X_n\subseteq \mat_n(\Q)$ and use the embedding
\[
 \jmath(M):=
 \operatorname{diag}
 \bigl(\underbrace{M,\ldots,M}_{n\text{ blocks}}\bigr)
 \in\mat_{n^2}(\overline{\Q}).
\]
Then $\bm{X}\jmath(M)
 =
 (\bm{x}_1M,\ldots,\bm{x}_nM)$.
Thus $\jmath(M)$ acts on every row $\bm{x}_i$ of the matrix $[x_{i,j}]_{1\leq i,j\leq n}$ of variables, and
 \[\mathrm{Det}_n \triangleleft \jmath(M)
 =
 \det
 \begin{pmatrix}
 \bm{x}_1M\\
 \vdots\\
 \bm{x}_nM
 \end{pmatrix}
 \qquad \text{and} \qquad
 \mathrm{Perm}_n \triangleleft \jmath(M)
 =
 \operatorname{per}
 \begin{pmatrix}
 \bm{x}_1M\\
 \vdots\\
 \bm{x}_nM
 \end{pmatrix}.
 \]

Let us first introduce the  generators~$X_n$.
Recall that $\bm e_j$ denote the $j$-th
standard basis vector of $\Q^n$, and let $\bm{1}$
be the row vector with all entries being ones.
For $2\leq j\leq n$, define
\[
 R_j:=(\bm e_j-\bm e_1)\bm{1}
 =
\begin{pmatrix}
-\bm 1\\
\bm 0_{(j-2)\times n}\\
\bm 1\\
\bm 0_{(n-j)\times n}
\end{pmatrix}\,,
\]
where the first row of $R_j$ consists entirely of $-1$,
the $j$-th row consists entirely of $1$, and  zero otherwise.
Since
$\bm{1}(\bm e_j-\bm e_1)=0$,
we have
$R_iR_j=\bm{0}_{n\times n}$ for all $2\leq i,j\leq n$.
Consider the finite set 
 \[X_n:=
 \{I_n,2I_n\}
 \cup
 \{I_n+R_j,R_j:2\leq j\leq n\}
 \subseteq\mat_n(\Q)\,.\]
 From this set, as mentioned above, we define 
 $Y_n:=\{\jmath(M):M\in X_n\}$.
Observe that  elements of $X_n$ and thus those of $Y_n$ commute.
Moreover, $\jmath$ preserves multiplication, so
$\langle Y_n\rangle
 =
 \{\jmath(M):M\in\langle X_n\rangle\}$.
 A simple computation shows that 
\begin{equation}
\label{eq-xn-2ndexample}
 \langle X_n\rangle
 =
 \left\{
 2^k\left(I_n+\sum_{j=2}^n m_jR_j\right)
 :k,m_j\in\N
 \right\}
\cup
 \{2^kR_j:k\in\N,\ 2\leq j\leq n\}
 \cup\{\bm{0}\}\,.
\end{equation}
Taking the Zariski closure in $\mat_n(\overline{\Q})$,
we obtain
\[
 \cl{\langle X_n\rangle}
 =
 \left\{
 tI_n+\sum_{j=2}^n u_jR_j:
 t,u_j\in\overline{\Q}
 \right\}.
\]
But then since $\jmath$ is a linear embedding $\cl{\langle Y_n\rangle}
 =
 \{\jmath(M):M\in\cl{\langle X_n\rangle}\}$ holds.

We work with the following congruence $\sim$.  Given $M,N\in\cl{\langle X_n\rangle}$, we define 
$M\sim N$ if and only if
$M=NU$ and 
 $N=MV$
for some $U,V\in\cl{\langle X_n\rangle}$.
We again note in passing that this is a congruence thanks to commutativity of $\langle X_n\rangle$.
We extend it to the image of $\jmath$, in the natural way, by defining
$\jmath(M)\sim\jmath(N)$
 if and only if 
 $M\sim N$.
Consequently, we can conclude  that 
\[
 \langle Y_n\rangle/{\sim}
 \cong\langle X_n\rangle/{\sim}
 \qquad \text{and} \qquad 
 \cl{\langle Y_n\rangle}/{\sim}
 \cong\cl{\langle X_n\rangle}/{\sim}.
\]
Observe that  
every group in $\cl{\langle X_n\rangle}$ lies in
one congruence class, as  the inverses provide the
multipliers required in both directions.
Theorem~\ref{TheoremMainQuotient} therefore applies
with $N=1$ and gives finiteness of
$\langle X_n\rangle/{\sim}$, and hence of
$\langle Y_n\rangle/{\sim}$.

In the sequel, 
we  consider the  notion of \emph{mutual reducibility} $\approx$
among polynomials. Formally, 
for all $f,g\in\overline{\Q}[\bm{X}]$, write $f\approx g$ if and only if
$f=g \triangleleft M$
 and
 $g=f \triangleleft N$
 for some 
 $M,N\in\cl{\langle Y_n\rangle}$. This relation is 
an equivalence relation on~$\overline{\Q}[\bm{X}]$. 
Crucially,  definitions given above imply that  
$M\sim N$ if and only if
 \[
f \triangleleft  \jmath(M)
 \approx
 f \triangleleft \jmath(N)\,
\]
holds for all $f\in \overline{\Q}[\bm{X}]$.
Therefore, finiteness of the matrix quotient implies that $f \triangleleft \langle Y_n\rangle $ has only finitely many mutual-reducibility classes, since
\[
\left|( f \triangleleft \langle Y_n\rangle )/{\approx}\right|
\leq \left|\langle X_n\rangle/{\sim}\right|<\infty.
\]
 It is worth noting that the converse implication need not hold, as
different matrix classes may become indistinguishable
when acting on a particular polynomial.
This is where the determinant $\mathrm{Det}_n$ and permanent $\mathrm{Perm}_n$ will differ.
Consider the  orbits of these two polynomials
$\mathcal D_n:=
 \mathrm{Det}_n \triangleleft \langle Y_n\rangle$
 and 
$\mathcal P_n:=
 \mathrm{Perm}_n \triangleleft \langle Y_n\rangle$, and 
 take the closures in the coefficient
space of homogeneous degree-$n$ polynomials over
$\overline{\Q}$. Then
\begin{itemize}
    \item For the determinant, observe that 
    $\det(M)=t^n$ for matrices~$M=tI_n+\sum_{j=2}^n u_jR_j$ in the closure~$\cl{\langle X_n\rangle}$. 
Hence, 
$\mathrm{Det}_n \triangleleft \jmath(M)
 = t^n\mathrm{Det}_n$.
Using Equation~\eqref{eq-xn-2ndexample} above,
we get
$\mathcal D_n
 =
 \{2^{kn}\mathrm{Det}_n:k\in\N\}\cup\{\bm{0}\}$, which simply gives
 $\overline{\mathcal D_n}
 =
 \overline{\Q}\,\mathrm{Det}_n$.
 This in turn proves that $\mathcal D_n$ and its closure $\cl{\mathcal D_n}$ have
exactly two mutual-reducibility classes, as 
(1)
all nonzero multiples of $\mathrm{Det}_n$ are
mutually reducible using scalar matrices from
$\cl{\langle Y_n\rangle}$ and (2)
the zero polynomial has its own class.

\item The permanent have  different number of the the mutual-reducibility classes;
let us go slowly here. 
For $2\leq j\leq n$, after applying $\jmath(R_j)$,
every column of the matrix of variables $[x_{i,j}]_{1\leq i,j\leq n}$ will become
the difference between columns $j$ and $1$. 
Hence
\[Q_j:=
 \mathrm{Perm}_n \triangleleft \jmath(R_j)
 =
 n!\prod_{i=1}^n(x_{i,j}-x_{i,1})\] is a nonzero polynomial. More generally,  we get 
\[
 \mathrm{Perm}_n \triangleleft \jmath(\sum_{j=2}^n u_jR_j)
 =
 n!\prod_{i=1}^n
 \left(\sum_{j=2}^n u_j(x_{i,j}-x_{i,1})\right).
\]
Moreover, for $M=tI_n+\sum_jv_jR_j$ in $\cl{\langle X_n\rangle}$, we have $M(\sum_{j=2}^n u_jR_j)=t(\sum_{j=2}^n u_jR_j)$, implying that 
\[
 \bigl( \mathrm{Perm}_n \triangleleft
 \jmath(\sum_{j=2}^n u_jR_j)\bigr)\triangleleft \jmath(M) 
 =
 \mathrm{Perm}_n \triangleleft t^n\bigl(\jmath(\sum_{j=2}^n u_jR_j)\bigr).
\]
We are now in a position to describe the mutual-reducibility classes
in $\mathcal P_n$.
Observe that   generated invertible matrices give polynomials
mutually reducible to $\mathrm{Perm}_n$, as the
inverses belong to the matrix closure. However, 
the nonzero singular generated matrices give
multiples of the $Q_j$ polynomial, which are pairwise incomparable with respect to~$\approx$.
Moreover, none of the $Q_j$ polynomial is comparable
to $\mathrm{Perm}_n$, as  $Q_j$ contains
the monomial $\prod_{i=1}^n x_{i,j}$, whereas the
permanent does not.
(More in detail, the  identities above show that
every further reduction of $Q_j$ by a matrix in
$\cl{\langle Y_n\rangle}$ is a scalar multiple of
$Q_j$. Since $\mathrm{Perm}_n$ is not such a
multiple, it cannot be obtained from $Q_j$.
Thus they are not mutually reducible.)
Together with the class of zero polynomial, we get 
 $n+1$ mutual-reducibility classes in
$\mathcal P_n$.
In contrast, we will show that $\overline{\mathcal P_n}$
has infinitely many classes.
For each $\lambda\in\overline{\Q}$, consider
\[
 g_\lambda:=
\mathrm{Perm}_n \triangleleft \jmath(R_2+\lambda R_3)
 =
 n!\prod_{i=1}^n
 \bigl(x_{i,2}-x_{i,1}
       +\lambda(x_{i,3}-x_{i,1})\bigr)\,,
\]
where  $\jmath(R_2+\lambda R_3)$ belongs to
$\cl{\langle Y_n\rangle}$.
Since the orbit map is polynomial, it follows that
$g_\lambda\in\overline{\mathcal P_n}$. We claim that these polynomials $g_\lambda$ are not multiple of each other. Assume otherwise  $g_\lambda=cg_\mu$ holds, then we can set
$x_{i,2}=1$ and all variables in the $i$-th row $i$
to zero, for every $i\geq2$.
The resulting equality is
\[
 x_{1,2}-x_{1,1}
 +\lambda(x_{1,3}-x_{1,1})
 =
 c\bigl(
 x_{1,2}-x_{1,1}
 +\mu(x_{1,3}-x_{1,1})
 \bigr).
\]
But comparing the coefficients of $x_{1,2}$ and $x_{1,3}$
gives $c=1$ and $\lambda=\mu$.
Hence, 
the $g_\lambda$  are pairwise non-equivalent
under~$\approx$. 
It follows that the polynomial $g_1$ is not mutually reducible to  polynomials in $\mathcal P_n$.
Consequently, $g_1\in\overline{\mathcal P_n}\setminus\mathcal P_n$ holds.
\end{itemize}

By the above, we can conclude that 
\[
 |\mathcal D_n/{\approx}|
 =
 |\overline{\mathcal D_n}/{\approx}|
 =2 \qquad \text{ and }
 \qquad |\mathcal P_n/{\approx}|
 =n+1,
 \qquad \text{ and }  \qquad
 |\overline{\mathcal P_n}/{\approx}|=\infty.
\]
This immediately  proves that 
\[
 \overline{\mathrm{Perm}_n \triangleleft \langle Y_n\rangle
 }
 \not\subseteq
 \overline{\mathrm{Det}_n \triangleleft \langle Y_n\rangle
 }.
\]
An explicit witness is $g_1$, as
it contains the monomial $\prod_{i=1}^n x_{i,2}$
with coefficient $n!$, whereas every polynomial
in~$\overline{\mathcal D_n}$
has coefficient zero at this monomial.

\section*{Acknowledgment}
We would like to thank Nikhil Balaji (IIT Delhi) and Sebastien Tavenas (CNRS) for helpful discussions about the border complexity and algebraic branching programs.

\bibliographystyle{unsrt}
\bibliography{biblio}
\newpage
\appendix

\section{Background on Exterior Algebra}
\label{app:exterior}

We recall the exterior-algebra background used in the proofs, following
the presentation in~\cite[Appendix~C]{OurVASS26}.
The encoding of subspaces as exterior powers, discussed below,
underlies the factorisation result used in
Theorem~\ref{theorem-1} and in the proofs of
Theorems~\ref{theorem2} and~\ref{TheoremMainQuotient}.
Throughout the appendix, matrices act on row vectors. An empty prefix
or suffix in a decomposition is interpreted as the identity matrix~$I_d$. 

\paragraph{Encoding subspaces.}
Let $V=\Q^d$ and $0\le r\le d$. The exterior power
$\bigwedge^r V$ is the vector space generated by expressions
$v_1\wedge\cdots\wedge v_r$, subject to multilinearity and the
alternating rule: a product with two equal vectors is zero.
In particular, interchanging two vectors changes the sign.
If $e_1,\ldots,e_d$ is the standard basis of $V$, then
\[
 \{e_{i_1}\wedge\cdots\wedge e_{i_r}:
       1\le i_1<\cdots<i_r\le d\}
\]
is a basis of $\bigwedge^r V$, so its dimension is $\binom dr$.
We use the convention $\bigwedge^0 V=\Q$, with empty wedge equal to~$1$.
The direct sum $\bigwedge V:=\bigoplus_{r=0}^d\bigwedge^r V$
is the exterior algebra of $V$ and has dimension~$2^d$.

The wedge $v_1\wedge\cdots\wedge v_r$ is nonzero exactly when
$v_1,\ldots,v_r$ are linearly independent. Thus an $r$-dimensional
subspace $U$ can be represented by a nonzero vector
\[
 u_U:=u_1\wedge\cdots\wedge u_r\in\bigwedge^r V,
\]
where $u_1,\ldots,u_r$ is any basis of $U$. Changing this basis
multiplies $u_U$ by a nonzero scalar. Two subspaces of dimension $r$
are equal exactly when their representing vectors are proportional.
Moreover, for subspaces $U,K\le V$, their representing wedges satisfy
\[
 u_U\wedge u_K\ne0\quad\Longleftrightarrow\quad U\cap K=\{0\}.
\]
Indeed, the wedge on the left is nonzero exactly when the two chosen
bases together are linearly independent.

\paragraph{Testing whether a product preserves rank.}
A \emph{linear functional} on a $\mathbb Q$-vector space is a linear map from
that space to~$\Q$. The following construction turns the intersection
test on subspaces, above, into evaluation of a linear functional
representing one subspace on a vector respresenting a second subspace.

\begin{claim}\label{clm:exterior-test}
Let $K\le V$ have dimension $d-r$. There is a linear functional
$\varphi_K:\bigwedge^r V\to\Q$ such that, for every
$r$-dimensional subspace $U$,
\[
 \varphi_K(u_U)\ne0\quad\Longleftrightarrow\quad U\cap K=\{0\}.
\]
\end{claim}
\begin{proof}
Choose a surjective linear map $\psi_K:V\to\Q^r$ with kernel $K$.
It induces the linear map
\[
 \bigwedge^r\psi_K:\bigwedge^r V\longrightarrow
                  \bigwedge^r\Q^r\cong\Q.
\]
Explicitly, after identifying the wedge of the standard basis of
$\Q^r$ with $1\in\mathbb Q$, this is the functional
\[
 \varphi_K(v_1\wedge\cdots\wedge v_r)
       =\det\!\begin{pmatrix}\psi_K(v_1)\\\vdots\\\psi_K(v_r)\end{pmatrix}.
\]
Multilinearity and alternation of the determinant make this a
well-defined linear map. For a basis $u_1,\ldots,u_r$ of $U$,
$\varphi_k(u_1 \wedge \cdots \wedge u_r)$
is nonzero exactly when $\psi_K$ restricts to an isomorphism on $U$,
equivalently when $U\cap K=\{0\}$. The assertion remains true when $r=0$, since the determinant of the
empty matrix is $1$.
\end{proof}

For rank-$r$ matrices $a,b$, restriction of $b$ to $\im(a)$ gives
\begin{equation}\label{eq:rank-intersection-appendix}
 \rk(ab)=r-\dim\bigl(\im(a)\cap\ker(b)\bigr).
\end{equation}
Consequently, the product $ab$ has rank $r$ exactly when
$\varphi_{\ker(b)}(u_{\im(a)})\ne0$.
We denote by $\End(W)$ the vector space of all linear maps
from a finite-dimensional vector space $W$ to itself with addition and
scalar multiplication defined pointwise.  Note that
$\dim\End(W)=(\dim W)^2$. 

\paragraph{Kernels and images of rank-preserving products.}
The following elementary facts are used in all three proofs.

\begin{claim}\label{clm:rank-identities}
If matrices $a_1,\ldots,a_k\in \mat_d(\mathbb Q)$ and their product $a_1\cdots a_k$ all have rank $r$, then,
for $1\le i\le j\le k$,
\[
 \rk(a_i\cdots a_j)=r,\qquad
 \ker(a_i\cdots a_j)=\ker(a_i),\qquad
 \im(a_i\cdots a_j)=\im(a_j).
\]
Conversely, if all $a_i$ have rank $r$ and
$\rk(a_i a_{i+1})=r$ for $1\le i<k$, then their product has rank~$r$.
In particular, any nonempty rank-stable set generates a rank-stable
subsemigroup. A matrix is rank-stable exactly when its kernel and
image have intersection $\{0\}$.
\end{claim}
\begin{proof}
The rank cannot increase under multiplication, so the rank of a factor
lies between that of the full product and that of each of its
constituent matrices. Thus every factor has rank $r$.
For row actions, $\ker(a_i)\subseteq\ker(a_i\cdots a_j)$ and
$\im(a_i\cdots a_j)\subseteq\im(a_j)$; equality of dimensions gives
the asserted identities.

For the converse, we use induction on the number of matrices. The partial
product $a_1\cdots a_i$ has image $\im(a_i)$, which is disjoint from
$\ker(a_{i+1})$ by~\eqref{eq:rank-intersection-appendix}.
The next multiplication therefore preserves rank.
If every pair of matrices in a set preserves their common rank,
this argument applies to every product over that set, including
products of two such products. Finally, apply
\eqref{eq:rank-intersection-appendix} with $b=a$ and use
$\dim\ker(a)+\dim\im(a)=d$.
\end{proof}

\section{Detailed Proof of Theorem~\ref{theorem-1}}
\label{app:theorem-1}\label{app:stable}

Fix a finite nonempty set $X\subseteq\mat_d(\Q)$ and a congruence
$\sim$ on $\langle X\rangle$. For the two claims below, assume that
every stable subsemigroup $S\le\langle X\rangle$ satisfies
$|S/{\sim}|\le N$. We first obtain the bound for products whose
factors and resulting matrix all have the same rank, and then use
it in the decreasing induction on the minimum rank of a class.

\FixedRankFactors*
\begin{proof}
By~\cite[Proposition~7]{OurVASS26}, with the definition of
factorisation tree in Section~3 of that paper, the product has a
factorisation tree of height at most $d+2$. At every node with at
least three children, the matrix labelling the node is rank-stable.

Along each path from the root to a leaf, stop at the first node
with at least three children and remove its descendants.
The remaining tree has at most two children per internal node and
height at most $d+2$. Hence it has at most $2^{d+2}$ leaves.
Each remaining leaf is either an original leaf, labelled by one
of the $a_i$, or a node at which we stopped, labelled by a
rank-stable matrix. These labels are products of consecutive
factors that partition $a_1\cdots a_k$, as required.
\end{proof}

\FixedRankClasses*
\begin{proof}
Write $t=|T|$ and $h=2^{d+2}$. If $T$ is empty there are no products
to count. Otherwise, every rank-$r$ product over $T$ has the kernel
of its first matrix and the image of its last matrix, by
Claim~\ref{clm:rank-identities}. There are therefore at most $t^2$
possible pairs $(K,U)$ of kernel and image.

For such a pair with $K\cap U=\{0\}$, consider
\[
 S_{K,U}:=\{a\in\langle T\rangle:\ker(a)=K,\ \im(a)=U\}.
\]
If this set is nonempty, it is a stable subsemigroup.
Indeed, for $a,b\in S_{K,U}$, the intersection
$\im(a)\cap\ker(b)=U\cap K$ is zero. Thus $ab$ has rank $r$,
kernel $K$, and image $U$. Our hypothesis gives
$|S_{K,U}/{\sim}|\le N$. Hence the congruence class of a rank-stable product \(a_1 \dots a_k\) is determined by the triple of (1) \(\ker(a_1)=:K\), (2) \(\im(a_k)=:U\) and (3) its induced element in \(S_{K,U}/{\sim}\). Therefore the rank-stable matrices of rank $r$ in $\langle T\rangle$ meet at most $Nt^2$ classes in total.

By Claim~\ref{clm:fixed-rank-factors}, every rank-$r$ product over
$T$ is a product of at most $h$ factors, each either in $T$ or a
rank-stable matrix of rank $r$. There are at most $t+Nt^2$ choices
for the class of each factor. Since~$\sim$ is a congruence, the
number of classes of the resulting products is at most
\[
 \sum_{j=1}^h(t+Nt^2)^j\le(1+t+Nt^2)^h. \qedhere
\]
\end{proof}

\StableTheoremQuo*
\begin{proof}
If the quotient is finite, take $N=|\langle X\rangle/{\sim}|$.
For the converse, assume the uniform bound $N$ on stable
subsemigroups. Write $m=|X|$, $h=2^{d+2}$, and, for each equivalence class $c$,
define
\[
 \rho(c):=\min\{\rk(a):a\in c\}.
\]
Choosing a minimum-rank representative of either of two classes
shows that
\begin{equation}\label{eq:rho-multiplication}
 \rho(bc)\le\min\{\rho(b),\rho(c)\}.
\end{equation}
In particular, every factor of a product whose class has minimum
rank $r$ belongs to a class with minimum rank at least $r$.

Let $M_r$ be the number of classes $c$ with $\rho(c)\ge r$.
We show by decreasing induction, for $1\le r\le d$, that these
numbers are finite and satisfy an explicit recurrence.
For the base case, the invertible matrices in $\langle X\rangle$
form a stable subsemigroup, if there are any. Every class with
minimum rank $d$ meets this subsemigroup, so $M_d\le N$.

Suppose that $r<d$ and that $M_{r+1}=n$ is finite.
Consider all classes with minimum rank $r$ of either of the forms
\[
 [x]_{\sim}\quad\hbox{or}\quad b[x]_{\sim},
 \qquad x\in X,\quad \rho(b)>r.
\]
There are at most $t:=m(n+1)$ such classes. Choose one rank-$r$
matrix in each and let $T\subseteq\langle X\rangle$ be the finite
set of these representatives. This set is fixed for the entire
induction step, and $|T|\le t$.

Let $a_1\cdots a_k$, with $a_i\in X$, represent any class whose
minimum rank is $r$. It admits a decomposition
\begin{equation}\label{eq:stable-decomposition-appendix}
 a_1\cdots a_k=(y_1c_1)\cdots(y_\ell c_\ell)y_{\ell+1},
\end{equation}
where $c_i\in X$, each $[y_ic_i]_{\sim}$ has minimum rank $r$,
and every $y_i$, including the final suffix, is either empty or
belongs to a class of minimum rank greater than $r$.
To obtain this decomposition, take the shortest initial factor
whose class has minimum rank $r$ and separate its last generator
$c_1$. Repeat on the remaining suffix for as long as its class
has minimum rank $r$. All factors have minimum rank at least $r$
by~\eqref{eq:rho-multiplication}, so the procedure has precisely
the stated properties. Since the full class has minimum rank $r$,
at least one factor $y_ic_i$ is selected.

Each class $[y_ic_i]_{\sim}$ has a representative $b_i\in T$.
Replacing these factors gives
\[
 a_1\cdots a_k\sim b_1\cdots b_\ell y_{\ell+1}.
\]
The product $b_1\cdots b_\ell$ has rank at most $r$ because each
$b_i$ has rank $r$. It cannot have smaller rank: multiplying it
by the remaining suffix, when present, would give a matrix of
rank less than $r$ in the original class. Thus its rank is exactly
$r$.
Claim~\ref{clm:fixed-rank-classes} bounds its possible classes by
$(1+t+Nt^2)^h$. There are at most $n+1$ choices for the suffix
class, including the empty suffix as one possibility.
Consequently,
\begin{equation}\label{eq:stable-count-recurrence}
 M_r\le n+(n+1)(1+t+Nt^2)^h,
 \qquad n=M_{r+1},\quad t=m(n+1).
\end{equation}
This completes the induction. There is at most one additional
class with minimum rank zero, since the zero matrix is unique.
Hence the whole quotient has at most $M_1+1$ classes.

We now verify the stated numerical bound. Put
$K_r=M_r+1$ and $A=(N+3)m^2$. Since $m\ge1$ and $K_{r+1}\ge1$,
\[
 2+mK_{r+1}+Nm^2K_{r+1}^2\le A K_{r+1}^2.
\]
Using $1+z^h\le(1+z)^h$ for $z\ge0$ in
\eqref{eq:stable-count-recurrence}, we obtain
\[
 K_r\le A^h K_{r+1}^{2h+1},
 \qquad
 \sqrt A\,K_r\le(\sqrt A\,K_{r+1})^{2h+1}.
\]
Iterating $d-1$ times and using $K_d\le N+1$ gives
\begin{align*}
 |\langle X\rangle/{\sim}|
 &\le K_1
 \le\bigl[\sqrt A\,(N+1)\bigr]^{(2h+1)^{d-1}}\\
 &\le\bigl[m(N+1)^2\bigr]^{(2^{d+3}+1)^{d-1}}
   =F_d(N,m),
\end{align*}
where the last inequality uses $\sqrt{N+3}\le N+1$ for $N\ge1$.
This also covers the case $d=1$, when no induction step is needed.

Finally, a product of minimum length in its class has pairwise
incongruent nonempty prefix products. Otherwise, the factor between
two congruent prefixes could be omitted to obtain a shorter
representative. Its length is therefore at most the cardinality
of the quotient. This proves the diameter bound as well.
Since $(2^{d+3}+1)^{d-1}=2^{O(d^2)}$, the asserted asymptotic form
follows.
\end{proof}

\section{Detailed Proof of Corollary~\ref{cor:closed-torsion-quotient}}
\label{app:closed-torsion-quotient}

\corclosedtorsionquotient*
\begin{proof}
By Theorem~1, it suffices to give a uniform bound for $|S/{\sim}|$, where 
$S\leq\langle X\rangle$ is a stable subgroup. The case that all matrices in $S$ have rank $0$ is
immediate. Otherwise, restriction of the elements of $S$ to their common image identifies $S$
with a subsemigroup of $\operatorname{GL}_r(\mathbb Q)$, where
$1\leq r\leq d$, and preserves relative closedness of the congruence.
Taking closures over $\overline{\mathbb Q}$ within the indicated groups, put
\[
 G=\overline S\leq\operatorname{GL}_r(\overline{\mathbb Q}),
 \qquad
 H=\overline{\{(a,b)\in S\times S:a\sim b\}}\leq G\times G.
\]
Being, respectively, closures of subsemigroups of $\mathrm{GL}_r(\overline{\mathbb Q})$
and $\mathrm{GL}_r(\overline{\mathbb Q})\times \mathrm{GL}_r(\overline{\mathbb Q})$,
$G$ and $H$ are algebraic groups by~\cite[Lemma 9]{OurVASS26}.
Since $H$ contains the diagonal of $G$, the subgroup
$N=\{g\in G:(g,1)\in H\}$ is closed and normal, and
$(g,h)\in H$ if and only if $gh^{-1}\in N$.
Relative closedness therefore shows that the quotient map
$q:G\to G/N$ induces an isomorphism $S/{\sim}\cong q(S)$.
The quotient $G/N$ is a linear algebraic group
\cite[Theorem~11.5]{humphreys1975linear}, and $q(S)$ is a torsion subgroup thereof.

The remaining problem is that while Burnside would now show finiteness of the quotient, we require a uniform bound \(N\). Hence we have to prove that (1) the quotient is uniformly torsion, and (2) that the quotient can be embedded into a not-too-high dimension \(d'\). In particular, while the quotient is again linear algebraic by \cite[Theorem~11.5]{humphreys1975linear}, it does not state a bound on the corresponding matrix dimension \(d'\), which we now have to prove in our setting.

To see (1), let
\[
 E_d=\operatorname{lcm}\{n\geq1:\varphi(n)\leq d!\},
\]
where $\varphi$ is Euler's totient function. This is finite by the well-known identity \(\varphi(n) \geq \sqrt{\frac{n}{2}}\), wherefore the lcm is taken over a finite set. For $a\in S$, let $C_a$
be the closure of the cyclic group generated by $a$. Let \(C_a^{\circ}\) be the identity component of \(C_a\). 
By \cite[Theorem~3.1 and Remark~2.3]{GaluppiStanojkovski},
the size of the quotient~$|C_a/C_a^\circ|$ equals the order of the torsion subgroup of the
multiplicative group generated by the eigenvalues of $a$.
This torsion subgroup is cyclic and lies in the splitting field of
the characteristic polynomial of $a$, whose degree is at most
$r!\leq d!$. Its order $m$ thus satisfies $\varphi(m)\leq d!$,
so $m\mid E_d$ and $a^{E_d}\in C_a^\circ$.
Since $q(a)$ has finite order, $q(C_a)$ is finite and
$q(C_a^\circ)=1$. Consequently $q(a)^{E_d}=1$ for every $a\in S$.
Burnside's theorem for linear groups of finite exponent
\cite[Theorem~8.1.11(i)]{Robinson} implies that $q(S)$ is finite.
As $q(S)$ is dense in $G/N$, it follows that $G/N=q(S)$.

To see (2): By \cite[Theorem~1.1]{Brion}, there is a finite subgroup $F\leq G$
with $q(F)=G/N$. Jordan's theorem
\cite[Theorem~1.1]{Breuillard} yields an abelian subgroup $A\leq F$
with $[F:A]\leq J(d)$, where $J(d)$ depends only on $d$.
Simultaneous diagonalization shows that $A$ has at most $r\leq d$
generators. Since in an abelian group every element can be identified with a word \(g_1^{n_1} \dots g_{r}^{n_r}\) where \(g_1, \dots, g_r\) are the generators and \(n_i \in \N\), we obtain $|q(A)|\leq E_d^r \leq E_d^d$, and
\[
 |S/{\sim}|=|G/N|
 \leq [F:A]\,|q(A)|
 \leq J(d)E_d^d.
\]
This bound depends only on $d$, so Theorem~1 now applies.
\end{proof}

\section{Detailed Proof of Theorem~\ref{theorem2}}
\label{app:rank}

Fix a finite nonempty set $X\subseteq\mat_d(\Q)$ and a congruence
$\sim$ on $\langle X\rangle$. Throughout the argument below,
assume that $|S/{\sim}|\le N$ for every rank-stable subsemigroup
$S\le\langle X\rangle$, and put
\[
 D=\binom d{\lfloor d/2\rfloor},\qquad
 H=\binom{N+D^2+1}{D^2}.
\]
We first prove the pruning claim from the overview. Its proof
uses the exterior-algebra background in Appendix~\ref{app:exterior}
to find sufficiently many mutually joined cuts. The uniform bound
on rank-stable subsemigroups then supplies two congruent prefixes.

\RankClique*

\paragraph{The cut graph and its independent sets.}
Let $r$ be the common rank of $a_1,\ldots,a_k$ and their product.
The cut graph has vertices $1,\ldots,k-1$. Distinct cuts $i,j$
are joined when
\begin{equation}\label{eq:cut-edges-appendix}
 \rk(a_i a_{j+1})=\rk(a_j a_{i+1})=r.
\end{equation}
Equivalently, both $\im(a_i)\cap\ker(a_{j+1})$ and
$\im(a_j)\cap\ker(a_{i+1})$ are zero. For $i=j$, the corresponding
rank equality also holds, because the original product preserves
rank; we do not add loops to the graph.

\IndependentCuts*
\begin{proof}
Put $W=\bigwedge^r\Q^d$. For each cut $i$, choose a wedge
$u_i\in W$ representing the $r$-dimensional space $\im(a_i)$.
By Claim~\ref{clm:exterior-test}, there is a linear functional
$\varphi_i:W\to\Q$ associated with $\ker(a_{i+1})$ such that
\begin{equation}\label{eq:cut-pairing}
 \varphi_i(u_j)\ne0
 \quad\Longleftrightarrow\quad
 \im(a_j)\cap\ker(a_{i+1})=\{0\}
 \quad\Longleftrightarrow\quad
 \rk(a_j a_{i+1})=r.
\end{equation}
In particular, $\varphi_i(u_i)\ne0$.

Define the rank-one map $T_i:W\to W$ and the linear functional
$f_i:\End(W)\to\Q$ by
\[
 T_i(z)=\varphi_i(z)u_i,\qquad
 f_i(T)=\varphi_i\bigl(T(u_i)\bigr).
\]
The map $T_i$ has image $\Q u_i$, and $f_i$ is linear because
evaluation at $u_i$ and the map $\varphi_i$ are both linear.
For all cuts $i,j$, we have
\begin{equation}\label{eq:cut-functional}
 f_i(T_j)=\varphi_j(u_i)\varphi_i(u_j).
\end{equation}
Thus $f_i(T_i)=\varphi_i(u_i)^2\ne0$, whereas for distinct
$i,j$, the value $f_i(T_j)$ is nonzero exactly when $i,j$ are joined.

Let $I$ be an independent set. We show that the maps $T_i$,
$i\in I$, are linearly independent in $\End(W)$.
Suppose that $\sum_{i\in I}\alpha_iT_i$ is the zero map, and fix
$j\in I$. Evaluating $f_j$ at this relation gives
\[
 0=f_j\left(\sum_{i\in I}\alpha_iT_i\right)
   =\sum_{i\in I}\alpha_i f_j(T_i)
   =\alpha_j\varphi_j(u_j)^2.
\]
All terms other than the $j$th vanish by independence of $I$.
Since $\varphi_j(u_j)\ne0$, we obtain $\alpha_j=0$.
This holds for every $j\in I$, proving linear independence.
Therefore
\[
 |I|\le\dim\End(W)=\binom dr^2\le D^2.
\]
The conventions in Appendix~\ref{app:exterior} cover the case $r=0$ as
well. In that case all $a_i$ are zero and the cut graph is complete.
\end{proof}

\begin{proof}[Proof of Claim~\ref{clm:rank-clique}]
Write $R(p,q)$ for the least integer such that every graph on that
many vertices has either a clique of size $p$ or an independent
set of size $q$. We use the following classical bound~\cite{ErdosSzekeres35,diestel2000graph} for this
\begin{gather}
 R(p,q)\le\binom{p+q-2}{p-1}.   
\label{eq:erdos}
\end{gather}

For completeness, the recurrence
$R(p,q)\le R(p-1,q)+R(p,q-1)$ follows by separating the neighbours
and non-neighbours of a vertex. Together with
$R(1,q)=R(p,1)=1$, induction and Pascal's identity give the
bound~\eqref{eq:erdos}.
In the cut graph the number of vertices is $k-1\ge H$ and
\[
 R(N+2,D^2+1)\le\binom{N+D^2+1}{D^2}=H.
\]
Claim~\ref{clm:independent-cuts} excludes an independent set of
size $D^2+1$. Hence there is a clique of size $N+2$, i.e., there are mutually joined cuts
$i_0<i_1<\cdots<i_{N+1}$.

For $1\le j\le N+1$, put
\[
 b_j:=a_{i_{j-1}+1}\cdots a_{i_j}.
\]
By Claim~\ref{clm:rank-identities}, each $b_j$ has rank $r$,
with kernel $\ker(a_{i_{j-1}+1})$ and image $\im(a_{i_j})$.
We check that the $b_j$ generate a rank-stable subsemigroup.
For any $p,q\in\{1,\ldots,N+1\}$, the intersection
\[
 \im(b_p)\cap\ker(b_q)
   =\im(a_{i_p})\cap\ker(a_{i_{q-1}+1})
\]
is zero: if the cuts $i_p$ and $i_{q-1}$ are distinct, this follows
from the existence of an edge between them; if they coincide, it follows from rank
preservation at that cut in the original product.
Thus all products $b_p b_q$ have rank $r$, and
Claim~\ref{clm:rank-identities} shows that
$S:=\langle b_1,\ldots,b_{N+1}\rangle$ is rank-stable.

By hypothesis, $|S/{\sim}|\le N$. Hence there are two non-empty
prefixes of $b_1,\ldots,b_{N+1}$ with congruent products, say
\[
 b_1\cdots b_p\sim b_1\cdots b_q,
 \qquad 1\le p<q\le N+1.
\]
Multiplying by the surrounding factors in the original product
of the $a_i$
gives
\[
 a_1\cdots a_k
   \sim a_1\cdots a_{i_p}\,a_{i_q+1}\cdots a_k.
\]
Taking $i=i_p$ and $j=i_q+1$ proves the claim.
\end{proof}

\RankCriterion*
\begin{proof}
The forward implication is immediate by taking the cardinality
of the quotient $\langle X \rangle / {\sim}$ as a uniform bound on rank-stable subsemigroups. For the converse, assume
a uniform bound $N$ on rank-stable subsemigroups.
Since $X$ is finite, to prove finiteness of the quotient, it suffices to bound the
length of every equivalence class.

Let $a_1\cdots a_k$, with $a_i\in X$, be a product whose length
$k$ equals the length of its equivalence class. Every nonempty factor of this
displayed product is itself shortest in its class since replacing a
factor by a shorter congruent product would contradict the
minimality of $k$.
Define
\[
 L_d=N,\qquad L_r=(H+1)L_{r+1}+H\quad(0\le r<d).
\]
We prove by decreasing induction on $r$ that every such product
with rank at least $r$ has $k\le L_r$. 

For the base case, if the product has rank $d$, all its nonempty
prefixes are invertible. The invertible matrices in
$\langle X\rangle$ form a rank-stable subsemigroup whose quotient
has at most $N$ classes. If $k>N$, two of the $k$ nonempty prefixes
would be congruent. Omitting the factor between them would give
a shorter representative of the same class. Hence $k\le N=L_d$.

For the induction step, products of rank greater than $r$ are
already covered by $L_{r+1}\le L_r$, so suppose the rank is $r<d$.
Decompose the product as
\begin{equation}\label{eq:rank-decomposition-appendix}
 a_1\cdots a_k=y_0(c_1y_1)\cdots(c_\ell y_\ell),
\end{equation}
where $c_i\in X$, each nonempty $y_i$ has rank greater than $r$,
and each $c_i y_i$ has rank $r$.
Such a decomposition can be obtained from right to left:
take the shortest suffix of rank $r$ and separate its first
generator; repeat on the remaining prefix while its rank is $r$.
All nonempty factors have rank at least $r$, so the procedure
ends with an empty prefix or one of rank greater than $r$.
The suffix after the separated generator has rank greater than
$r$, or is empty, by the choice of a shortest suffix.

Every nonempty $y_i$ is shortest in its own class, so the
induction hypothesis bounds its number of generators by
$L_{r+1}$. There is at least one factor $c_i y_i$, and their
product has rank exactly $r$: its rank is at most $r$, while
multiplication on the left by $y_0$ gives the original rank-$r$
product. If $\ell>H$, then Claim~\ref{clm:rank-clique} would remove a nonempty consecutive list of
these factors. Substituting in~\eqref{eq:rank-decomposition-appendix}
would give a shorter congruent product over $X$, a contradiction.
Thus $\ell\le H$, and
\[
 k\le L_{r+1}+\ell(L_{r+1}+1)
    \le L_{r+1}+H(L_{r+1}+1)=L_r.
\]
This proves the induction. Since
$L_r+1=(H+1)(L_{r+1}+1)$, it yields
\[
 L_0=(N+1)(H+1)^d-1=G_d(N).
\]
Every class has length at most $G_d(N)$, proving both the diameter
bound and finiteness.

To verify the two asymptotic estimates, put $M=D^2\le4^d$.
Counting multisets by ordered lists gives
\[
 H=\binom{M+N+1}{N+1}\le(M+1)^{N+1},
 \qquad
 H=\binom{M+N+1}{M}\le(N+2)^M.
\]
The first inequality implies
$\log_2(G_d(N)+1)=O(d^2N)$; the second implies
$\log_2(G_d(N)+1)=O(d4^d\log_2(N+1))$.
Therefore
\[
 G_d(N)\le\min\{2^{O(d^2N)},\ (N+1)^{2^{O(d)}}\}.
\]
\end{proof}

\section{Detailed Proof of Theorem~\ref{TheoremMainQuotient}}
\label{app:closure}

In this section, all Zariski closures are taken over $\overline\Q$.
Put $C:=\cl{\langle X\rangle}\subseteq\mat_d(\overline\Q)$,
where $X\subseteq\mat_d(\Q)$ is finite and nonempty.
Kernels and images of rational matrices can be computed over $\Q$
and then extended to $\overline\Q$; their ranks and the condition
that an intersection of two subspaces is zero are unchanged.
For the absorption claim and the final theorem proof, $\sim$ is
defined on all of $C$, and every group $G\le C$ satisfies
$|G/{\sim}|\le N$.

\paragraph{Groups in the closure and the absorption claim.}
We first justify precisely which group in the closure contains
a stable subsemigroup. In general, the identity of such a group
need not be $I_d$ but is rather an idempotent
matrix.

\begin{lemma}\label{lem:group-envelope-revised}
Let $K,U\le\overline\Q^d$ satisfy $\overline\Q^d=K\oplus U$.
The matrices with kernel $K$ and image $U$ form an algebraic group
\[
 \mathcal H(K,U):=
 \{a\in\mat_d(\overline\Q):\ker(a)=K,\ \im(a)=U\}.
\]
The identity of this group
is the projection $e$ onto $U$ along $K$.
If $S$ is a nonempty subsemigroup of
$C\cap\mathcal H(K,U)$, its Zariski closure inside
$\mathcal H(K,U)$ is a group contained in $C$.
\end{lemma}
\begin{proof}
Restriction to $U$ identifies $\mathcal H(K,U)$ with $\gl(U)$:
an invertible linear map on $U$ extends uniquely by zero on $K$.
In a basis adapted to $K\oplus U$, these are precisely the matrices
with zero block on $K$ and invertible block on $U$.
This also shows that $\mathcal H(K,U)$ is locally closed in matrix
space and that its group operations are regular.
The identity extends the identity on $U$ by zero on $K$.
When $U=\{0\}$, this is the one-element group $\{0\}$.

We use the fact that a nonempty Zariski-closed subsemigroup of an
algebraic group is a group; see also~\cite[Lemma~9]{OurVASS26}.
%
By continuity of matrix multiplication, 
the closure of a subsemigroup is again a subsemigroup, both in
matrix space and inside an algebraic group. To see this, write
$\overline S$ for the closure in the space concerned. For $s\in S$,
continuity of left multiplication and $sS\subseteq S$ give
$s\overline S\subseteq\overline S$. For $b\in\overline S$,
we then have $Sb\subseteq\overline S$; continuity of right
multiplication gives $\overline S b\subseteq\overline S$.
In particular, $C$ is a semigroup.
The closure of $S$ inside $\mathcal H(K,U)$ is thus a closed
subsemigroup of an algebraic group, and hence a group.
Since $C\cap\mathcal H(K,U)$ is relatively closed and contains
$S$, it contains this relative closure as well.
\end{proof}

\ClosureAbsorption*
\begin{proof}
First let $c\in\langle X\rangle$ be rank-stable. Its kernel and
image have intersection zero, so Lemma~\ref{lem:group-envelope-revised}
applies to $\langle c\rangle$ inside
$\mathcal H(\ker(c),\im(c))$.
It gives a group $G\le C$ containing $c$, with identity $e_c$,
the projection onto $\im(c)$ along $\ker(c)$.
The restriction of a semigroup congruence to a group has a group
as its quotient: the classes of the identity and inverses yield
the identity and inverses in the quotient.
Thus $G/{\sim}$ is a finite group of order at most $N$.
The order of the class of $c$ gives an integer $1\le q\le N$
such that
\begin{equation}\label{eq:local-power-identity}
 c^q\sim e_c.
\end{equation}

Now suppose $\im(a)=\im(ab)$ and put $U=\im(a)$.
The equality says $Ub=U$, so $b$ restricts to a surjective,
and hence invertible, linear map on $U$. Iterating gives
\[
 U=Ub^d\subseteq\im(b^d).
\]
The matrix $c=b^d$ is rank-stable. Indeed, the descending sequence
of images of powers of $b$ stabilises in at most $d$ steps.
Hence $\im(b^d)=\im(b^{2d})$.
The projection $e_c$ fixes $\im(b^d)$ pointwise, and therefore
fixes $U$. Multiplying~\eqref{eq:local-power-identity} by $a$ gives
\[
 ab^{dq}=ac^q\sim ae_c=a.
\]
This proves the claim, with exponent $dq\ge1$.
\end{proof}

\paragraph{Short products with the same kernel and image.}
We next make precise the exterior-algebra step used to construct
insertions at a cut. The following is the form of
\cite[Lemma~8]{Finiteness26} needed here. We include its proof,
which also gives the slightly stronger bound $\binom dr+1$.

\begin{lemma}\label{lem:short-kernel-image}
Let $a_1,\ldots,a_\ell\in\mat_d(\Q)$ and their product $a$
all have rank $r$. There are indices
$1=i_1<\cdots<i_t=\ell$, with $t\le\binom dr+1\le2^d+1$,
such that $a_{i_1}\cdots a_{i_t}$ has the same kernel and image
as $a$. For $\ell=1$, take the single index $i_1=1$.
\end{lemma}
\begin{proof}
Choose a shortest subsequence that includes both endpoints and whose
product still has rank $r$; denote the matrices in this subsequence by
$b_1,\ldots,b_t$. Such a subsequence exists because the full
sequence has rank $r$. The product $b_1\cdots b_t$ has kernel $\ker(a_1)$ and image
$\im(a_\ell)$, i.e., the kernel and image are those of $a$ by
Claim~\ref{clm:rank-identities}. It remains to bound $t$.
There is nothing to prove if $t=1$.

For $1\le i<t$, choose a wedge $u_i$ representing $\im(b_i)$,
and let $\varphi_i$ be the functional associated with
$\ker(b_{i+1})$ by Claim~\ref{clm:exterior-test}.
Since the selected product preserves rank,
$\varphi_i(u_i)\ne0$.
For $1\le i<j<t$, we must have $\varphi_j(u_i)=0$.
Otherwise, $b_i b_{j+1}$ would have rank $r$. All other adjacent
joins in
\[
 b_1\cdots b_i b_{j+1}\cdots b_t
\]
already preserve rank, so Claim~\ref{clm:rank-identities} would
make this a shorter rank-$r$ product retaining both endpoints,
contrary to the choice of the subsequence.

The square matrix $(\varphi_j(u_i))_{1\le j,i<t}$ is therefore
triangular with nonzero diagonal, and is invertible. In particular,
$u_1,\ldots,u_{t-1}$ are linearly independent: applying the
$\varphi_j$ to a relation among these vectors gives an invertible
linear system for its coefficients. Since they lie in
$\bigwedge^r\Q^d$, we obtain $t-1\le\binom dr$, as required.
\end{proof}

\paragraph{Insertions and regrouping at one node.}
For the next lemma and the forest construction, only the conclusion
of Claim~\ref{clm:closure-absorption} is needed. Thus we assume
that $\sim$ is a congruence on $\langle X\rangle$ satisfying
\begin{equation}\label{eq:absorption-appendix}
 \im(p)=\im(pb)\quad\Longrightarrow\quad
 p\sim pb^n\text{ for some }n\ge1,
 \qquad p,b\in\langle X\rangle.
\end{equation}
In the theorem, we apply this to the restriction of the congruence
on $C$. The construction follows the strategy of
\cite[Proposition~9]{Finiteness26}, replacing equalities obtained
from finiteness there by the congruences in
\eqref{eq:absorption-appendix}.

\begin{lemma}\label{lem:local-regrouping}
Let $a_1,\ldots,a_s\in\langle X\rangle$, where $s\ge3$, and put
$a=a_1\cdots a_s$. Suppose
\[
 \rk(a_1)=\cdots=\rk(a_s)=\rk(a)=\rk(a^2).
\]
There are matrices $u_1,\ldots,u_t\in\langle X\rangle$, with
$t\ge3$, such that $a\sim u_1\cdots u_t$, every $u_j$ has the
kernel and image of $a$, and each $u_j$ is a product of at most
$2^{d+2}$ matrices chosen, with repetition allowed, from
$a_1,\ldots,a_s$. In particular, the $u_j$ generate a stable
subsemigroup.
\end{lemma}
\begin{proof}
Write $r=\rk(a)$, $K=\ker(a)$, and $U=\im(a)$.
Since $a$ is rank-stable, $K\cap U=\{0\}$.
At each cut $i\in\{1,\ldots,s-1\}$, let
\[
 p_i=a_1\cdots a_i,\qquad q_i=a_{i+1}\cdots a_s.
\]
Apply Lemma~\ref{lem:short-kernel-image} to $p_i$ and $q_i$.
It supplies products $y_i$ and $x_i$, respectively, each using
at most $2^d+1$ of the corresponding child matrices, such that
\begin{align}
 \ker(y_i)&=K,& \im(y_i)&=\im(a_i),\label{eq:aux-prefix}\\
 \ker(x_i)&=\ker(a_{i+1}),& \im(x_i)&=U.\label{eq:aux-suffix}
\end{align}
These auxiliary products reproduce the kernels and images;
they need not be congruent to the prefixes or suffixes.

\begin{claim}\label{clm:valid-insertion}
For every cut $i$, $\im(p_i x_i y_i)=\im(p_i)$.
\end{claim}
\begin{proof}
The product $p_i x_i$ has rank $r$, since
$\im(p_i)=\im(a_i)$ and $\ker(x_i)=\ker(a_{i+1})$,
and the original join $a_i a_{i+1}$ preserves rank.
Its image is $U$. The next multiplication by $y_i$ also preserves
rank, since $U\cap\ker(y_i)=U\cap K=\{0\}$.
The resulting image is $\im(y_i)=\im(a_i)=\im(p_i)$.
\end{proof}

By~\eqref{eq:absorption-appendix}, choose $n_i\ge1$ with
$p_i\sim p_i(x_i y_i)^{n_i}$.
We insert these powers in order from cut $s-1$ to cut $1$.
When processing cut $i$, every power already inserted is to its
right, so the product still begins with the original prefix
$p_i$. Multiplication by the remaining suffix therefore preserves
the congruence used for that insertion. After all cuts have been
processed, we obtain
\begin{equation}\label{eq:all-insertions}
 a\sim a_1(x_1y_1)^{n_1}a_2(x_2y_2)^{n_2}
                 \cdots a_{s-1}(x_{s-1}y_{s-1})^{n_{s-1}}a_s.
\end{equation}

Regroup the product on the right into the following ordered list
of factors. Start with $a_1x_1$. For $i=1,\ldots,s-1$, append
$n_i-1$ copies of $y_i x_i$, followed by $y_i a_{i+1}x_{i+1}$
if $i<s-1$, and by $y_{s-1}a_s$ if $i=s-1$.
When $n_i=1$, no copy of $y_i x_i$ is appended.
This list has $t=s+\sum_{i=1}^{s-1}(n_i-1)\ge s$ factors, and
its product is exactly the right-hand side of
\eqref{eq:all-insertions}, using
$(x_i y_i)^{n_i}=x_i(y_i x_i)^{n_i-1}y_i$ as a regrouping identity.

\begin{claim}\label{clm:regrouped-kernel-image}
Every factor in this list has rank $r$, kernel $K$, and image $U$.
\end{claim}
\begin{proof}
For $a_1x_1$, the join preserves rank by
$\ker(x_1)=\ker(a_2)$, and the endpoints give kernel $K$ and
image $U$.
For $y_i x_i$, the join is the original cut $i$ expressed through
\eqref{eq:aux-prefix}--\eqref{eq:aux-suffix}, so it preserves rank
and has the same endpoints $K,U$.
For $y_i a_{i+1}x_{i+1}$, the first join preserves rank because
$\im(y_i)=\im(a_i)$, and the second because
$\ker(x_{i+1})=\ker(a_{i+2})$. Its kernel is that of $y_i$ and
its image that of $x_{i+1}$, again $K,U$.
The final factor $y_{s-1}a_s$ has the required properties by a
similar argument.
\end{proof}

Call these factors $u_1,\ldots,u_t$.
As $K$ and $U$ are complementary, products of the $u_j$ also have
kernel $K$, image $U$, and rank $r$. They therefore generate a
stable subsemigroup.
Each $u_j$ consists of at most two auxiliary products and at most
one original child matrix. Thus it uses at most
\[
 2(2^d+1)+1\le2^{d+2}
\]
matrices among $a_1,\ldots,a_s$, completing the proof.
\end{proof}

\paragraph{Modifying the whole tree.}
Recall that a stable $\sim$-factorisation tree has its root labelled
by the given matrix, its leaves have labels in $X$, and every internal label
is congruent to the product of its children's labels. An internal
node has at least two children; if it has at least three, their
labels generate a stable subsemigroup. Its reduced height is the
maximum number of nodes with at least three children on a path
from the root to a leaf.

\StableForest*
\begin{proof}
For each $a\in\langle X\rangle$, choose a product over $X$
equal to $a$. By~\cite[Theorem~7]{Finiteness26}, it has a rank-stable
factorisation tree of height at most $d(d+3)$ and reduced height
at most $d$ such that, at every node with at least three children,
the child labels $a_1,\ldots,a_s$ and the parent label $a$ satisfy
\[
 a=a_1\cdots a_s,\qquad
 \rk(a_1)=\cdots=\rk(a_s)=\rk(a)=\rk(a^2).
\]
This is the rank-stable factorisation tree of that paper.
The extra rank equalities on the children are part of this
starting-tree result; they do not assert stability of the children.

We process the tree from the leaves upwards, keeping the root
label of each processed subtree unchanged.
A leaf needs no modification.
At a binary node, attach the already modified child subtrees to
the same parent. Their root labels remain the original two
matrices, whose product is the parent label.
At a node with at least three children, apply
Lemma~\ref{lem:local-regrouping} to its original child labels.
Replace its children by $u_1,\ldots,u_t$. Their product is
congruent to the unchanged parent label, and they generate a
stable subsemigroup.

For each $u_j$, choose the product of at most $2^{d+2}$ original
child matrices given by the lemma. Assemble this product using
a binary tree of height at most $d+2$, labelling its internal
nodes by the exact products of their children.
For each occurrence of $a_i$ among its leaves, attach a copy of
the already modified subtree rooted at $a_i$.
If the product has one matrix, attach that subtree directly.
All eventual leaves belong to $X$. Every new binary node satisfies
equality with its child product, and every other internal node
satisfies the required congruence. Thus the result is a stable
$\sim$-factorisation tree rooted at $a$.

We verify the height bounds. At a modified node, the new top node
replaces the original node with at least three children. Below
it, only binary levels are inserted before reaching a copy of
one of the original child subtrees. Consequently the modification
adds at most $d+3$ levels and introduces no additional node with
at least three children along a path. Every path in the modified
tree follows a corresponding sequence of original child subtrees,
even when some of them have been copied. The starting path has
length at most $d(d+3)$ and passes through at most $d$ nodes with
at least three children. Hence the new height is at most
\[
 d(d+3)+d(d+3)=2d(d+3),
\]
and its reduced height is at most $d$.
Choosing such a tree for each $a$ gives the asserted forest.
\end{proof}

\TheoremNameMainQuotient*
\begin{proof}
Claim~\ref{clm:closure-absorption} shows that the restriction of
$\sim$ to $\langle X\rangle$ satisfies the hypothesis of
Lemma~\ref{lem:forest}. For every $a\in\langle X\rangle$, choose
the resulting stable $\sim$-factorisation tree. Its height is at
most $h:=2d(d+3)$ and its reduced height at most $d$.
We prune this tree while preserving the class represented at
each surviving node.

Consider a node with at least three children. The child labels
generate a stable subsemigroup $S\le\langle X\rangle$ with some
common  kernel and image $K,U$ satisfying $K\cap U=\{0\}$.
By Lemma~\ref{lem:group-envelope-revised}, the relative closure
of $S$ inside $\mathcal H(K,U)$ is a group $G\le C$.
Therefore $|S/{\sim}|\le|G/{\sim}|\le N$.
If the node has more than $N$ children, two of their nonempty
prefix products are congruent. We can delete the intervening
children and their subtrees without changing the class of the
parent product. Repeating this operation leaves at most $N$
children. All retained child labels still
belong to $S$, so the same argument remains applicable.
Perform this pruning at each such node.
Unary nodes may remain when $N=1$; this is harmless for counting
leaves and for the congruence between a label and its child product.

For clarity, mark the nodes that originally had at least three
children, even if pruning leaves them with one or two.
Every unmarked internal node has at most two children; every
marked node has at most $N$ children. The height is at most $h$,
and every path contains at most $d$ marked nodes.
A tree of height at most $h$ with at most $q$ marked nodes on each
path has at most $2^hN^q$ leaves. This follows by induction on
height: at an unmarked root sum at most two bounds
$2^{h-1}N^q$, and at a marked root sum at most $N$ bounds
$2^{h}N^{q-1}$; a leaf contributes one.
Applying this with $q=d$ bounds the number of leaves by
\begin{gather}
 2^{2d(d+3)}N^d.
 \label{eq:BOUND}
\end{gather}

Every remaining leaf is labelled by a matrix in $X$.
Induction from the leaves upwards, using the congruence at each
internal node, shows that their ordered product is congruent to
the root label $a$. Thus every class has length at most~\eqref{eq:BOUND}. This proves the diameter estimate and, since
$X$ is finite, the finiteness of $\langle X\rangle/{\sim}$.
\end{proof}

\section{Detailed Proof of Theorem~\ref{theo:homogeneous-abp-separation}}
\label{app:abp-separation}

\HomogeneousABPSeparation*

\begin{proof}
Recall that $v(\bm\theta)$ lists all parameter monomials
obtained by choosing at most one parameter from each layer.
Translating such a monomial produces a linear combination of
monomials in the same list. This proves the existence and
uniqueness of the matrix $\rho_{\bm t}$ satisfying
\[
 v(\bm\theta)\rho_{\bm t}=v(\bm\theta+\bm t).
\]
Its entries are polynomials in $\bm t$ with integer
coefficients. Moreover, the entry in the row indexed by
the constant monomial and the column indexed by $\theta_i$
is $t_i$.
Thus the parameter values can be recovered from its entries.
Composing translations gives
\[
 \rho_{\bm s}\rho_{\bm t}=\rho_{\bm s+\bm t},
 \qquad \rho_{\bm0}=I_{L_\theta},
 \qquad \rho_{\bm t}^{-1}=\rho_{-\bm t}.
\]

Recall also the coefficient-evaluation identity from the
overview:
\begin{equation}\label{eq:abp-appendix-coefficient-evaluation}
 v(\bm0)\rho_{\bm t}C_U
 =\operatorname{coeff}\bigl(U(\bm t,\bm x)\bigr).
\end{equation}
We record the multiplication rules used below.
Block multiplication, together with
Equation~\eqref{eq:abp-appendix-coefficient-evaluation}, gives
\[
\begin{aligned}
 \Gamma_{\bm s}\Gamma_{\bm t}&=\Gamma_{\bm s+\bm t},
 &\Gamma_{\bm t}V_0&=V_0,
 &C\Gamma_{\bm t}&=C,\\
 V_0^2&=\bm0,
 &C^2&=\bm0,
 &CV_0&=\bm0,
\end{aligned}
\]
and
\begin{equation}\label{eq:abp-appendix-output-product}
 (V_0\Gamma_{\bm s})(\Gamma_{\bm t}C)
 =\iota\bigl(U(\bm s+\bm t,\bm x)\bigr).
\end{equation}
For every homogeneous polynomial $f$ of degree $d$, we also have
\[
 \Gamma_{\bm t}\iota(f)=\iota(f)\Gamma_{\bm t}=\iota(f).
\]
Multiplication of $\iota(f)$, on either side, by
$V_0\Gamma_{\bm t}$, $\Gamma_{\bm t}C$, or another
encoded polynomial gives $\bm0$.
Among two factors of the forms $V_0\Gamma_{\bm s}$ and
$\Gamma_{\bm t}C$, the only potentially nonzero product
is the one in Equation~\eqref{eq:abp-appendix-output-product}.

These rules show that every element of $\langle X\rangle$
has one of the forms
\begin{equation}\label{eq:abp-appendix-normal-forms}
 \Gamma_{\bm n},\qquad V_0\Gamma_{\bm n},\qquad
 \Gamma_{\bm n}C,\qquad
 \iota\bigl(U(\bm n,\bm x)\bigr),
 \qquad \bm n\in\Z^\nu.
\end{equation}
Conversely, each matrix in this list belongs to
$\langle X\rangle$. Indeed, the generators
$\Gamma_{\bm e_i}$ and $\Gamma_{-\bm e_i}$ produce
every integer translation. They also produce the identity,
and $V_0,C$ are generators.
The last family is obtained by multiplying $V_0$, an integer
translation, and $C$.
The zero matrix is included, since $U(\bm0,\bm x)=0$.

\ABPClosureDescription*
\begin{proof}
We take the closures of the four families in
Equation~\eqref{eq:abp-appendix-normal-forms}.
First, $\Z^\nu$ is Zariski dense in
$\overline{\Q}^{\,\nu}$: a polynomial vanishing on all
integer tuples is zero, by applying the corresponding
univariate fact successively to its variables.
It follows that the integer points in each of our
parameterized families have the same closure as the family
obtained using all algebraic parameter values.
More explicitly, any polynomial in matrix entries which
vanishes on the integer-parameter family pulls back to a
polynomial vanishing on $\Z^\nu$, and hence on the entire
parameter space.

The family $\{\Gamma_{\bm t}:\bm t\in\overline{\Q}^{\,\nu}\}$
is closed. Its parameters occur as matrix entries, and all
remaining entries are polynomial functions of them.
It is therefore a graph of polynomial functions, with
additional fixed coordinates.
The same reasoning applies to $V_0\Gamma_{\bm t}$,
whose nonzero block is $v(\bm t)$ and contains every
coordinate $t_i$.
It also applies to $\Gamma_{\bm t}C$:
its block in block row $3$ and block column $4$ is
$(\bm0,\rho_{\bm t})$, so the parameters are again
recoverable from matrix entries.

For the output family, the map
$\bm t\mapsto U(\bm t,\bm x)$ is polynomial and has
image $\mathcal A_{w,d}$.
The map $\iota$ identifies the coefficient space
$\overline{\Q}[\bm x]_d$ with a linear subspace of the
matrix space. Its inverse on that subspace reads off the
coefficient row, so this identification preserves Zariski
closures. Consequently,
\[
 \cl{\{\iota(U(\bm n,\bm x)):\bm n\in\Z^\nu\}}
 =\iota\bigl(\overline{\mathcal A_{w,d}}\bigr).
\]

The closure of a finite union is the union of its closures,
which gives the claimed equality.
The four families are disjoint, as can be seen from their
blocks. A translation has top-left entry $1$.
A matrix $V_0\Gamma_{\bm t}$ has the nonzero block
$v(\bm t)$ in position $(1,2)$.
A matrix $\Gamma_{\bm t}C$ has the nonzero block
$(\bm0,\rho_{\bm t})$ in position $(3,4)$.
An encoded polynomial has zero entries outside block $(1,4)$.
These observations also justify the disjoint-union notation
in the claim.
\end{proof}

Recall that $\sim$ is the least congruence on
$\cl{\langle X\rangle}$ satisfying
$\Gamma_{\bm t}\sim I_{L_{\mathrm{tot}}}$ for every
$\bm t\in\overline{\Q}^{\,\nu}$.
We describe all its classes in order to prove the next two
claims.

\UniversalABPCongruence*
\begin{proof}
The defining relations imply
\[
 V_0\Gamma_{\bm t}\sim V_0,
 \qquad \Gamma_{\bm t}C\sim C,
 \qquad
 \iota\bigl(U(\bm t,\bm x)\bigr)
 =V_0\Gamma_{\bm t}C\sim V_0C=\bm0.
\]
Since every polynomial in $\mathcal A_{w,d}$ is an output
of the universal ABP, this proves
$\iota(\mathcal A_{w,d})\subseteq[\bm0]_\sim$.

To prove the reverse inclusion, consider the equivalence
relation $\approx$ whose classes are
\[
\begin{gathered}
 \{\Gamma_{\bm t}:\bm t\in\overline{\Q}^{\,\nu}\},
 \qquad
 \{V_0\Gamma_{\bm t}:\bm t\in\overline{\Q}^{\,\nu}\},\\
 \{\Gamma_{\bm t}C:\bm t\in\overline{\Q}^{\,\nu}\},
 \qquad \iota(\mathcal A_{w,d}),
\end{gathered}
\]
together with the singleton $\{\iota(f)\}$ for each
$f\in\overline{\mathcal A_{w,d}}\setminus\mathcal A_{w,d}$.
Claim~\ref{clm:abp-closure-description} shows that these
sets partition the entire matrix closure.

This partition defines a semigroup congruence.
Indeed, the multiplication rules above determine the family
of a product from the families of its factors.
The only product of two nontranslation auxiliary families
which may be nonzero is
$(V_0\Gamma_{\bm s})(\Gamma_{\bm t}C)$;
its output always belongs to $\iota(\mathcal A_{w,d})$.
Translations fix every encoded polynomial, and all other
matrices annihilate encoded polynomials on either side.
Thus replacing either factor by an $\approx$-equivalent
matrix preserves the $\approx$-class of the product.

Furthermore, $\approx$ satisfies
$\Gamma_{\bm t}\approx I_{L_{\mathrm{tot}}}$.
By the minimality of $\sim$, we have
$\sim\,\subseteq\,\approx$.
In particular, the $\sim$-class of zero is contained
in its $\approx$-class, namely $\iota(\mathcal A_{w,d})$.
This proves the desired equality.

The relations at the beginning of the proof also show that
each of the four displayed sets lies in one $\sim$-class.
Since the remaining $\approx$-classes are singletons,
we in fact have $\sim\,=\,\approx$.
In particular, every encoded polynomial outside
$\iota(\mathcal A_{w,d})$ forms a singleton class.
\end{proof}

\UniversalABPQuotient*
\begin{proof}
By Equation~\eqref{eq:abp-appendix-normal-forms}, every
element of $\langle X\rangle$ belongs to one of the
four classes described above.
Their representatives
$I_{L_{\mathrm{tot}}},V_0,C,\bm0$ all belong to
$\langle X\rangle$: the identity is a product of
opposite translations, and $\bm0=V_0^2$.
The four classes are distinct by the explicit description
of $\sim$. Hence the quotient has exactly four elements.
\end{proof}

We also verify the application of
Theorem~\ref{TheoremMainQuotient} in the overview.
The translations form a group. Every other matrix in the
closure has square zero, by its block form.
If a matrix group contains such a matrix, it contains
$\bm0$ and must be the singleton group $\{\bm0\}$.
To see the last assertion, if its identity is $e$, the
inverse of $\bm0$ in the group would give
$e=\bm0$, after which every group element equals
$\bm0$.
Thus every group in the closure is contained in the
translation group or is $\{\bm0\}$.
Its quotient under $\sim$ is therefore a singleton.
The theorem applies with $N=1$ and gives finiteness of
$\langle X\rangle/{\sim}$, consistently with the exact
count just proved.

For clarity, write
\[
 G:=\begin{pmatrix}p&q\\r&s\end{pmatrix},
 \qquad \Delta:=ps-qr,
 \qquad
 \binom{\xi_k}{\eta_k}:=G^k\binom{x}{y}.
\]
We had defined the polynomials~$F_{w,d}$ as follows.
\[
\begin{aligned}
 F_{2,2k+3}
 &=a\xi_k^2+b\eta_k^2+z\Delta^k(ux+vy),\\
 F_{2,2k+4}
 &=a\xi_{k+1}\xi_k+b\eta_{k+1}\eta_k
   +\Delta^{k+1}(ux+vy),
\end{aligned}
\]
and
\[
 F_{w,d}=F_{2,d}+\sum_{i=1}^{w-2}c_i z_i^{d-1}.
\]
These polynomials use $2w+7$ variables (When more variables are available,
we regard these polynomials as independent of the additional
variables).

\ABPApproximation*
\begin{proof}
We first construct the approximations at width two.
Introduce the two matrices
\[
 D:=\begin{pmatrix}a&0\\0&b\end{pmatrix},
 \qquad
 J:=\begin{pmatrix}0&1\\-1&0\end{pmatrix}.
\]
For any $2\times2$ matrix $M$ over a commutative ring,
direct multiplication gives
\begin{equation}\label{eq:abp-alternating-form}
 M^{\mathsf T}JM=(\det M)J.
\end{equation}
We also have
\begin{equation}\label{eq:abp-endpoint-cancellation}
 (x+\varepsilon v,\ y-\varepsilon u)J\binom{x}{y}
 =\varepsilon(ux+vy).
\end{equation}
These identities explain the cancellation of the negative
powers of $\varepsilon$ in both approximations.

Suppose first that $d=2k+3$.
The product in the overview is
\[
 (x+\varepsilon v,\ y-\varepsilon u)
 (G^{\mathsf T})^k
 (D+\varepsilon^{-1}zJ)
 G^k\binom{x}{y}.
\]
Its part involving $D$ equals
\[
 a\xi_k^2+b\eta_k^2
 +\varepsilon(v,-u)(G^{\mathsf T})^kDG^k\binom{x}{y}.
\]
By Equations~\eqref{eq:abp-alternating-form}
and~\eqref{eq:abp-endpoint-cancellation}, its remaining
part is exactly $z\Delta^k(ux+vy)$.
The output is therefore
$F_{2,2k+3}+\varepsilon H_{2,2k+3}$, where
\[
 H_{2,2k+3}
 :=(v,-u)(G^{\mathsf T})^kDG^k\binom{x}{y}.
\]

Now suppose that $d=2k+4$.
The middle matrix of the approximation can be written as
\[
 \begin{pmatrix}
 a+r/\varepsilon&s/\varepsilon\\
 -p/\varepsilon&b-q/\varepsilon
 \end{pmatrix}
 =D+\varepsilon^{-1}\operatorname{adj}(G^{\mathsf T})J,
\]
where $\operatorname{adj}$ denotes the adjugate matrix.
The identity
$G^{\mathsf T}\operatorname{adj}(G^{\mathsf T})=\Delta I_2$
gives
\[
 (G^{\mathsf T})^{k+1}\operatorname{adj}(G^{\mathsf T})JG^k
 =\Delta^{k+1}J.
\]
Consequently, the part with coefficient
$\varepsilon^{-1}$ contributes
$\Delta^{k+1}(ux+vy)$.
The part involving $D$ contributes
\[
 a\xi_{k+1}\xi_k+b\eta_{k+1}\eta_k
 +\varepsilon(v,-u)(G^{\mathsf T})^{k+1}DG^k\binom{x}{y}.
\]
Thus the output is
$F_{2,2k+4}+\varepsilon H_{2,2k+4}$, where
\[
 H_{2,2k+4}
 :=(v,-u)(G^{\mathsf T})^{k+1}DG^k\binom{x}{y}.
\]

For every nonzero algebraic $\varepsilon$, all factors in
these products have homogeneous linear entries.
When matrix powers are expanded into successive factors,
the odd-degree product has $1+k+1+k+1=2k+3$ factors,
and the even-degree product has
$1+(k+1)+1+k+1=2k+4$ factors.
Powers with exponent zero contribute no factors.
Both products therefore define ABPs of width at most two
and exactly the required degree.

For general $w$, add $w-2$ paths, sharing only the source
and sink with the width-two ABP and with each other.
The $i$-th path has $d$ edges labelled
$c_i,z_i,\ldots,z_i$.
The resulting ABP has width at most $w$ and output
\[
 F_{w,d}+\varepsilon H_{2,d}.
\]
This also covers $w=2$, when there are no additional paths.

The coefficients of this output are polynomials in
$\varepsilon$. For every nonzero algebraic
$\varepsilon$, the output lies in $\mathcal A_{w,d}$.
Since $\overline{\Q}\setminus\{0\}$ is Zariski dense
in the affine line, its value at $\varepsilon=0$ lies
in $\overline{\mathcal A_{w,d}}$.
That value is $F_{w,d}$, proving the claim.
\end{proof}

We now establish the algebraic property used for the lower
bound. The aim is to control the common zero sets of the
first and last layers of a putative ABP.
We begin with two linear-algebra calculations and then
apply them to the polynomial $F_{w,d}$.

\begin{claim}\label{clm:abp-singular-matrix-space}
Let $U$ be a linear subspace of $\mat_2(\overline{\Q})$
consisting of singular matrices. Then $\dim U\le2$.
If $\dim U=2$, then $U$ contains a rank-one idempotent.
\end{claim}
\begin{proof}
If $U$ is nonzero, choose a nonzero matrix in $U$.
It has rank one. After invertible changes of row and
column bases, we may assume that it is
$\left(\begin{smallmatrix}1&0\\0&0\end{smallmatrix}\right)$.
For any $B=(b_{ij})\in U$, the determinant of
\[
 \begin{pmatrix}1&0\\0&0\end{pmatrix}+tB
\]
vanishes for every $t$. The coefficient of $t$ shows
that $b_{22}=0$. Since $B$ itself is singular,
we also have $b_{12}b_{21}=0$.

The two entries $b_{12}$ and $b_{21}$ are linear
functions on $U$. Their product is identically zero,
so one of them is identically zero.
Thus every matrix in $U$ has either its second row zero
or its second column zero, with the same alternative
holding throughout $U$. In particular, $\dim U\le2$.

Returning to the original bases, this says that $U$
is contained in a space of one of the forms
\[
 \{u\varphi:\varphi\in\overline{\Q}^{\,1\times2}\},
 \qquad
 \{u\varphi:u\in\overline{\Q}^{\,2\times1}\},
\]
where the column $u$ is fixed and nonzero in the first
case, and the row $\varphi$ is fixed and nonzero in
the second.
If $\dim U=2$, it equals the corresponding displayed
space. We may then choose $u,\varphi$ with
$\varphi u=1$. The matrix $u\varphi$ has rank one and
satisfies $(u\varphi)^2=u\varphi$, as required.
\end{proof}

The polynomial $F_{2,d}$ is linear in the four variables
$a,b,u,v$. Write
\begin{equation}\label{eq:abp-four-coefficients}
 F_{2,d}=aQ_a+bQ_b+uQ_u+vQ_v,
\end{equation}
where the four coefficient polynomials depend only on
$x,y,z,p,q,r,s$.
For $d=2k+3$, they are
\[
 Q_a=\xi_k^2,\qquad Q_b=\eta_k^2,
 \qquad Q_u=z\Delta^kx,\qquad Q_v=z\Delta^ky.
\]
For $d=2k+4$, they are
\[
 Q_a=\xi_{k+1}\xi_k,\qquad Q_b=\eta_{k+1}\eta_k,
 \qquad Q_u=\Delta^{k+1}x,\qquad Q_v=\Delta^{k+1}y.
\]
All four are homogeneous of degree $d-1$.

\begin{claim}\label{clm:abp-core-restriction}
Let $L\subseteq\overline{\Q}^{\,7}$ be a linear
subspace in the coordinates $x,y,z,p,q,r,s$, with
$\dim L\ge5$. If
\[
 \dim\operatorname{span}_{\overline{\Q}}
 \{Q_a|_L,Q_b|_L,Q_u|_L,Q_v|_L\}
 \le\dim L-5,
\]
then $L=\{x=y=0\}$.
\end{claim}
\begin{proof}
The restrictions in the claim are polynomial functions
on $L$. We consider its three possible dimensions.

If $\dim L=7$, then $L$ is the whole space.
The polynomials $Q_a$ and $Q_b$ are linearly independent:
setting $G=I_2$ gives $x^2$ and $y^2$, respectively.
Both have degree two in the pair of variables $x,y$,
whereas the nonzero polynomial $Q_u$ has degree one
in that pair. Hence these three polynomials are
linearly independent. This contradicts the assumed
bound $\dim L-5=2$.

Suppose next that $\dim L=6$, so that $L$ is a
hyperplane. Write $Q_u=\gamma x$ and $Q_v=\gamma y$,
where
\[
 \gamma=
 \begin{cases}
 z\Delta^k,&d=2k+3,\\
 \Delta^{k+1},&d=2k+4.
 \end{cases}
\]
If $\gamma|_L\ne0$ and $x|_L,y|_L$ are linearly
independent, then $Q_u|_L,Q_v|_L$ are linearly
independent. Indeed, the coordinate ring of a linear
subspace is an integral domain, so multiplication by
the nonzero polynomial $\gamma|_L$ preserves linear
independence.

If $\gamma|_L\ne0$ but $x|_L,y|_L$ are dependent,
then $L$ is defined by a nonzero linear equation
in $x,y$ alone. One independent coordinate remains
from this pair, and all the other coordinates are free.
At least one of $Q_a|_L,Q_b|_L$ is nonzero, as is seen
by setting $G=I_2$.
It has degree two in the remaining coordinate from
$x,y$, whereas a nonzero member of
$Q_u|_L,Q_v|_L$ has degree one in it.
These two restrictions are therefore linearly independent.

Finally, suppose that $\gamma|_L=0$.
The polynomial $\Delta=ps-qr$ has no linear factor:
viewed as a polynomial in $p$, its coefficients $s$
and $-qr$ have no common nonconstant divisor.
Thus $\Delta$ cannot vanish identically on a hyperplane.
The only remaining possibility is odd degree and
$L=\{z=0\}$. On this hyperplane $Q_a,Q_b$ are still
linearly independent, again by setting $G=I_2$.
In every case the span has dimension at least two,
contradicting the assumed bound $\dim L-5=1$.

It remains to consider $\dim L=5$.
The dimension bound now says that all four
coefficient polynomials vanish identically on $L$.
Suppose first that $\Delta|_L\ne0$.
At every point of $L$ where $G$ is invertible, the
odd-degree equalities $\xi_k^2=\eta_k^2=0$ force
$x=y=0$.
In even degree, the equalities
$\Delta^{k+1}x=\Delta^{k+1}y=0$ have the same consequence.
Such points form a dense subset of $L$, so the linear
functions $x$ and $y$ vanish on all of $L$.
Both $L$ and $\{x=y=0\}$ have dimension five,
and therefore they are equal.

Suppose instead that $\Delta|_L=0$.
Project $L$ onto the four coordinates of $G$, and
denote the image by $U$.
The kernel of this projection has dimension at most
three, since only $x,y,z$ are forgotten.
Thus $\dim U\ge2$.
Every matrix in $U$ is singular, so
Claim~\ref{clm:abp-singular-matrix-space} gives
$\dim U=2$ and a rank-one idempotent $E\in U$.
The projection kernel has dimension three and hence
contains all choices of $x,y,z$ with $G=\bm0$.
It follows that, for each fixed $G\in U$, all values
of $x,y,z$ occur in $L$.

Choose a nonzero column $\binom{x}{y}$ fixed by $E$,
and take $G=E$. Then
$G^j\binom{x}{y}=\binom{x}{y}$ for every $j\ge0$.
In both parity cases this gives $Q_a=x^2$ and
$Q_b=y^2$, which are not both zero at this point.
This contradicts their vanishing on $L$.
The claim follows.
\end{proof}

We now return to all $2w+7$ variables of $F_{w,d}$.
Let $W_0$ be the linear subspace defined by
\[
 x=y=0,\qquad c_i=z_i=0\quad(1\le i\le w-2).
\]
It has dimension nine; its free coordinates are
$a,b,u,v,z,p,q,r,s$.

\begin{claim}\label{clm:abp-zero-subspaces-appendix}
Every linear subspace of codimension at most $w$
on which $F_{w,d}$ vanishes contains $W_0$.
\end{claim}
\begin{proof}
It is enough to consider codimension exactly $w$.
Indeed, a subspace of smaller codimension contains
one of codimension $w$, and vanishing is preserved
when passing to a smaller subspace.
Let $W$ have codimension $w$, so that $\dim W=w+7$.

Project $W$ by forgetting the coefficient variables
$a,b,u,v,c_1,\ldots,c_{w-2}$, and write $L$ for its
image. These are $w+2$ forgotten coordinates.
The projection kernel has dimension $w+7-\dim L$.
Each vector in this kernel gives a linear relation,
with constant coefficients, among the following
polynomial functions on $L$:
\[
 Q_a|_L,\ Q_b|_L,\ Q_u|_L,\ Q_v|_L,
 \ z_1^{d-1}|_L,\ldots,z_{w-2}^{d-1}|_L.
\]
To see this, take any point of $W$ and add a kernel
vector. Only the coefficient variables change.
Subtracting the two zero values of $F_{w,d}$ gives
the indicated relation. Every point of $L$ has a
lift to $W$, so the relation holds throughout $L$.
Independent kernel vectors give independent relation
vectors. Consequently,
\begin{equation}\label{eq:abp-restriction-dimension}
 \dim\operatorname{span}
 \{Q_a|_L,Q_b|_L,Q_u|_L,Q_v|_L,
       z_1^{d-1}|_L,\ldots,z_{w-2}^{d-1}|_L\}
 \le\dim L-5.
\end{equation}
All spans in this proof are over $\overline{\Q}$.

Next project $L$ onto the seven coordinates
$x,y,z,p,q,r,s$. Write $L_0$ for its image and $T$
for the projection kernel.
The functions $Q_a,Q_b,Q_u,Q_v$ vanish on $T$,
since all their variables are zero there.
The coordinate functions $z_i$ span the dual of $T$.
Choose $\dim T$ of them which are linearly independent.
Their $(d-1)$-st powers are also linearly independent:
in these coordinates they are distinct monomials.
Thus restriction to $T$ maps the span in
Equation~\eqref{eq:abp-restriction-dimension} onto
a space of dimension at least $\dim T$.
The kernel of this restriction contains the span of
$Q_a|_L,Q_b|_L,Q_u|_L,Q_v|_L$.
Since these functions depend only on the projection
to $L_0$, their span has the same dimension as the
span of their restrictions to $L_0$.
We obtain
\[
 \dim T+
 \dim\operatorname{span}\{Q_a|_{L_0},Q_b|_{L_0},
                              Q_u|_{L_0},Q_v|_{L_0}\}
 \le\dim L-5.
\]
Using $\dim L=\dim T+\dim L_0$, this becomes
\[
 \dim\operatorname{span}\{Q_a|_{L_0},Q_b|_{L_0},
                              Q_u|_{L_0},Q_v|_{L_0}\}
 \le\dim L_0-5.
\]
In particular, $\dim L_0\ge5$.
Claim~\ref{clm:abp-core-restriction} therefore gives
$L_0=\{x=y=0\}$. Hence $x=y=0$ on $W$.

On $W$, the polynomial $F_{w,d}$ now reduces to
$\sum_{i=1}^{w-2}c_i z_i^{d-1}$.
Let $V$ be the projection of $W$ onto the coordinates
$(c_i,z_i)_{i=1}^{w-2}$.
We show that $\dim V\le w-2$.
Project $V$ onto its $z_i$-coordinates, and let $j$
be the dimension of this image.
The kernel has dimension $\dim V-j$ and supplies
that many independent linear relations among the
restrictions of $z_1^{d-1},\ldots,z_{w-2}^{d-1}$
to the image.
Their span therefore has dimension at most
$(w-2)-(\dim V-j)$.
On the other hand, $j$ independent coordinate
functions have independent $(d-1)$-st powers,
so the same span has dimension at least $j$.
Thus
\[
 j\le (w-2)-(\dim V-j),
 \qquad\text{and hence}\qquad \dim V\le w-2.
\]
For $w=2$, these coordinate spaces are zero-dimensional
and the same conclusion holds.

It follows that the kernel of the projection
$W\to V$ has dimension at least
$(w+7)-(w-2)=9$.
We already know that $x=y=0$ on $W$.
In this kernel all $c_i,z_i$ also vanish, so the
kernel is contained in the nine-dimensional space
$W_0$. It must therefore equal $W_0$.
In particular, $W_0\subseteq W$, as required.
\end{proof}

We now prove that $F_{w,d}$ cannot be computed by ABPs of width $w$ and degree $d$.  

\ABPWitnessClass*
\begin{proof}
Suppose that $\iota(F_{w,d})\sim\bm0$.
By Claim~\ref{clm:universal-abp-congruence} and the
injectivity of $\iota$, the polynomial $F_{w,d}$ is
computed by an ABP of width at most $w$ and degree $d$.
If $m>2w+7$, set all the unused variables to zero.
The resulting ABP still computes $F_{w,d}$, has
exactly $d$ homogeneous linear matrix factors, and
has width at most $w$.
We therefore work in the $2w+7$ variables used in
its definition.

Let the first and last layers consist of the linear
forms $\ell_1,\ldots,\ell_w$ and
$r_1,\ldots,r_w$, allowing zero forms if necessary.
The polynomial $F_{w,d}$ vanishes on the common
zero set of the $\ell_i$, and also on the common
zero set of the $r_i$.
Each of these zero sets is a linear subspace of
codimension at most $w$.
Claim~\ref{clm:abp-zero-subspaces-appendix} implies
that both contain $W_0$.
Hence every $\ell_i$ and every $r_i$ vanishes on
$W_0$. Since these forms are linear, each belongs
to the linear span of
\[
 x,y,c_1,z_1,\ldots,c_{w-2},z_{w-2}.
\]

Every path in the ABP contains one first-layer form
and one last-layer form. After expansion, each of its
monomials therefore has total degree at least two
in the displayed variables.
Taking the sum over all paths cannot create a monomial
of smaller degree in those variables.

However, in $F_{w,d}$ the sum of the monomials having
degree exactly one in those variables is
\[
 \begin{cases}
 z\Delta^k(ux+vy),&d=2k+3,\\
 \Delta^{k+1}(ux+vy),&d=2k+4.
 \end{cases}
\]
It is nonzero in both cases.
The summands involving $a,b$ have degree two in
$x,y$, and each $c_i z_i^{d-1}$ has degree $d$ in
the displayed variables. Thus none of them can
cancel this degree-one part.
This contradicts the ABP representation and proves
$\iota(F_{w,d})\not\sim\bm0$.
\end{proof}

By Claim~\ref{clm:universal-abp-congruence}, the last
claim gives $F_{w,d}\notin\mathcal A_{w,d}$.
Claim~\ref{clm:abp-approximation} gives
$F_{w,d}\in\overline{\mathcal A_{w,d}}$.
This proves Theorem~\ref{theo:homogeneous-abp-separation}.

For completeness, the description of $\sim$ also determines
the number of classes in the matrix closure.
For every nonzero $\lambda\in\overline{\Q}$,
the polynomial $\lambda F_{w,d}$ is still in
$\overline{\mathcal A_{w,d}}\setminus\mathcal A_{w,d}$.
Indeed, scalar multiplication preserves the ABP class
by scaling its first layer, and hence preserves its
closure; an ABP for $\lambda F_{w,d}$ could be scaled
back to one for $F_{w,d}$.
The distinct matrices $\iota(\lambda F_{w,d})$
therefore give infinitely many singleton classes.
Consequently,
\[
 |\langle X\rangle/{\sim}|=4,
 \qquad
 |\cl{\langle X\rangle}/{\sim}|=\infty.
\]
\end{proof}

\section{Detailed Proofs of the Examples}
\label{app:lower}

We collect here the calculations and proofs for the examples in
the preliminaries and overview. The final example gives an
infinite Zariski closure to which
Theorem~\ref{TheoremMainQuotient} applies.

\subsection{Example~\ref{example1prelim}}

For the matrices in that example, direct multiplication gives
$A_xA_y=A_y$ for all $x,y$. All these matrices have rank one, so every pair
preserves rank, including when $x=y$.
Under the row-vector convention,
\[
 \ker(A_x)=\{(0,t):t\in\Q\},\qquad
 \im(A_x)=\operatorname{Span}_{\Q}\{(1,x)\}.
\]
The images are distinct when $x\ne y$, so $\{A_x,A_y\}$ is
rank-stable but not stable. Finally, $A$ has rank one and
$A^2=0$. It is therefore not rank-stable, and the singleton
$\{A\}$ is not a stable set.

\subsection{Example~\ref{exm:generator-lower}}

Recall that $N,m\ge2$, $B=m^2+1$, and
\[
 A_i=\begin{pmatrix}B&0&Bi\\0&1&0\\0&i&0\end{pmatrix}
 \quad(1\le i\le m),\qquad
 W=(A_1\cdots A_m)^N.
\]
We first verify uniqueness of representation and identify the
kernels and images of products.

\begin{claim}\label{clm:generator-encoding}
For $i_1,\ldots,i_k\in\{1,\ldots,m\}$,
\begin{equation}\label{eq:generator-product-formula}
 A_{i_1}\cdots A_{i_k}=
 \begin{pmatrix}
 B^k&\displaystyle\sum_{t=1}^{k-1}B^t i_ti_{t+1}&B^k i_k\\
 0&1&0\\
 0&i_1&0
 \end{pmatrix}.
\end{equation}
The displayed matrix determines the entire sequence
$i_1,\ldots,i_k$. Every such matrix has rank two, kernel
$\operatorname{Span}_{\Q}\{(0,-i_1,1)\}$, and image
$\{(x,y,i_kx):x,y\in\Q\}$.
\end{claim}
\begin{proof}
The formula holds for $k=1$, with empty sum zero.
Multiplication on the right by $A_j$ multiplies the first entry
by $B$, adds $B^k i_kj$ to the second entry of the first row,
and gives last entry $B^{k+1}j$. The other two rows remain as
displayed. This proves the formula by induction.

The first diagonal entry determines $k$, and the entry in
position $(3,2)$ determines $i_1$.
For $k\ge2$, the base-$B$ digits of the entry in position $(1,2)$
determine all numbers $i_ti_{t+1}$, because
$1\le i_ti_{t+1}\le m^2<B$.
Starting with $i_1$, successive division by the already recovered
$i_t$ determines $i_{t+1}$. This also covers uniqueness for $k=1$
using $i_1$ alone.
Finally, the second and third rows are dependent and the first
row is independent of them, proving rank two. Solving for a row
vector annihilated by the matrix gives the stated kernel.
The first row divided by $B^k$, together with $(0,1,0)$, spans
exactly the stated image.
\end{proof}

Let $I$ be the set of matrices in $\langle X_m\rangle$ that do
not occur as factors of the displayed product $W$.
Uniqueness of representation implies that $I$ is a two-sided
ideal. Indeed, if a product containing the unique representation
of some $z\in I$ occurred as a factor of $W$, the representation
of $z$ would itself occur as a factor of $W$, a contradiction.
Thus multiplying a matrix of $I$ on either side by any matrix
of $\langle X_m\rangle$ keeps it in $I$.
Collapsing $I$ to one class and leaving every matrix outside $I$
in its own singleton class is consequently a semigroup congruence.

\begin{claim}\label{clm:generator-stable-bound}
Every stable subsemigroup meets at most $N+1$ classes of this
congruence, and one meets exactly $N+1$ classes.
\end{claim}
\begin{proof}
By Claim~\ref{clm:generator-encoding}, a common kernel fixes the
first generator index $i$, and a common image fixes the last
generator index $j$. For fixed $i,j$, a factor of $W$ is determined
by its length: the indices follow the periodic sequence
$1,\ldots,m,1,\ldots,m,\ldots$.
Its length lies in one residue class modulo $m$, as its first
and last indices are fixed. Among $1,\ldots,Nm$, there are
exactly $N$ lengths in each residue class, so there are at most
$N$ distinct such factors of $W$.
Any stable subsemigroup therefore meets at most $N$ singleton
classes and possibly the large class $I$.

For equality, put $C_0=A_1\cdots A_m$ and consider
$\langle C_0\rangle$.
All of its matrices have the same kernel and image, by
\eqref{eq:generator-product-formula}, and all products have rank
two. Thus it is stable. The matrices $C_0,\ldots,C_0^N$ are
distinct factors of $W$, and $C_0^{N+1}\in I$.
It follows that this stable subsemigroup meets exactly $N+1$
classes.
\end{proof}

The class of $W$ is a singleton and $W$ has length $Nm$, by
uniqueness of representation. Every other singleton class is
represented by a factor of $W$ of length at most $Nm$.
The large class contains $A_1^2$: since $m\ge2$, the periodic
product $W$ never has two consecutive occurrences of $A_1$.
That class has length at most $2\le Nm$.
Hence the diameter of the quotient is exactly $Nm=N|X_m|$,
while its smallest uniform stable bound is $N+1$.

\subsection{Example~\ref{exm:shift-separation}}

The generators act on row vectors by
\[
 (x_1,\ldots,x_d)A_i=(x_1,x_3,\ldots,x_d,ix_1).
\]
Iterating this action, for $1\le k\le d-1$, gives
\[
 (x_1,\ldots,x_d)A_{i_1}\cdots A_{i_k}
   =(x_1,x_{k+2},\ldots,x_d,i_1x_1,\ldots,i_kx_1),
\]
where the middle list is empty when $k=d-1$.
The rank is therefore $d-k$ in this range.
Every generator preserves the first coordinate, whereas every
product of $d-1$ generators depends only on that coordinate.
Consequently, for $k\ge d-1$,
\[
 a_1\cdots a_k=a_{k-d+2}\cdots a_k,
\]
and its rank is one. This proves the rank formula
$\max\{1,d-k\}$.

Let $S$ be the set of rank-one matrices in $\langle X\rangle$.
They are precisely the $m^{d-1}$ distinct matrices
\[
 P(i_1,\ldots,i_{d-1})=
 \begin{pmatrix}
 1&i_1&\cdots&i_{d-1}\\
 0&0&\cdots&0\\
 \vdots&\vdots&&\vdots\\
 0&0&\cdots&0
 \end{pmatrix},\qquad 1\le i_j\le m.
\]
For $P,Q\in S$, multiplication gives $PQ=Q$.
Hence $S$ is a rank-stable subsemigroup, and every matrix in it
is idempotent.
If a product $a$ of $k$ generators has rank greater than one,
then $k<d-1$, and
\[
 \rk(a^2)=\max\{1,d-2k\}<d-k=\rk(a).
\]
It cannot belong to any rank-stable subsemigroup. Thus all
rank-stable subsemigroups are contained in $S$.
For the equality congruence, the smallest uniform bound on their
quotient sizes is exactly $|S|=m^{d-1}$.

The images of the matrices in $S$ are the lines spanned by
$(1,i_1,\ldots,i_{d-1})$.
These lines are pairwise distinct: proportional vectors whose
first coordinate is one must be equal.
Every stable subsemigroup is contained in $S$, so the common-image
condition forces it to be a singleton. Conversely, every
singleton in $S$ is stable, since its matrix is idempotent.
The smallest uniform stable bound is therefore exactly one.

\subsection{Example~\ref{prop:counter}}

Index the rows and columns of a $2\times2$ matrix by $0,1$.
Write $E_{uv}$ for the matrix with a single $1$ in position
$(u,v)$ and zeros elsewhere. Thus
$E_{uv}E_{st}=E_{ut}$ when $v=s$, and is zero otherwise.
Recall that $X_k=\{A_1,\ldots,A_k\}$, where the $j$th diagonal
block of $A_i$ is $E_{10}$ for $j<i$, $E_{01}$ for $j=i$,
and $I_2$ for $j>i$.
The target is $T_k=\operatorname{diag}(E_{01},\ldots,E_{01})$.

Every diagonal block of every product belongs to
\[
 \mathcal U=\{0,I_2,E_{00},E_{01},E_{10},E_{11}\},
\]
which is closed under multiplication. All off-diagonal blocks
are zero. Thus $\langle X_k\rangle$ is finite, with at most $6^k$
matrices, and all entries of all products belong to $\{0,1\}$.

\begin{claim}\label{clm:counter-length-revised}
The matrix $T_k$ has length exactly $2^k-1$ with respect to $X_k$.
\end{claim}
\begin{proof}
Consider any product representing $T_k$. Starting with
$e_0=(1,0)$ in each diagonal block, track its image under the
successive generators. At the end, each tracked row equals
$e_1=(0,1)$. None can become zero at an intermediate step,
since subsequent multiplication cannot restore a zero row.
At every step the tracked rows therefore determine a bit pattern
$(b_1,\ldots,b_k)$, where the row in block $j$ is $e_{b_j}$.
Initially all bits are zero.

Multiplication by $A_i$ keeps all tracked rows nonzero exactly
when $b_1=\cdots=b_{i-1}=1$ and $b_i=0$.
It then changes the first $i-1$ bits to zero, bit $i$ to one,
and leaves the higher bits unchanged. Its index must therefore
be the least index whose bit is zero.
The integer $\sum_{j=1}^k2^{j-1}b_j$ increases by exactly one
at each step. Reaching the all-one pattern from the all-zero
pattern requires $2^k-1$ steps. This proves the lower bound.

To exhibit a product of that length, define matrices recursively
by $C_1=A_1$ and $C_j=C_{j-1}A_jC_{j-1}$ for $2\le j\le k$.
Using $E_{01}E_{10}E_{01}=E_{01}$, induction shows that $C_j$
has its first $j$ diagonal blocks equal to $E_{01}$ and its
remaining blocks equal to $I_2$. Hence $C_k=T_k$.
Expanding the recursion into generators gives lengths
$n_1=1$ and $n_j=2n_{j-1}+1$, so $n_k=2^k-1$.
\end{proof}

\begin{claim}\label{clm:counter-stability-revised}
Every rank-stable subsemigroup of $\langle X_k\rangle$ is a
singleton.
\end{claim}
\begin{proof}
Among the matrices in $\mathcal U$, the only ones whose rank drops
on squaring are $E_{01}$ and $E_{10}$.
Ranks add across diagonal blocks and cannot increase on squaring.
Consequently every rank-stable matrix in $\langle X_k\rangle$
is diagonal, with only zeros and ones on its diagonal.
For two such matrices $p,q$, let their supports be the sets of
diagonal positions containing a one. Their ranks are the sizes
of their supports, and $\rk(pq)$ is the size of the intersection.
Thus $\rk(p)=\rk(q)=\rk(pq)$ forces the supports, and hence the
matrices, to be equal. Every rank-stable subsemigroup is therefore
a singleton. Such subsemigroups exist, for example
$\{A_1^2\}$, since $A_1^2=\operatorname{diag}(0,I_2,\ldots,I_2)$
is idempotent.
\end{proof}

With equality as congruence, the smallest uniform bound on
rank-stable subsemigroups is thus $N=1$.
By Claim~\ref{clm:counter-length-revised}, the diameter is at least
$2^k-1=2^{d/2}-1$ in dimension $d=2k$, as asserted.

\subsection{Example~\ref{exam:counter-cont}}

We first consider the finite-closure case.
The semigroup $\langle X_k\rangle$ is finite, so it is Zariski
closed in $\mat_{2k}(\overline\Q)$: singletons are closed, and
a finite union of closed sets is closed.
Every matrix group is rank-stable. Indeed, if its identity is
$e$, then $g=eg=ge$ implies $\rk(g)\le\rk(e)$, and
$e=gg^{-1}$ gives the reverse inequality; the same applies to
every product in the group.
Claim~\ref{clm:counter-stability-revised} therefore implies that
every group in this closure is trivial.
The equality congruence satisfies the hypothesis of
Theorem~\ref{TheoremMainQuotient} with $N=1$, while $T_k$ still
has length $2^k-1$. In dimension $d=2k$, the diameter is therefore
at least $2^{d/2}-1$.

For the infinite-closure case, add one coordinate to each
generator and put
\[
 B_i=\operatorname{diag}(2,A_i),\qquad
 Y_k=\{B_1,\ldots,B_k\}
       \subseteq\mat_{2k+1}(\Q).
\]
Write $C_k:=\cl{\langle Y_k\rangle}$ for the closure in
$\mat_{2k+1}(\overline\Q)$.
Every product of $\ell$ generators from $Y_k$ has the form
\[
 B_{i_1}\cdots B_{i_\ell}
   =\operatorname{diag}(2^\ell,A_{i_1}\cdots A_{i_\ell}).
\]
In particular, powers of any $B_i$ are distinct, so
the generated semigroup and its closure are infinite.
Let $S_k=\langle X_k\rangle$, which is finite. The set
\[
 Z_k:=\{\operatorname{diag}(t,B):t\in\overline\Q,\ B\in S_k\}
\]
is a finite union of affine lines, and hence is Zariski closed.
It contains $\langle Y_k\rangle$, so $C_k\subseteq Z_k$.
Projection onto the lower-right block defines a semigroup
homomorphism $\pi:C_k\to S_k$.
Define $a\sim b$ if $\pi(a)=\pi(b)$. This is a congruence on
the entire closure, identifying matrices exactly when their
lower-right blocks are equal.

If $G\le C_k$ is a group, its homomorphic image $\pi(G)$ is a
matrix group in $S_k$. Every such group is trivial, as proved
above. Thus $|G/{\sim}|=|\pi(G)|=1$.
The restriction of $\pi$ to $\langle Y_k\rangle$ is onto $S_k$,
and induces an isomorphism
\[
 \langle Y_k\rangle/{\sim}\ \cong\ S_k
\]
that sends the class of $B_i$ to $A_i$.
Representations in these corresponding generators have the same
lengths. The class mapping to $T_k$ therefore has length exactly
$2^k-1$. Thus Theorem~\ref{TheoremMainQuotient} applies with
$N=1$ and infinite closure, while in dimension $d=2k+1$ the
quotient has diameter at least $2^{(d-1)/2}-1$.

In fact this closure contains an infinite group whose quotient
is a singleton. Put
$E=A_1^2=\operatorname{diag}(0,I_2,\ldots,I_2)$.
For every $n\ge2$, the matrix $B_1^n$ equals
$\operatorname{diag}(2^n,E)$. These infinitely many points are
Zariski dense in the line
$\{\operatorname{diag}(t,E):t\in\overline\Q\}$, because a nonzero
polynomial in one variable has only finitely many roots.
Hence $C_k$ contains the group
\[
 \{\operatorname{diag}(t,E):t\in\overline\Q\setminus\{0\}\},
\]
whose identity is $\operatorname{diag}(1,E)$ and whose quotient
under $\sim$ is a singleton.

Both cases show that exponential dependence on $d$ is necessary.
These lower bounds do not establish the optimality of the
exponent $O(d^2)$ in the theorem's upper bound.

\end{document}